\documentclass[10pt]{article}
\usepackage{fullpage,citesort,epsfig,graphics,amsbsy}
\usepackage{psfrag}
\usepackage[normal,small]{caption}
\newcommand{\beq}{\begin{equation}}
\newcommand{\eeq}{\end{equation}}
\newcommand{\bea}{\begin{eqnarray}}
\newcommand{\eea}{\end{eqnarray}}

\newcommand{\be}{\begin{equation}}
\newcommand{\ee}{\end{equation}}
\newcommand{\bq}{\begin{eqnarray}}
\newcommand{\eq}{\end{eqnarray}}

\def\math{\mathsurround=0pt }
\def\leftrightarrowfill{$\math \mathord\gets \mkern-6mu \cleaders\hbox{$\mkern-2mu \mathord- \mkern-2mu$}\hfill
 \mkern-6mu \mathord\to$}
\def\overleftrightarrow#1{\vbox{\ialign{##\crcr
     \leftrightarrowfill\crcr\noalign{\kern-1pt\nointerlineskip}
     $\hfil\displaystyle{#1}\hfil$\crcr}}}

\newcommand{\bfs}{\boldsymbol}

\let\l=\lambda

\def\nn{\nonumber} \def\bd{\begin{document}} \def\ed{\end{document}}
\def\ds{\documentstyle} \let\fr=\frac \let\bl=\bigl \let\br=\bigr
\let\Br=\Bigr \let\Bl=\Bigl
\let\bm=\bibitem
\let\na=\nabla
\let\pa=\partial \let\ov=\overline
\def\ft#1#2{{\textstyle{{\scriptstyle #1}\over {\scriptstyle #2}}}}
\def\fft#1#2{{#1 \over #2}}
\def\vp{\varphi}
\def\sst#1{{\scriptscriptstyle #1}}
\def\oneone{\rlap 1\mkern4mu{\rm l}}
\def\td{\tilde}
\def\wtd{\widetilde}
\def\dalemb#1#2{{\vbox{\hrule height .#2pt
        \hbox{\vrule width.#2pt height#1pt \kern#1pt
                \vrule width.#2pt}
        \hrule height.#2pt}}}
\def\square{\mathord{\dalemb{6.8}{7}\hbox{\hskip1pt}}}
\def\wtd{\widetilde}
\def\R{\rlap{\rm I}\mkern3mu{\rm R}}
\def\im{{\rm i}}
\def\tilg{\tilde{g}}
\def\tilF{\tilde{F}}
\def\tilA{\tilde{A}}
\def\varf{\varphi}
\def\tilf{\tilde{\phi}}
\def\tilh{\tilde{h}}
\def\rme{{\rm e}}
\def\ep{\epsilon}
\def\0{{(0)}}
\def\9{{(9)}}
\def\8{{(8)}}
\def\7{{(7)}}
\def\6{{(6)}}
\def\5{{(5)}}
\def\4{{(4)}}
\def\3{{(3)}}
\def\2{{(2)}}
\def\1{{(1)}}
\newcommand{\trace}{{\rm Tr}}
\newcommand{\ub}{\overline{U}}
\newcommand{\vb}{\overline{V}}
\newcommand{\uh}{\widehat{U}}
\newcommand{\vh}{\widehat{V}}
\newcommand{\ubh}{\overline{\widehat{U}}}
\newcommand{\vbh}{\overline{\widehat{V}}}
\newcommand{\lb}{\bar{\l}}
\newcommand{\Fb}{\overline{F}}
\newcommand{\Fh}{\widehat{F}}
\newcommand{\Fbh}{\overline{\widehat{F}}}
\newcommand{\Ab}{\overline{A}}
\newcommand{\Ah}{\widehat{A}}
\newcommand{\Abh}{\overline{\widehat{A}}}
\newcommand{\Gb}{\overline{G}}
\newcommand{\Gh}{\widehat{G}}
\newcommand{\Gbh}{\overline{\widehat{G}}}
\newcommand{\Pb}{\overline{P}}
\newcommand{\Ph}{\widehat{P}}
\newcommand{\Pbh}{\overline{\widehat{P}}}
\newcommand{\Qb}{\overline{Q}}
\newcommand{\Qh}{\widehat{Q}}
\newcommand{\Qbh}{\overline{\widehat{Q}}}
\newcommand{\Bb}{\overline{B}}
\newcommand{\Bh}{\widehat{B}}
\newcommand{\Bbh}{\overline{\widehat{B}}}
\newcommand{\fhns}{\hat{F}^{\rm (NS)}}
\newcommand{\fhrr}{\hat{F}^{\rm (RR)}}
\newcommand{\ahns}{\hat{A}^{\rm (NS)}}
\newcommand{\ahrr}{\hat{A}^{\rm (RR)}}
\newcommand{\hhrr}{\hat{H}^{\rm (RR)}}
\newcommand{\hchi}{\hat{\chi}}
\newcommand{\hphi}{\hat{\phi}}
\newcommand{\htau}{\hat{\tau}}
\newcommand{\cG}{{\cal G}}
\newcommand{\cGb}{\overline{{\cal G}}}
\newcommand{\cH}{{\cal H}}
\newcommand{\cP}{{\cal P}}
\newcommand{\cPb}{\overline{{\cal P}}}
\newcommand{\cQ}{{\cal Q}}
\newcommand{\cQb}{\overline{{\cal Q}}}
\newcommand{\cM}{{\cal M}}
\newcommand{\cN}{{\cal N}}
\newcommand{\cO}{{\cal O}}
\newcommand{\cD}{{\cal D}}
\newcommand{\cL}{{\cal L}}
\newcommand{\cA}{{\cal A}}
\newcommand{\cB}{{\cal B}}
\newcommand{\hg}{\hat{g}}
\newcommand{\cE}{{\cal E}}

\newcommand{\vpp}{\mbox{$\langle{\scriptstyle++}\rangle$}}
\newcommand{\vmp}{\mbox{$\langle{\scriptstyle-+}\rangle$}}
\newcommand{\vppp}{\mbox{$\langle{\scriptstyle+++}\rangle$}}
\newcommand{\vmpp}{\mbox{$\langle{\scriptstyle-++}\rangle$}}
\newcommand{\vpmp}{\mbox{$\langle{\scriptstyle+-+}\rangle$}}
\newcommand {\Kftw} {K^\land_{43}}
\newcommand {\Kttw} {K^\land_{32}}
\newcommand {\Ktow} {K^\land_{21}}
\newcommand {\Kofw} {K^\land_{14}}
\newcommand {\Kftv} {K^\lor_{43}}
\newcommand {\Kttv} {K^\lor_{32}}
\newcommand {\Ktov} {K^\lor_{21}}
\newcommand {\Kofv} {K^\lor_{14}}
\newcommand {\Po} {k^+_1}
\newcommand {\Pt} {k^+_0}
\newcommand {\Pth} {k^+_3}
\newcommand {\Pf} {k^+_2}
\newcommand {\Kp} {q^+}
\newcommand {\T} {T_1}
\newcommand {\Uu} {T_0}
\newcommand {\eS} {T_3}
\newcommand {\V} {T_2}
\newcommand {\invt} {(-t)}
\newcommand {\pref} {\frac{1}{8\pi^2}}
\newcommand {\rleft} {\Pt<\Kp<\Pth}
\newcommand {\rmid} {\Pth<\Kp<\Po}
\newcommand {\rright} {\Po<\Kp<\Pf}
\newcommand {\Anei} {A_{\land\land\lor\lor}}
\newcommand {\Aalt} {A_{\land\lor\land\lor}}
\newcommand {\Nc} {}

\begin{document}

\setlength{\captionmargin}{20pt}
\begin{titlepage}
\begin{flushright}
UFIFT-HEP-06-01\\
\end{flushright}

\vskip 3cm

\begin{center}
\begin{Large}
{\bf Scattering of Glue by Glue on the Light-cone Worldsheet II:
Helicity Conserving Amplitudes\footnote{Supported 
in part by the Department
of Energy under Grant No. DE-FG02-97ER-41029.}}
\end{Large}

\vskip 2cm
{\large 
D. Chakrabarti\footnote{E-mail  address: {\tt dipankar@phys.ufl.edu}}, 
J. Qiu\footnote{E-mail  address: {\tt jqiu@phys.ufl.edu}}, 
and C. B. Thorn\footnote{E-mail  address: {\tt thorn@phys.ufl.edu}}
}
\vskip0.20cm
{\it Institute for Fundamental Theory\\
Department of Physics, University of Florida,
Gainesville FL 32611}

\vskip 1.0cm
\end{center}

\begin{abstract}\noindent
This is the second of a pair of articles on scattering of glue by glue,
in which we give the light-cone gauge calculation of the one-loop on-shell 
helicity conserving scattering
amplitudes for gluon-gluon scattering (neglecting
quark loops). The $1/p^+$ factors
in the gluon propagator are regulated by  replacing $p^+$ integrals with
discretized sums omitting the $p^+=0$ terms in each sum. We also employ
a novel ultraviolet regulator that is
convenient for the light-cone worldsheet description
of planar Feynman diagrams. The helicity
conserving scattering amplitudes are divergent
in the infra-red. The infrared divergences in the
elastic one-loop amplitude are shown to cancel,
in their contribution to cross sections, against ones in the
cross section for unseen bremsstrahlung gluons. We include here the
explicit calculation of the latter, because it assumes an
unfamiliar form due to the peculiar way discretization of $p^+$
regulates infrared divergences. In resolving the infrared
divergences we employ a covariant definition of jets, which
allows a transparent demonstration of the Lorentz invariance of 
our final results. Because we use an explicit cutoff of the ultraviolet 
divergences in exactly 4 space-time dimensions, we must introduce
explicit counterterms to achieve this final covariant
result. These counter-terms are polynomials in the external momenta of
the precise order dictated by power-counting. We discuss the
modifications they entail for the light-cone worldsheet action
that reproduces the ``bare'' planar diagrams of the gluonic sector
of QCD. The simplest way to do this is to interpret the QCD string
as moving in six space-time dimensions. 
\end{abstract}
\vfill
\end{titlepage}
\section{Introduction}
This is the second of a pair of articles on scattering of glue by
glue to one loop in the language of the lightcone worldsheet
\cite{thooftlargen,bardakcit,thornsheet}. The first
article (I) \cite{chakrabartiqt} worked out the (finite) one loop amplitudes 
for helicity nonconserving processes. In this
article we extend the calculation to the helicity conserving
case, for which it is necessary to deal with ultraviolet and
infrared divergences.

We refer the reader to the
introduction of I for the detailed motivation and background
for this work. Here we briefly mention the highlights. The goal of the
program to give a worldsheet description of the sum of planar diagrams
is to shed light on Field/String duality from the ``field'' side
at weak coupling. This is just the perturbation expansion of the field theory.
The mapping of the sum of planar diagrams
to a worldsheet system \cite{bardakcit,thornsheet} in essence allows one to
read off the worldsheet dynamics for scalar field theory and
Yang-Mills theory in this weak coupling limit. A serious limitation of these
initial articles, however, is that they transcribed the ``bare'' Feynman 
diagrams, without including any of the counterterms necessary to 
maintain gauge invariance. In the context of dimensional regularization
this limitation is innocuous, because dimensional regularization
automatically includes them. However, we want the worldsheet
formalism to work in four dimensions, and so we must have
in the worldsheet action the flexibility to include 
counterterms that go beyond the initial input Lagrangian.
In \cite{thornscalar} the ultraviolet structure of $\phi^3$
theory was analyzed on the lightcone and it was shown, to 
all orders in perturbation theory, that 
two new counterterms in addition to those associated with mass, 
wave function, and coupling renormalization were necessary
and sufficient. Happily, a local modification of the
``bare'' worldsheet action allowed for these new terms. 

The aim of I and the present article is to execute the
same program for Yang-Mills theory in light-cone gauge. 
Because the corresponding analysis is
considerably more complex, we have limited their scope to one loop. 
In I we focused on one-loop helicity-violating amplitudes for which
the on-shell tree diagrams vanished. As a consequence the
one loop amplitudes are finite in both the ultraviolet and infrared.
We could therefore confirm that the worldsheet description 
produces the correct known answers without dealing with collinear
and soft gluon emission processes.

In contrast, the helicity-conserving processes studied in the present
article display the full infrared divergence structure of 
non-abelian gauge theory. Just as in I, our infrared
regulator is discretization of the $p^+$ integrals
omitting the $p^+=0$ terms. But for the one-loop amplitudes
we deal with here, this is essentially equivalent to simply
reserving the $p^+$ integrations to last. This is because all
of the {\it artificial} $p^+=0$ divergences actually cancel algebraically
if the integrands from all diagrams, with momentum
routed appropriately, are combined before the
loop integrations are performed \cite{berndk}. However, the true
infrared divergences, for which $p^+$ discretization
provides a temporary regulator, are not cancelled in this combination, but
rather are organized to cancel against divergences due to
the absorption and emission of
real soft and collinear gluons according to the Lee-Nauenberg theorem
\cite{leenauenberg}.
In addition, of course, these amplitudes
have ultraviolet divergences which are taken care of by 
the renormalization program \cite{thornfreedom,perryfreedom}. 
In this article we give a complete lightcone gauge calculation
of the scattering of glue by glue, including the processes 
involving extra gluons that resolve the infrared divergences.
We work in four dimensions using the worldsheet friendly ultra-violet
cutoff employed in \cite{thornscalar,chakrabartiqt}.
We organize the Feynman diagrams of the $SU(N_c)$ Yang-Mills theory
according to 't Hooft's large $N_c$ expansion \cite{thooftlargen}, 
and we calculate the one-loop planar diagrams surviving the $N_c\to\infty$
limit. The 't Hooft limit suppresses diagrams with quark loops, so 
they are not included here.

We remark here that our non-traditional methods
of regulating and dealing with divergences are not sheer perversities
on our part, but rather they are guided by the desire to
fit these calculations into the framework of the lightcone
worldsheet formalism. We hope the reader will bear
with us on  this point, and we assure her that, although
unconventional, we have been extremely careful with the
well-known subtleties and pitfalls of this difficult subject
\cite{leibbrandt,irdivergencepapers}.

To keep this paper reasonably self-contained, we repeat two short
sections from I that summarize the lightcone Feynman rules 
(Section 2) and some useful identities (Section 3). A brief
Section 4 lists all the four gluon trees. Then in Section
5 we discuss bremsstrahlung processes. We use a covariant
definition of jets that proves to be very helpful in
achieving nice compact results for these processes. We deal
with initial state collinear (mass) divergences as in the 
original Lee-Nauenberg paper \cite{leenauenberg}, where
they are cancelled by including extra near collinear gluons
in the initial state.
This is in contrast to the standard technique, used
in analysing jets experimentally, that absorbs the initial
state collinear divergences 
into the initial state parton distribution function
\cite{irdivergencepapers}.
Section 6
briefly summarizes the results for triangle and swordfish 
diagrams obtained in \cite{thornlcnotes}. Section 7 contains the
meat of this paper, the calculation of box diagrams.
The reduction procedure developed in \cite{chakrabartiqt}
for the helicity violating box diagrams is helpful for this.
Section 8 describes the results of calculating the remaining quartic triangle
and double quartic diagrams, and Section 9 finally puts everything together,
and explains the cancellation of infrared divergences in cross
sections. Section 10 discusses the problem of giving a
local worldsheet description of all the counterterms necessary for
Lorentz invariance. We find this can be simply done by interpreting
the ``string'' dynamics of the worldsheet as occurring in six dimensional
spacetime. The two added dimensions are holographically generated
on the ``string'' side of Field/String duality and
play no role on the ``field'' side.
A final Section 11 wraps up the paper with a brief
look at issues still to resolve in the future.

The reader who does not wish to follow the technical details of
this work may get a glimpse of the main results and
their impact on the lightcone worldsheet dynamics by simply reading
sections 2, 9, 10 and 11. We particularly would like to draw his attention
to the nice and compact final results for the 
infrared finite and Lorentz covariant
probabilities (more precisely, the unnormalized
squared amplitudes for jet-jet scattering) 
Eqs.~(\ref{jetprobfinal}), (\ref{jetprobfinal2}) of Section 9.2. 
The implications of the necessary counterterms
for the worldsheet dynamics, which is
the subject of Section 10, should also be amusing for this reader.  

\section{Feynman Rules for Light-cone gauge Yang-Mills}
Here, we use the notation and conventions in Ref.~\cite{beringrt},
according to which the values of the 
non-vanishing three transverse gluon vertices are:
\bea
{{}\atop\mbox{\includegraphics[width=1.2cm]
{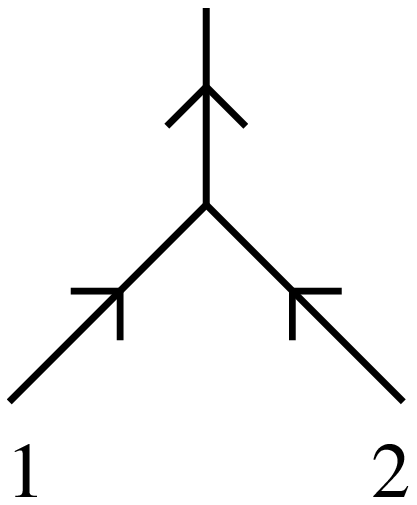}}}
\displaystyle\quad&=&{2gp_3^+\over p_1^+p_2^+}
\left(p_1^+{p_2^\land}-p_2^+{p_1^\land}\right)
={2gp_3^+\over p_1^+p_2^+}K^\land_{12}
\label{upupdown}\\
{{}\atop\mbox{\includegraphics[width=1.2cm]
{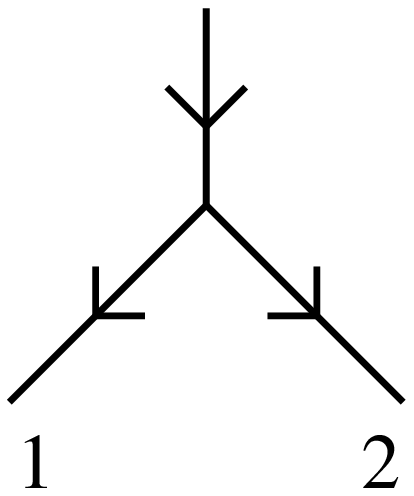}}}
\displaystyle\quad&=&{2gp_3^+\over p_1^+p_2^+}
\left(p_1^+{p_2^\lor}-p_2^+{p_1^\lor}\right)
={2gp_3^+\over p_1^+p_2^+}K^\lor_{12}
\label{downdownup}
\eea 
The quartic vertices in this helicity basis are given by
\bea
{{}\atop\mbox{\includegraphics[width=1.2cm]
{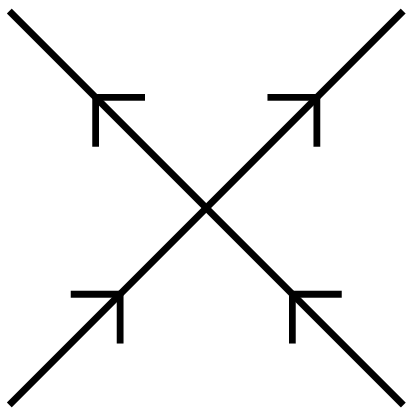}}}
\displaystyle\quad&=&-2{{g^2}{p^+_1p^+_3+p^+_2p^+_4\over
(p^+_1+p^+_4)^2}
\label{upupdowndown}}\\
{{}\atop\mbox{\includegraphics[width=1.2cm]
{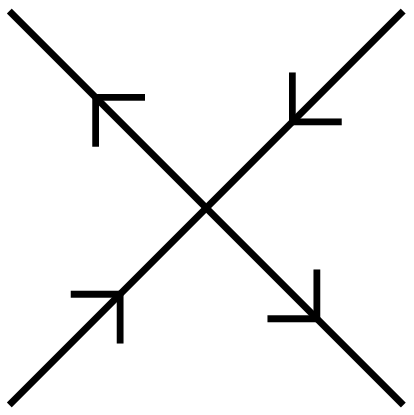}}}
\displaystyle\quad&=&+2{{g^2}\left({p^+_1p^+_2+p^+_3p_4^+\over
(p^+_1+p^+_4)^2}+{p^+_1p^+_4+p_2^+p^+_3\over
(p^+_1+p^+_2)^2}\right)}
\label{updownupdown}
\eea
In these expressions, $p_k^\land=(p_k^x+ip_k^y)/\sqrt2$, 
$p_k^\lor=(p_k^x-ip_k^y)/\sqrt2$, 
and $p_k^+=(p^0_k+p^z_k)/\sqrt{2}$  
are momenta {\it entering} the diagram on leg $k$,
and $g$ is proportional to the conventional QCD coupling $g_s$. 
Note that these are light-cone gauge ($A_-=0$) expressions
and include the contributions that arise when the longitudinal
field $A_+$ is eliminated from the formalism.\footnote{
These vertex rules are convenient
for the mixed $\tau=ix^+,p^+,{\bfs p}$ representation used in the
imaginary $x^+$ worldsheet formalism \cite{bardakcit}, in
which an $i$ from each vertex has been absorbed in each $dx^+$:
$idx^+=d\tau$ and the propagator is $(2p^+)^{-1}e^{-\tau{\bfs p}^2/2p^+}$. 
It will sometimes be convenient in this paper to
return to full Minkowski momentum space $p^-,p^+,{\bfs p}$. 
Then with the vertex rules given here and no $i$'s in the
momentum space propagators $1/(p^2-i\epsilon)$ with $p^2={\bfs p}^2-2p^+p^-$, 
each Minkowski loop momentum integral
should be accompanied by a $-i$: $-id^4q_M$. With a further Wick rotatation
to Euclidean space $d^4q_M=id^4q_E$ the $i$'s would disappear entirely.}
We also should point out that we are giving these rules in the context of the 
't Hooft's $1/N_c$ expansion at fixed $N_cg^2_s$. 
Then the {\it planar} diagrams of the $SU(N_c)$ theory
are correctly given if we take $g\equiv g_s\sqrt{N_c/2}$. Non-planar diagrams
with this definition of $g$ must be accompanied by appropriate
powers of $1/N^2_c$, depending on the
number of ``handles'' in the diagram. We have not spelled
the details out here, because our focus will be on the planar diagrams
in this article. The results we obtain should therefore be compared
to the limit $N_c\to\infty$, fixed $g_s^2N_c$ of those in
the literature. In making such comparisons, note that 
our definition of $g$ multiplies conventionally
defined $n$-gluon tree amplitudes by a factor $N_c^{n/2-1}\to N_c$ for
$n=4$, so for each gluon scattering process we remove this factor
before comparing to the literature.

\section{K Identities}
As we have seen, the quantities
\bea
K^\mu_{ij}&\equiv& p^+_ip^\mu_j-p^+_jp^\mu_i
\eea
play a central role in the cubic Yang-Mills vertex. In fact,
we shall find that the simplest forms of the various helicity
amplitudes are achieved by expressing them as functions of the $K$'s.
These simple forms are in fact identical to those achieved by
Parke and Taylor using a bispinor
representation of polarization vectors as the
now famous Parke-Taylor amplitudes \cite{parketaylor}. 
For us the role of the spinor matrix 
elements in those formulae will be played exclusively by $K_{ij}^\land$
and $K_{ij}^\lor$.

In order to reduce the expressions for the helicity amplitudes to the 
Parke-Taylor form, we will need a number of identities enjoyed by the $K$'s.
For a general $n$-gluon amplitude we can form $K_{ij}$ for each pair
of gluons $(ij)$, where $i,j=1,\ldots,n$ distinguish the different gluons.
By momentum conservation, it is immediate that
\bea
\sum_j K^\mu_{ij}&=&0.
\eea
From the fact that $K$ is an anti-symmetric product we have Bianchi-like
identities
\bea
p^+_iK^\mu_{jk}+p^+_kK^\mu_{ij}+p^+_jK^\mu_{ki}&=&0\\
K^\land_{li}K^\land_{jk}+K^\land_{lk}K^\land_{ij}
+K^\land_{lj}K^\land_{ki}&=&0
\eea
Finally, the most powerful type of identity follows from
a very simple calculation
\bea
\sum_j{K^\land_{ij}K^\lor_{jk}\over p^+_j}&=&-p^+_ip^+_k\sum_j 
{p_j^2\over2p^+_j}
\eea
which seems like a complicated non-linear relation. However 
when we are considering scattering amplitudes, the momenta 
all satisfy $p_i^2=0$ so the right side is zero! This identity plays a
central role in showing that trees with all
but one like helicities vanish. (Trees with all like helicity 
can't even be drawn.) They are also crucial for
reducing the complexity of the helicity amplitudes that don't vanish.

\section{Summary of Tree Amplitudes}

In this section we simply list the four point trees amplitudes on and
off shell obtained in \cite{chakrabartiqt}.
There are no tree diagrams for $\land\land\land\land$
or $\lor\lor\lor\lor$ polarizations.
The off-shell $\land\land\land\lor$
four-point tree is given by,
omitting the coupling factor $2g$ for each vertex,
\bea
A^{tree}_{\land\land\land\lor}&=&-{p^+_4\over p^+_1p^+_2p^+_3}
\left[{K^\land_{32}K^\land_{14}\over(p_2+p_3)^2}+
{K^\land_{43}K^\land_{21}\over(p_1+p_2)^2}\right]\nonumber\\ 
&=& 
-{p_4^+(K_{43}^\land K_{32}^\land p_1^2
+K_{14}^\land K_{43}^\land p_2^2+K_{21}^\land K_{14}^\land p_3^2
+K_{32}^\land K_{21}^\land p_4^2)\over p_1^+p_2^+p_3^+(p_1+p_2)^2
(p_2+p_3)^2}\ \to\ 0 \qquad{\rm On~shell}
\eea
The notation here is that $\land$ denotes an incoming arrow representing
helicity $+1$, while $\lor$ denotes an outgoing arrow representing
helicity $-1$.

The only non-zero four point trees are those with two of each helicity.
There are two distinct helicity patterns. The amplitude
for adjacent helicity $\land\land\lor\lor$ is given by
\bea
A^{tree}_{\land\land\lor\lor}&=&-{1\over(p_1^++p_4^+)^2}
\left[{p_1^+p_3^+\over p_2^+p_4^+}{K_{14}^\lor K_{32}^\land\over
(p_1+p_4)^2} +{p_2^+p_4^+\over p_1^+p_3^+}{K_{14}^\land K_{32}^\lor\over
(p_1+p_4)^2}+{p_1^+p_3^++p_2^+p_4^+\over 2}\right]\nonumber\\
&&\hskip.5in-{(p_1^++p_2^+)^2K_{21}^\land K_{43}^\lor
\over p_1^+p_2^+p_3^+p_4^+(p_1+p_2)^2}
\label{uuddamp}
\eea
When some legs are off-shell, we use the shorthand notation
$p_i^*\equiv p_i^2/p_i^+$ to simplify the writing. 
\bea
A^{tree}_{\land\land\lor\lor}
&=&-{2K_{21}^{\land2} K_{43}^{\lor2}
\over p_1^+p_2^+p_3^+p_4^+(p_1+p_2)^2(p_1+p_4)^2}\nonumber\\
&&+{(p_1^++p_2^+)(p_1^*+p_2^*-p_3^*-p_4^*)\over2(p_1+p_4)^2}
-{p_2^+p_4^+(p_1^*-p_3^*)+p_1^+p_3^+(p_4^*-p_2^*)
\over2(p_1^++p_4^+)(p_1+p_4)^2}\\
&&+{(p_1^++p_2^+)^2(p_1^*+p_2^*)(p_3^*+p_4^*)
\over2(p_1+p_2)^2(p_1+p_4)^2}+{K_{21}^\land K_{43}^\lor
(p_1^++p_2^+)[p_1^+p_3^+(p_1^*-p_3^*)+p_2^+p_4^+(p_2^*-p_4^*)]
\over p_1^+p_2^+p_3^+p_4^+(p_1+p_2)^2(p_1+p_4)^2}\nonumber\\
&\to&-{2K_{21}^{\land2} K_{43}^{\lor2}
\over p_1^+p_2^+p_3^+p_4^+(p_1+p_2)^2(p_1+p_4)^2}\hskip2in{\rm (On~Shell)}
\label{uuddtree}
\eea

The other distinct helicity arrangement for four gluon
scattering is alternating helicity $\land\lor\land\lor$: 
\bea
A^{tree}_{\land\lor\land\lor}&=&-{1\over(p_1^++p_4^+)^2}
\left[{p_1^+p_2^+\over p_3^+p_4^+}{K_{14}^\lor K_{32}^\land\over
(p_1+p_4)^2} +{p_3^+p_4^+\over p_1^+p_2^+}{K_{14}^\land K_{32}^\lor\over
(p_1+p_4)^2}-{p_1^+p_2^++p_3^+p_4^+\over 2}\right]\nonumber\\
&&-{1\over(p_1^++p_2^+)^2}
\left[{p_1^+p_4^+\over p_2^+p_3^+}{K_{43}^\land K_{21}^\lor\over
(p_1+p_2)^2} +{p_2^+p_3^+\over p_1^+p_4^+}{K_{43}^\lor K_{21}^\land\over
(p_1+p_2)^2}-{p_1^+p_4^++p_2^+p_3^+\over 2}\right]
\label{ududamp}
\eea
where the quartic vertex contribution has been split between the
last terms in each of the square brackets. Notice that the second line on the
right side is obtained from the first line with the relabeling 
substitutions $1\to2\to3\to4\to1$ and $\land\to\lor\to\land$.
Furthermore the first line can be obtained from the first line on the
right of (\ref{uuddamp}) by interchanging $2\leftrightarrow3$ and
multiplying by the factor $-1$. Thus by inspection we immediately obtain 
the simplifications
\bea
A^{tree}_{\land\lor\land\lor}&=&
-{2K_{31}^{\land2} K_{42}^{\lor2}\over p_1^+p_2^+p_3^+p_4^+
p_{12}^2p_{14}^2}-{p_{13}^+p_{24}^+(p_1^*+p_3^*)(p_2^*+p_4^*)\over
2p_{12}^2p_{14}^2}+{p_{13}^2[p_4^+p_2^*+p_2^+p_4^*+p_3^+p_1^*+p_1^+p_3^*]
\over2p_{12}^2p_{14}^2}\nonumber\\
&&-K_{31}^\land K_{42}^\lor{
p_3^+p_4^+(p_1^2+p_2^2)+p_1^+p_2^+(p_3^2+p_4^2)
-p_2^+p_3^+(p_1^2+p_4^2)-p_1^+p_4^+(p_2^2+p_3^2)\over p_1^+p_2^+p_3^+p_4^+
(p_1+p_4)^2(p_1+p_2)^2}\nonumber\\
&&+{p_3^+p_4^+(p_1^*-p_2^*)+p_1^+p_2^+(p_4^*-p_3^*)\over
2p_{14}^2p_{14}^+}+{p_1^+p_4^+(p_2^*-p_3^*)+p_2^+p_3^+(p_1^*-p_4^*)\over
2p_{12}^2p_{12}^+}\nonumber\\
&\to&
-{2K_{31}^{\land2} K_{42}^{\lor2}\over p_1^+p_2^+p_3^+p_4^+
p_{12}^2p_{14}^2}\hskip1.5in{\rm (On~Shell)}       
\eea
Here and in the following we use the shorthand notation
$p_{i,i+1}=p_i+p_{i+1}$, with $i=0,1,2,3$ and $p_4\equiv p_0$.

\section{Gluon Bremsstrahlung and Jets}
The consistent resolution of infrared divergences in loop corrections 
to scattering amplitudes involves a cancellation against corresponding
infrared divergences in the cross section for the emission (or
absorption) of an extra gluon, whose momentum is either collinear
with one of the gluons in the core process or ``soft''. If the
core process is scattering of glue by glue, the associated
bremsstrahlung amplitudes are 5 gluon amplitudes.

\begin{figure}[ht]
\begin{center}
\psfrag{'1'}{$1$}
\psfrag{'2'}{$2$}
\psfrag{'3'}{$3$}
\psfrag{'4'}{$4$}
\includegraphics[width=4in]{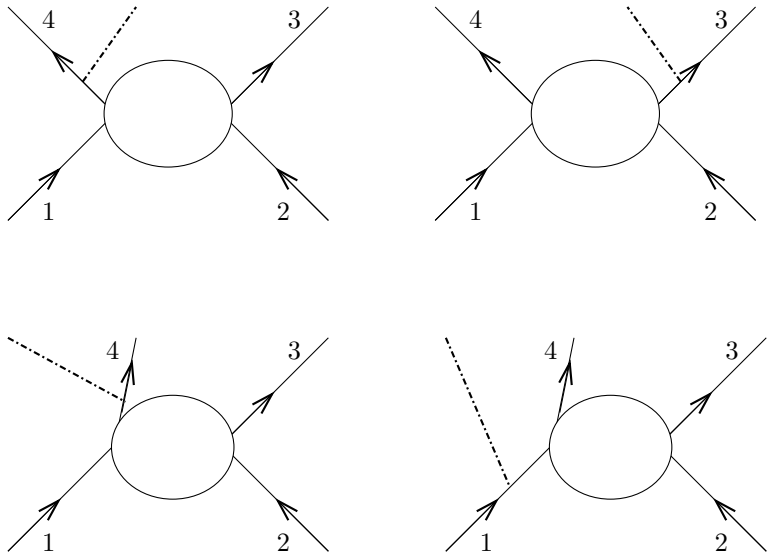}
\caption{The bremsstrahlung diagrams associated
with glue-glue scattering involving leg 4. At $N_c=\infty$
the sum of the diagrams on each line may be independently squared to give the
leading contribution to the cross section. Similar pairs of diagrams involving
each of the other legs must also be included.}
\label{brem}
\end{center}
\end{figure}

In the context of the large $N_c$ limit it is necessary to
combine coherently only the bremsstrahlung diagrams 
with the same cyclic ordering.
So, for example, in the diagrams shown in Fig.~\ref{brem} at $N_c=\infty$
it is only necessary to square the sum of the two diagrams on each line 
and combine the results on different lines incoherently.
Because $N_c=\infty$
suppresses nonplanar diagrams, a gluon line attached between two 
outgoing gluons
(as with the diagrams on the first line of Fig.~\ref{brem}) 
must be outgoing when $N_c=\infty$\footnote{At first glance the reader might
think that the distinction between an incoming or outgoing line
has nothing to do with planarity. But a well-defined large $N_c$ limit only
makes sense for physical quantities that are singlets under the gauge group.
We specify our planar scattering amplitudes by imagining 
that the external gluons are all attached to a huge connected 
Wilson loop including 
a single connected portion at late times and another 
single connected portion at early times. 
The large $N_c$ limit of such a quantity
has the properties we describe in this paragraph.}. 
Similarly a gluon line attached between two incoming gluons must be incoming. 
On the other hand, a gluon attached between an incoming and an outgoing
gluon (as with the diagrams on the second line of Fig.~\ref{brem})
may be either incoming or outgoing.

As is well known infrared and collinear divergences are present 
only when the bremsstrahlung
gluon attaches to external legs. For example if the brem 
gluon is collinear
with $p_4$, there is a collinear divergence in the phase-space integral
of the square of the diagrams where the gluon is emitted from or absorbed by
leg 4. Calling the brem gluon's four-momentum $k$, for fixed $k^+$ the
collinear point is ${\bfs k}=k^+{\bfs p}_4/p_4^+$. Then it is convenient
to write
\bea
{\bfs k}=k^+{{\bfs p}_4\over p_4^+}+{\hat{\bfs k}}
\eea 
and examine the phase space integral for $|{\hat{\bfs k}}|$ 
in a neighborhood of zero. Here we assume $k^+=O(1)$ so the
brem gluon is not soft. In effect,
rather than measuring a single gluon, we insist that we measure a ``jet''
of total transverse momentum ${\bfs P}\approx
(k^++p_4^+){\bfs p}_4/p_4^+$ within a resolution $\Delta$ \cite{stermanw}.
A simple calculation shows that
\bea
{(p_4^+{\bfs k}-k^+{\bfs p}_4)^2\over|k^+||p_4^+|}=2|{\vec k}||{\vec p}_4|
(1-\cos\theta)
\eea
where the overarrow denotes the 3 spatial components of a four-vector
and $\cos\theta={\vec k}\cdot{\vec p}_4/|{\vec k}||{\vec p}_4|$.
The left side is thus a nice measure of the angular size of
a jet, so we define the phase space of a jet of resolution $\Delta$
by the restriction
\bea
{(p_4^+{\bfs k}-k^+{\bfs p}_4)^2\over|k^+||p_4^+|}<\Delta^2
\eea
This translates to ${\hat{\bfs k}}^2<|k^+|\Delta^2/|p_4^+|$.

The amplitudes for the emission of a hard 
brem gluon from the right of leg 4 (as in the first diagram on
the first line of Fig.~\ref{brem}) are given, for the two polarizations, by
\bea
A_{\rm Brem}^\lor&=&-2g{k^++p_4^+\over k^+p_4^+}
{K^\lor_{k,4}A_{\rm Core}(p_1,p_2,p_3,k+p_4)\over
(k+p_4)^2},\qquad {\rm Outgoing~helicity}\\
A_{\rm Brem}^\land&=&
-2g{p_4^+\over k^+(k^++p_4^+)}{K^\land_{k,4}
A_{\rm Core}(p_1,p_2,p_3,k+p_4)\over
(k+p_4)^2},
\qquad {\rm Incoming~helicity}
\eea
When the brem gluon (with momentum $k$) is emitted from the left of leg 4, 
the amplitudes are the
same except that ${\bfs K}_{4,k}$ appears instead of ${\bfs K}_{k,4}$. Thus the
amplitudes for emission from left and right 
have opposite signs. The amplitudes do not cancel, however,
because they have different gauge group structure. As already mentioned
at $N_c=\infty$ the two terms enter the cross section incoherently. When the
brem gluon has the same helicity as leg 4 and is collinear with $p_4$, 
it and gluon 4 are distinguished only by their $p^+$ values. Then we
arbitrarily call the one with smaller $|p^+|$ the brem gluon.

Now it is easy to see that
\bea
{\bfs K}_{k,4}&=&-p_4^+{\hat{\bfs k}},\qquad
(k+p_4)^2=-p_4^+{\hat{\bfs k}}^2/k^+=-2p_4^+{\hat{k}}^\land{\hat{k}}^\lor/k^+
\eea
Then we have immediately,
\bea
&&{d{\bfs p}_4\over 2|p_4^+|}{d{\bfs k}\over 2|k^+|(2\pi)^3}
(|A^\lor|^2+|A^\land|^2)
=\nonumber\\
&&\hskip+.75in{d{\bfs P}\over 2|P^+|}
{d{\hat{\bfs k}}\over|k^+| (2\pi)^3}{p_4^{+}
\over k^++p_4^+}\left({(k^++p_4^+)^2
\over p_4^{+2}}+{p_4^{+2}
\over (k^++p_4^+)^2}\right)
{g^2\over {\hat{\bfs k}}^2}|A_{\rm Core}|^2
\eea
The collinear divergence is now transparent in the integration over
${\hat{\bfs k}}$ near zero. It is the coefficient of the
phase space factor ${d{\bfs P}/2|P^+|}$
that we should compare to the square of the
tree amplitudes with self-energy corrections on external
lines.

We now show that the collinear divergence just isolated, when summed
over all possible $k^+$ is canceled
by a corresponding divergence in the self-energy correction to
leg 4. This cancellation is a consequence of the Lee-Nauenberg
theorem \cite{leenauenberg}, 
which stipulates that all collinear states with the same
energy be included. The total energy of the brem gluon and the
gluon represented by leg 4 is in the collinear limit
\bea
k^-+p_4^-={{\bfs k}^2\over 2k^+}+{{\bfs p}_4^2\over2p_4^+}
={{\bfs p}_4^2\over2p_4^+}(1+k^+/p_4^+)={{\bfs P}^2\over
2(k^++p_4^+)}
\eea
which makes it clear that we should include all $k^+,p_4^+$ consistent with
fixed $P^+=k^++p_4^+$.
We temporarily regulate the divergence by giving the
collinear gluons a small mass $\mu$. Since lightcone phase space is 
mass independent, $\mu$ only appears in the above analysis in the
expression for $(k+p_4)^2$:
\bea
(k+p_4)^2\to-{p_4^+\over k^+}\left(
{\hat{\bfs k}}^2+\mu^2{(k^++p_4^+)^2\over p_4^{+2}}\right)
\eea
With this regulator the transverse momentum integral we need is just
\bea
\int_{0<{\hat{\bfs k}}^2|p_4^+|<|k^+|\Delta^2} d{\hat{\bfs k}}
{{\hat{\bfs k}}^2\over [{\hat{\bfs k}}^2
+\mu^2{(k^++p_4^+)^2/p_4^{+2}}]^2}=\pi\ln{|k^+||p_4^+|\Delta^2\over
|k^++p_4^+|^2\mu^2e}
\eea
Then the coefficient of the jet phase space factor is
\bea
\int_\Delta{d{\bfs k}\over 2|k^+|(2\pi)^3}(|A^\lor|^2+|A^\land|^2)
={1\over|k^+|}{g^2\over 8\pi^2} 
\left({(k^++p_4^+)\over p_4^{+}}+{p_4^{+3}
\over (k^++p_4^+)^3}\right)|A_{\rm Core}|^2\ln{|k^+||p_4^+|\Delta^2\over
|k^++p_4^+|^2\mu^2e}
\eea
The blowup as $\mu\to0$ is the collinear divergence we are seeking
to resolve.
According to the Lee-Nauenberg theorem, to get an infrared safe
quantity we must sum over all $k^+$ in the range $0<|k^+|<|P^+|$.
And we must also include brem emission from the
left of leg 4. The first term represents the emission of a brem gluon
with identical helicity to leg 4, so when we sum that term over the whole range
of $k^+$ we have included emission from both the left and right
of leg 4. However the second term, represents the emission of a brem gluon with
opposite helicity, and when summed over the whole range
gives only brem gluon emission from the right of leg 4. The emission
of an opposite helicity gluon (with momentum $k$) from the left has the
same squared amplitude, but it is convenient to switch the roles of
$k$ and $p_4$, so $k$ always refers to the right gluon.
Then the total emission rate is given by
\bea
\sum_{0<|k^+|<|P^+|}
\int_\Delta{d{\bfs k}\over 2|k^+|(2\pi)^3}(|A^\lor|^2+|A_R^\land|^2
+|A_L^\land|^2)
=\nonumber\\
&&\hskip-3in\sum_{k^+}{g^2\over 8\pi^2} 
\left({|P^+|\over|k^+(P^+-k^+)|}+{|P^+-k^+|^{3}
\over |k^+P^{+3}|}+{|k^+|^{3}
\over |(P^+-k^+)P^{+3}|}\right)|A_{\rm Core}|^2
\ln{|k^+(P^+-k^+)|\Delta^2\over|P^+|^2\mu^2e}
\eea
Calling $x=|k^+|/|P^+|$, $|P^+|$ times
 the quantity in parentheses can be rearranged 
\bea
{1\over x(1-x)}+{(1-x)^3\over x}     
+{x^3\over1-x}&=&2\left(x(1-x)+{x\over 1-x}+{1-x\over x}\right)\nonumber\\
&=&2\left({|k^+||P^+-k^+|\over P^{+2}}+{|k^+|\over |P^+-k^+|}
+{|k^+|\over |P^+-k^+|}\right)
\eea
So with this notation the squared amplitude for jet production
along gluon 4 is
\bea
\sum_{0<|k^+|<|P^+|}
\int_\Delta{d{\bfs k}\over 2|k^+|(2\pi)^3}(|A^\lor|^2+|A_R^\land|^2
+|A_L^\land|^2)
=\nonumber\\
&&\hskip-2in{g^2\over 4\pi^2}{|A_{\rm Core}|^2\over|P^+|}\sum_{k^+} 
\left(x(1-x)+{x\over 1-x}+{1-x\over x}\right)
\ln{x(1-x)\Delta^2\over\mu^2e}
\label{collinearbrem}
\eea
We still have to include the self energy corrections on the external lines.

In Ref.~\cite{chakrabartiqt} we obtained 
\bea
\Pi^{\land\lor}={g^2\over 4\pi^2}{p^2\over|p^+|}\sum_{k^+} 
\left(x(1-x)+{x\over 1-x}+{1-x\over x}\right)
\ln(x(1-x)p^2\delta e^{\gamma})
\eea
for the gluon self-energy for an off-shell gluon of
momentum $p$ after subtraction of the counterterms 
necessary to keep the gluon massless. Redoing the calculation
with a mass $\mu$ for the particles circulating in the loop
yields the modification
\bea
\Pi_\mu^{\land\lor}&=&{g^2\over 4\pi^2}{1\over|p^+|}\sum_{k^+}(\mu^2+x(1-x)p^2) 
\left(1+{1\over (1-x)^2}+{1\over x^2}\right)
\ln((\mu^2+x(1-x)p^2)\delta e^{\gamma})\\
&\to& {\rm Constant} +{p^2\over|p^+|}{g^2\over 4\pi^2}\sum_{k^+}
\left(x(1-x)+{x\over 1-x}+{1-x\over x}\right)\ln(\mu^2\delta e^{\gamma+1})
\eea
in the on-shell limit $p^2\to0$. 
The constant (which is of order $\mu^2$) 
must be absorbed in a mass counterterm to keep the gluon massless.
On the external line the $p^2$ is canceled by the extra gluon
propagator, and the effect of the correction is just the wave function
renormalization
\bea
Z=1+{g^2\over 4\pi^2}{1\over|p^+|}\sum_{k^+}
\left(x(1-x)+{x\over 1-x}+{1-x\over x}\right)\ln(\mu^2\delta e^{\gamma+1})
\eea
Of course the tree amplitude is multiplied by $\sqrt{Z}$ for each leg
and the squared tree amplitude by a factor of $Z$. Thus the
correction on leg 4 is just
\bea
(Z-1)|A_{\rm Core}|^2={g^2\over 4\pi^2}
{|A_{\rm Core}|^2\over|p^+|}
\sum_{k^+}
\left(x(1-x)+{x\over 1-x}+{1-x\over x}\right)\ln(\mu^2\delta e^{\gamma+1})
\label{selfenergy}
\eea
Combining this self energy correction with (\ref{collinearbrem}) gives for the
complete jet production, identifying $P=p$, 
\bea
\langle |{\cal M}|^2\rangle_{\rm jet}
={g^2\over 4\pi^2}{|A_{\rm Core}|^2\over|p^+|}\sum_{k^+} 
\left(x(1-x)+{x\over 1-x}+{1-x\over x}\right)
\ln{[x(1-x)\Delta^2\delta e^\gamma]}
\label{jetproduction}
\eea
We see that the collinear divergence problem, which in lightcone
gauge is confined to the self energy insertions on external lines, 
is resolved provided we interpret scattering amplitudes in terms of jets.

However, this is not the end of the story because we still see UV
divergences ($\ln\delta$) and IR divergences due to $x$ near 0 or 1.
As explained in \cite{chakrabartiqt} the latter divergences 
are regulated by discretization of $p^+$. The UV divergence from
this calculation is to be combined with the UV divergences from 
triangle, box, and internal line self-energy diagrams to give
the appropriate scale dependent coupling. Part of the IR divergences
here will be canceled by soft gluon bremsstrahlung which we
discuss next. In particular, we must find that the dependence
on the resolution $\Delta$ is finite.
 But there will be residual IR divergences that are to 
be canceled by IR divergences from the other one loop diagrams
(box, triangle, etc.).

Our discussion of jet production only included the diagram
with brem gluon attached the external leg identified with the jet 
(for definiteness we chose leg 4), neglecting its interference with
diagrams with the brem gluon attached to other lines. 
This approximation is only valid, however, when the resolution is
smaller than the momentum $k^+{\bfs p}_4/p_4^+$ of the gluon in the
jet. We can stipulate that all of the
momenta $p_i$ are of order $O(1)$, but even so
for small enough $k^+$ it is essential that we include the
other diagrams.
In this case all components of the brem gluon momentum are small
and we must combine gluon emission from all legs coherently.
In the large $N_c$ limit life is simpler because we only need to
include coherent emission from neighboring lines as already discussed.

For definiteness, focus first on the coherent emission of a gluon
between legs $3$ and $4$, both of which we assume to have outgoing helicity.
Then the emission amplitudes are
\bea
A^\lor&=& -2gA_{\rm Core}\left[{k^++p_4^+\over k^+p_4^+}{K_{k,4}^\lor
\over (k+p_4)^2}+{k^++p_3^+\over k^+p_3^+}{K_{3,k}^\lor
\over (k+p_3)^2}\right]\\
A^\land&=& -2gA_{\rm Core}\left[{p_4^+\over k^+(k^++p_4^+)}{K_{k,4}^\land
\over (k+p_4)^2}+{p_3^+\over k^+(k^++p_3^+)}{K_{3,k}^\land
\over (k+p_3)^2}\right]
\eea
In these formulas we have assumed that $A_{\rm Core}$ is the same in 
both terms, which is approximately true when all components of
$k$ are small. The case where $k$ collinear with one of the external
momenta but not small will not introduce errors, because in that
case the interference between different terms is negligible. 
For this we need to insist that no two external legs are collinear,
which we do. 
The squared amplitudes are:
\bea
|A^\lor|^2&=&4g^2|A_{\rm Core}|^2
\bigg\{{(k^++p^+_4)^{2}{\bfs K}_{k,4}^2\over2k^{+2}p_4^{+2}(k+p_4)^4}
+{(k^++p^+_3)^{2}{\bfs K}_{k,3}^2\over2k^{+2}p_3^{+2}(k+p_3)^4}
-{(k^++p^+_4)(k^++p^+_3){\bfs K}_{k,3}\cdot{\bfs K}_{k,4}\over
k^{+2}p_3^+p_4^+(k+p_3)^2(k+p_4)^2}\bigg\}\nonumber\\
|A^\land|^2&=&4g^2|A_{\rm Core}|^2
\bigg\{{p_4^{+2}{\bfs K}_{k,4}^2\over2k^{+2}(k^++p^+_4)^{2}(k+p_4)^4}
+{p_3^{+2}{\bfs K}_{k,3}^2\over2k^{+2}(k^++p^+_3)^{2}(k+p_3)^4}
\nonumber\\&&\hskip2in
-{p_3^+p_4^+{\bfs K}_{k,3}\cdot{\bfs K}_{k,4}\over
k^{+2}(k^++p^+_4)(k^++p^+_3)(k+p_3)^2(k+p_4)^2}\bigg\}\nonumber
\eea
These expressions contribute to the jet cross section for jets both
along $p_3$ and $p_4$. Let us define ${\bfs v}_i\equiv {\bfs p}_i/p_i^+$.
 Then they should  be integrated over
the union of the two domains ${\cal D}_4:
({\bfs k}-k^+{\bfs v}_4)^2<k^+\Delta^2/p_4^+$
and ${\cal D}_3: ({\bfs k}-k^+{\bfs v}_3)^2<k^+\Delta^2/p_3^+$. In order not
to double count we integrate over the entire first domain, but 
we integrate only over the part of the second domain satisfying
$({\bfs k}-k^+{\bfs v}_4)^2>k^+\Delta^2/p_4^+$. The necessary integrals 
over ${\cal D}_4$ can be
found in Appendix A. Define ${\hat{\bfs k}}\equiv {\bfs k}-k^+{\bfs v}_4$,
and give all of the gluons 3,4 and $k$ a small mass $\mu$.
Then we have
\bea
{1\over2k^{+2}}
\int_{{\cal D}_4} d{\hat{\bfs k}}\ {{\bfs K}_{k,4}^2\over(k+p_4)^4}&=&
{1\over2}\int_{{\cal D}_4} d{\hat{\bfs k}}\ 
{{\hat{\bfs k}}^2\over({\hat{\bfs k}}^2+\mu^2(k^++p_4^+)^2/p_4^{+2})^2}
={\pi\over2}\ln{k^+p_4^+\Delta^2\over (k^++p_4^+)^2\mu^2e}\\
{1\over2k^{+2}}\int_{{\cal D}_4} d{\hat{\bfs k}}\ {{\bfs K}_{k,3}^2\over(k+p_3)^4}&=&{1\over2}
\int_{{\cal D}_4} d{\hat{\bfs k}}\ 
{k^{+2}({\hat{\bfs k}}-k^+{\bfs v}_{34})^2\over
(({\hat{\bfs k}}-k^+{\bfs v}_{34})^2+\mu^2(k^++p_3^+)^2/p_3^{+2})^2}
\nonumber\\
&=&{\pi\over2}\cases{\ln{\displaystyle k^+p_3^{+2}(
\Delta^2-k^+p_4^+{\bfs v}^2_{34})\over\displaystyle 
 p_4^+(k^++p_3^+)^2\mu^2e},
& \qquad$|k^+|<{\Delta^2\over|p_4^+|{\bfs v}^2_{34}}$ \cr
\ln{\displaystyle k^+p_4^+{\bfs v}^2_{34}\over\displaystyle
k^+p_4^+{\bfs v}^2_{34}-\Delta^2},
& \qquad$|k^+|>{\Delta^2\over|p_4^+|{\bfs v}^2_{34}}$ \cr}\\
{1\over k^{+2}}
\int_{{\cal D}_4} d{\hat{\bfs k}}\ {{\bfs K}_{k,3}\cdot{\bfs K}_{k,4}
\over(k+p_3)^2(k+p_4)^2}&=&
\int_{{\cal D}_4} d{\hat{\bfs k}}\ 
{{\hat{\bfs k}}\cdot({\hat{\bfs k}}-k^+{\bfs v}_{34})\over
({\hat{\bfs k}}^2+\mu^2(k^++p_4^+)^2/p_4^{+2})
(({\hat{\bfs k}}-k^+{\bfs v}_{34})^2+\mu^2(k^++p_3^+)^2/p_3^{+2})}
\nonumber\\
&=&{\pi\over2}\cases{\ln{\displaystyle 
\Delta^4\over\displaystyle 
 k^{+2}p_4^{+2}{\bfs v}^4_{34}},
& \qquad$|k^+|<{\Delta^2\over|p_4^+|{\bfs v}^2_{34}}$ \cr
0,
& \qquad$|k^+|>{\Delta^2\over|p_4^+|{\bfs v}^2_{34}}$ \cr}
\eea  
where the final forms are valid as $\mu\to0$. We  have used the identity
\bea
(k+p_i)^2 &=& -{{\bfs K}_{k,i}^2+\mu^2(k^++p_i^+)^2\over k^+p_i^+}
=-{p_i^{+2}({\bfs k}-k^+{\bfs v}_i)^2+\mu^2(k^++p_i^+)^2\over k^+p_i^+}
\label{kplusp2}
\eea
to simplify the integrands.

In assembling these
contributions we write separate equations for small and large $k^+$,
simplifying the coefficients in the first case:
\bea
\int_{{\cal D}_4} {d{\hat{\bfs k}}}\ 
{|A^\lor|^2+|A^\land|^2\over16|k^+|\pi^3}&\approx&
{g^2|A_{\rm Core}|^2\over4\pi^2|k^+|}\ln{{k^{+4}p_3^{+2}p_4^{+2}{\bfs v}_{34}^4
(1-k^+p_4^+{\bfs v}^2_{34}/\Delta^2)\over (k^++p_4^+)^2(k^++p_3^+)^2\mu^4e^2}}
\qquad{\rm for}~|k^+|<{\Delta^2\over|p_4^+|{\bfs v}^2_{34}}\\
\int_{{\cal D}_4} d{\hat{\bfs k}}\ {|A^\lor|^2+|A^\land|^2\over
16|k^+|\pi^3}&=&
{g^2|A_{\rm Core}|^2\over8\pi^2}\bigg\{\left({(k^++p_4^+)^2\over p_4^{+2}}
+{p_4^{+2}\over(k^++p_4^+)^2}\right)
\ln{k^+p_4^+\Delta^2\over (k^++p_4^+)^2\mu^2e}\nonumber\\
&&\hskip-.25in-\left({(k^++p_3^+)^2\over p_3^{+2}}
+{p_3^{+2}\over(k^++p_3^+)^2}\right)\ln\left(1-{\Delta^2\over
k^+p_4^+{\bfs v}^2_{34}}\right)\bigg\}
\qquad{\rm for}~|k^+|>{\Delta^2\over|p_4^+|{\bfs v}^2_{34}}\nonumber\\
&\approx&
{g^2|A_{\rm Core}|^2\over8\pi^2}\bigg\{\left({(k^++p_4^+)^2\over p_4^{+2}}
+{p_4^{+2}\over(k^++p_4^+)^2}\right)
\ln{k^+p_4^+\Delta^2\over (k^++p_4^+)^2\mu^2e}\nonumber\\
&&\hskip1.4in-2\ln\left(1-{\Delta^2\over
k^+p_4^+{\bfs v}^2_{34}}\right)\bigg\}
\qquad{\rm for}~|k^+|>{\Delta^2\over|p_4^+|{\bfs v}^2_{34}}
\eea
where the approximation in the last line is valid because the
logarithm factor cuts off large $k^+$. We still need to add the
contribution of the part of the domain ${\cal D}_3$ that doesn't
intersect ${\cal D}_4$.

To handle the double constraint on the domain of integration,
it is convenient to divide the contribution into two contributions,
\bea
&I:& {k^+\over p_4^+}\Delta^2<({\bfs k}-k^+{\bfs v}_4)^2
<{k^+\over p_4^+}\Delta_0^2;\qquad ({\bfs k}-k^+{\bfs v}_3)^2
<{k^+\over p_3^+}\Delta^2\\
&II:& {k^+\over p_4^+}\Delta_0^2<({\bfs k}-k^+{\bfs v}_4)^2;
\qquad ({\bfs k}-k^+{\bfs v}_3)^2
<{k^+\over p_3^+}\Delta^2
\eea
where we choose $\Delta_0\gg\Delta$ large enough so that in region
II we only need to include the diagram with the brem gluon attached
to leg 3. It is not hard to show that in region II $k^+$ necessarily 
satisfies
\bea
k^+>{1\over v_{34}^2}\left({\Delta_0\over\sqrt{|p_4^+|}}-
{\Delta\over\sqrt{|p_3^+|}}\right)^2.
\eea
Unfortunately the converse is not quite true: the condition
on $k^+$ that implies ${\bfs k}$ is in region II is slightly
more strict:
\bea
k^+>{1\over v_{34}^2}\left({\Delta_0\over\sqrt{|p_4^+|}}+
{\Delta\over\sqrt{|p_3^+|}}\right)^2.
\eea
But at least the contribution from the
part of region II that satisfies this last constraint
is simply given:
\bea
2\pi g^2|A_{\rm Core}|^2
\left({(k^++p_3^+)^2\over p_3^{+2}}
+{p_3^{+2}\over(k^++p_3^+)^2}\right)
\ln{k^+p_3^+\Delta^2\over (k^++p_3^+)^2\mu^2e},\qquad
{\rm for}~|k^+|>{1\over v_{34}^2}\left({\Delta_0\over\sqrt{|p_4^+|}}+
{\Delta\over\sqrt{|p_3^+|}}\right)^2
\label{II}
\eea
There remains the narrow window in $k^+$
\bea
{1\over v_{34}^2}\left({\Delta_0\over\sqrt{|p_4^+|}}-
{\Delta\over\sqrt{|p_3^+|}}\right)^2<k^+<
{1\over v_{34}^2}\left({\Delta_0\over\sqrt{|p_4^+|}}+
{\Delta\over\sqrt{|p_3^+|}}\right)^2 .
\eea
which contains a mixture of region I and II. But by taking $\Delta_0
\gg\Delta$ the contribution of this window can be made arbitrarily small,
and at the same time $\Delta$ can be neglected
in the lower limit on (\ref{II})
\bea
\int_{{\cal D}_3^{II}} d{\hat{\bfs k}}\ {|A^\lor|^2+|A^\land|^2\over16|k^+|\pi^3}&\approx&
{g^2|A_{\rm Core}|^2\over8\pi^2|k^+|}\left({(k^++p_3^+)^2\over p_3^{+2}}
+{p_3^{+2}\over(k^++p_3^+)^2}\right)
\ln{k^+p_3^+\Delta^2\over (k^++p_3^+)^2\mu^2e},\nonumber\\
&&\hskip2.5in\qquad{\rm for}~
|k^+|>{\Delta_0^2\over|p_4^+|v_{34}^2}
\label{jetleg3}
\eea
We limit the size of $\Delta_0$ so that all $k^+$ contributing to
region $I$ are negligible compared to the external $p_i^+$. In that
case the integrand for region $I$ simplifies to
\bea
|A^\lor|^2+|A^\land|^2&\approx&4g^2|A_{\rm Core}|^2
\bigg\{{{\bfs K}_{k,4}^2\over k^{+2}(k+p_4)^4}
+{{\bfs K}_{k,3}^2\over k^{+2}(k+p_3)^4}
-2{{\bfs K}_{k,3}\cdot{\bfs K}_{k,4}\over
k^{+2}(k+p_3)^2(k+p_4)^2}\bigg\}
\eea
In region $I$ one can show that $\sqrt{|k^+|}>\Delta(\sqrt{|p_3^+|}
-\sqrt{|p_4^+|})/\sqrt{|p_3^+p_4^+|}v_{34}$ so we stipulate that
$|p_3^+|>|p_4^+|$ so that $k^+$ stays away from 0 and it is safe to
take the continuum limit of the $k^+$ sums. 
(To deal with the case $|p_3^+|<|p_4^+|$
we just switch the roles of legs 3 and 4 in the calculation).
Now we  use the identity (\ref{kplusp2}) and
\bea
{\bfs K}_{k,3}\cdot{\bfs K}_{k,4}&=&k^{+2}p_3\cdot p_4-{1\over2}k^+p_3^+
(k+p_4)^2-{1\over2}k^+p_4^+(k+p_3)^2-\mu^2(p_3^+p_4^++k^+p_3^++k^+p_4^+)
\eea
to simplify the integrand even further
\bea
|A^\lor|^2+|A^\land|^2&\approx&4g^2|A_{\rm Core}|^2
\bigg\{{-p_4^+\over k^{+}(k+p_4)^2}
+{-p_3^+k^+(k+p_3)^2-\mu^2p_3^{+2}\over k^{+2}(k+p_3)^4}\nonumber\\
&&-{2k^{+2}p_3\cdot p_4+k^+p_3^+(k+p_4)^2+k^+p_4^+(k+p_3)^2\over
k^{+2}(k+p_3)^2(k+p_4)^2}\bigg\}\nonumber\\
&\approx&4g^2|A_{\rm Core}|^2
\bigg\{
{-\mu^2p_3^{+2}\over k^{+2}(k+p_3)^4}
-{2p_3\cdot p_4\over(k+p_3)^2(k+p_4)^2}\bigg\}
\eea
We have dropped $\mu^2$ terms in the numerators when the denominators
are prevented from vanishing strongly enough by virtue of being in region
$I$. Since only $(k+p_3)^2$ is allowed to get small in region $I$,
we only needed to keep $\mu^2$ when that factor appears 
squared in the denominator. The first term in braces will only
receive contributions in the integration for ${\hat k}^2=O(\mu^2)$
as $\mu\to0$. Thus the second constraint defining region $I$ collapses to
a constraint on $k^+$ only: $\Delta^2<k^+p_4^+v_{34}^2<\Delta_0^2$,
so the ${\hat{\bfs k}}$ integration can be freely done:
\bea
2\pi\int_0^{k^+\Delta^2/p_3^+} {\hat k}d{\hat k}{-\mu^2\over
({\hat{\bfs k}}^2+\mu^2)^2}=-\pi+\pi{\mu^2\over \mu^2+k^+\Delta^2/p_3^+}
\to -\pi=\pi\ln{1\over e},\qquad{\rm for}~\Delta^2<k^+p_4^+v_{34}^2<\Delta_0^2
\eea 

The second term in braces is not only Lorentz invariant but only involves
the variables constrained in defining region $I$:
\bea
-2k\cdot p_3<\Delta^2,\qquad \Delta^2<-2k\cdot p_4<\Delta_0^2
\label{covconstraints}
\eea
(In this form the constraints coincide with the earlier ones only when $\mu=0$.
But the effect of the change is $O(\mu^2)$ and negligible in
region $I$.)
Furthermore the measure for $k$ integration $dk^+d{\bfs k}/2|k^+|
=d^4k\delta(k^2+\mu^2)$ is also invariant. Thus it can be evaluated in any
convenient frame (for instance, one in which ${\bfs p}_3={\bfs p}_4=0$.)
The result is, assuming $\mu\ll\Delta,\Delta_0\ll p_i$, 
\bea
2\int_{{\cal D}_3^{I}} d^4k {\delta(k^2+\mu^2)\over (k+p_3)^2(k+p_4)^2}
\approx -{\pi\over p_{34}^2}\ln{\Delta_0^2\over\Delta^2}\ 
\ln{\Delta_0\Delta^3\over -p_{34}^2\mu^2}
\eea
Of course we can't directly compare this to our previous results
because they have not yet been integrated over $k^+$. But it is
amusing to compare it with the integral of 
(\ref{jetleg3}) over the missing range $\Delta^2<k^+p_4^+v_{34}^2<\Delta_0^2$:
\bea
{g^2|A_{\rm Core}|^2\over4\pi^2}\int {d|k^+|\over |k^+|}\ln{k^+\Delta^2\over p_3^+
\mu^2e}&\approx&{g^2|A_{\rm Core}|^2\over4\pi^2}\ln{\Delta_0^2\over\Delta^2}
\left[\ln{\Delta^2\over |p_3^+|\mu^2e} + \ln{\Delta_0\Delta
\over v_{34}^2|p_4^+|}\right]\nonumber\\
&=&{g^2|A_{\rm Core}|^2\over4\pi^2}\ln{\Delta_0^2\over\Delta^2}
\ln{\Delta_0\Delta^3\over -p_{34}^2\mu^2e} 
\eea
because $v_{34}^2=({\bfs v}_3-{\bfs v}_4)^2=-{p_{34}^2/ p_3^+p_4^+}$.
Remarkably, this integral exactly matches the entire effect from region $I$.
The upshot is, that even though we didn't do the calculation
this way, we can summarize the complete answer by quoting the
following ``results'' for the ${\bfs k}$ integrals
\bea
\int_{{\cal D}_{34}} d{\hat{\bfs k}}\ 
{|A^\lor|^2+|A^\land|^2\over16|k^+|\pi^3}&\approx&
{g^2|A_{\rm Core}|^2\over4\pi^2|k^+|}\ln{{k^{+4}{\bfs v}_{34}^4
(1-k^+p_4^+{\bfs v}^2_{34}/\Delta^2)\over \mu^4e^2}}
\qquad{\rm for}~|k^+|<{\Delta^2\over|p_4^+|{\bfs v}^2_{34}}\\
\int_{{\cal D}_{34}} d{\hat{\bfs k}}\ {|A^\lor|^2+|A^\land|^2\over16|k^+|\pi^3}&\approx&
{g^2|A_{\rm Core}|^2\over8\pi^2|k^+|}\bigg\{\left({(k^++p_4^+)^2\over p_4^{+2}}
+{p_4^{+2}\over(k^++p_4^+)^2}\right)
\ln{k^+p_4^+\Delta^2\over (k^++p_4^+)^2\mu^2e}\nonumber\\
&&\hskip-1.7in+\left({(k^++p_3^+)^2\over p_3^{+2}}
+{p_3^{+2}\over(k^++p_3^+)^2}\right)
\ln{k^+p_3^+\Delta^2\over (k^++p_3^+)^2\mu^2e}-2\ln\left(1-{\Delta^2\over
k^+p_4^+{\bfs v}^2_{34}}\right)\bigg\}
\qquad{\rm for}~|k^+|>{\Delta^2\over|p_4^+|{\bfs v}^2_{34}}
\eea
We must also remember that in executing the sum over $k^+$, the
phase space measure is treated differently for the jet 
associated with each leg. Namely, the $k^+$ sum associated with
the jet along leg $i$ is taken holding $k^++p_i^+\equiv P^+_i$ fixed and
there is an additional factor $|p_i^+|/|k^++p_i^+|=|P_i^+-k^+|/|P_i^+|$
arising from transforming $dp_i^+d{\bfs p}_i/p_i^+=(|P_i^+-k^+|/|P_i^+|)
dP_i^+d{\bfs P}_i/P_i^+$. Fortunately these subtle modifications
are only significant when $k^+=O(p_i^+)$, i.e. 
for a hard brem gluon whose contribution is dominated by a
single diagram. Putting everything together we can write
\bea
|{\cal M}^{\rm Brem}_{34}|^2&=&
{g^2\over8\pi^2}\sum_{i=3,4}|A^i_{\rm Core}|^2
\sum_{|k^+|>\Delta^2/|P^+_4|v_{34}^2}{1\over|k^+|}\left({P_i^+\over P_i^+-k^+}
+{(P_i^+-k^+)^3\over P_i^{+3}}\right)
\ln{k^+(P_i^+-k^+)\Delta^2\over P_i^{+2}\mu^2e}
\nonumber\\
&&+{g^2|A_{\rm Core}|^2\over4\pi^2}
\sum_{|k^+|<\Delta^2/|P^+_4|v_{34}^2}{1\over|k^+|}\ln{{k^{+4}{\bfs v}_{34}^4
\over \mu^4e^2}}\label{34brem}
\eea
where we have used the cancellation
\bea
{g^2|A_{\rm Core}|^2\over4\pi^2}\int_0^1 {dt\over t}\ln(1-t)
-{g^2|A_{\rm Core}|^2\over4\pi^2}\int_1^\infty {dt\over t}\ln(1-1/t)=0
\eea
valid after the (safe for these terms) limit of continuous $k^+$.

It is illuminating to rewrite (\ref{34brem}) in a way that makes the
symmetry under $3\leftrightarrow4$ manifest.
\bea
|{\cal M}^{\rm Brem}_{34}|^2&=&
{g^2\over8\pi^2}\sum_{i=3,4}|A^i_{\rm Core}|^2
\sum_{|k^+|}{1\over|k^+|}\left({P_i^+\over P_i^+-k^+}
+{(P_i^+-k^+)^3\over P_i^{+3}}\right)
\ln{k^+(P_i^+-k^+)\Delta^2\over P_i^{+2}\mu^2e}
\nonumber\\
&&+{g^2|A_{\rm Core}|^2\over4\pi^2}
\sum_{|k^+|<\Delta^2/|P^+_4|v_{34}^2}
{1\over|k^+|}\ln{{k^{+2}{\bfs v}_{34}^4|p_3^+p_4^+|
\over \Delta^4}}\\
&\approx&
{g^2\over8\pi^2}\sum_{i=3,4}|A^i_{\rm Core}|^2
\sum_{|k^+|}{1\over|k^+|}\left({P_i^+\over P_i^+-k^+}
+{(P_i^+-k^+)^3\over P_i^{+3}}\right)
\ln{k^+(P_i^+-k^+)\Delta^2\over P_i^{+2}\mu^2e}
\nonumber\\
&&+{g^2|A_{\rm Core}|^2\over4\pi^2}
\left[\sum_{|k^+|<A}
{1\over|k^+|}\ln{k^{+2}{\bfs v}_{34}^4|P_3^+P_4^+|
\over \Delta^4}-\ln{\Delta^2\over A|P_4^+|v_{34}^2}\
\ln{\Delta^2\over A|P_3^+|v_{34}^2}\right]
\label{34brem2}
\eea
Here we have picked $A\gg m$, the $k^+$ discretization unit, and have 
approximated
\bea
\sum_{A<|k^+|<\Delta^2/|P^+_4|v_{34}^2}
{1\over|k^+|}\ln{{k^{+2}{\bfs v}_{34}^4|P_3^+P_4^+|
\over \Delta^4}}&\approx&\int_A^{\Delta^2/|P^+_4|v_{34}^2}{dt\over t}
\ln{{t^{2}{\bfs v}_{34}^4|P_3^+P_4^+|
\over \Delta^4}}\nonumber\\
&=&
-\ln{\Delta^2\over A|P_4^+|v_{34}^2}\
\ln{\Delta^2\over A|P_3^+|v_{34}^2}
\eea
In the form (\ref{34brem2}) 
the symmetry $3\leftrightarrow4$ is transparent, but
the finiteness of the $\Delta$ dependence, manifest in (\ref{34brem}),
has been obscured: the coefficient of $\ln\Delta^2$ has a small
$k^+$ divergence on the first line that is canceled by a small
$k^+$ divergence on the second line. Another advantage of
(\ref{34brem2}) is that the first line just gives the 34
contribution to the production cross sections of jet 3 and jet 4
that matches our earlier discussion. In particular the $\mu\to0$
divergence is now transparently canceled by the wave
function renormalization.

A virtually identical calculation applies to the 
absorption of extra gluons in the initial state by 
the right of leg 1 and by the left of leg 2. Outgoing brem gluons
emitted between legs 1 and 2 are suppressed at $N_c=\infty$ just
as were incoming unseen gluons absorbed between legs 3 and 4
were suppressed. The result for soft gluon absorption between legs 1 and 2
is obtained from (\ref{34brem}) by substituting $1,2$ for $3,4$.
In this case $k^+,p_1^+,p_2^+$ are all positive so the many absolute
value signs can be dropped. We note that this treatment of the
initial state uses the Lee-Nauenberg procedure as a model
of incoming legs as incoming jets, so the four legs of the
core process are treated in a parallel fashion. This is in contrast to 
the by now standard procedure of absorbing the initial sate collinear
divergences in the initial state parton distribution functions.
This standard procedure is indeed approriate in interpreting
collider experiments, where the gluonic process describes
the scattering of the hard constituents of incoming hadrons. The
Lee-Nauenberg procedure we follow is more general and works even in
theories, such as ${\cal N}=4$ supersymmetric Yang-Mills, in
which hadron-like bound states of constituents don't form.

The situation for unseen gluon absorption and bremsstrahlung radiation
on the left and right, either between legs 1 and 4 or between legs 2 and
3, is more complicated even at large $N_c$,
because extra gluons in the initial state, the final
state and both must be taken into account. This is because all these processes
are now allowed at $N_c=\infty$.
For definiteness let's focus on the region between
1 and 4: gluons emitted or absorbed by the left of leg 4 and the left of leg 1.
The diagrams for emission of a single unobserved gluon are shown on the
second line of Fig.~\ref{brem}. The squared amplitudes are
\bea
|A^\lor|^2+|A^\land|^2&=& 4g^2|A_{\rm Core}|^2
\bigg\{\left[{P_4^{+2}\over2(P_4^{+}-k^+)^2}
+{(P_4^{+}-k^+)^2\over2P_4^{+2}}\right]
{({\bfs k}-k^+{\bfs v}_4)^2\over[({\bfs k}-k^+{\bfs v}_4)^2
+\mu^2P_4^{+2}/(P_4^{+}-k^+)^2]^2}\nonumber\\
&&+\left[{P_1^{+2}\over2(P_1^{+}-k^+)^2}
+{(P_1^{+}-k^+)^2\over2P_1^{+2}}\right]
{({\bfs k}-k^+{\bfs v}_1)^2\over[({\bfs k}-k^+{\bfs v}_1)^2
+\mu^2P_1^{+2}/(P_1^{+}-k^+)^2]^2}\nonumber\\
&&-\left[{P^+_4(P_1^+-k^+)\over P^+_1(P_4^+-k^+) }+{ P^+_1(P_4^+-k^+)
\over P^+_4(P_1^+-k^+) }\right]\nn\\
&&\times {({\bfs k}-k^+{\bfs v}_1)\cdot({\bfs k}-k^+{\bfs v}_4)\over
[({\bfs k}-k^+{\bfs v}_1)^2
+\mu^2P_1^{+2}/(P_1^{+}-k^+)^2][({\bfs k}-k^+{\bfs v}_4)^2
+\mu^2P_4^{+2}/(P_4^{+}-k^+)^2]
}\bigg\}\nonumber
\eea
here we have used $P_i^+=p_i^++k^+$, expressed the
$(k+p_i)^2$ in terms of ${\bfs K}_{ij}$ which
we have written out explicitly. The main differences with the
34 contribution are that the helicities of the two legs
are opposite and $p_1^+>0$ while $p_4^+,k^+<0$.
Similarly with an extra unobserved soft gluon in the
initial state the relevant diagrams have a gluon {\it absorbed} on the
left of leg 1 or leg 4, but the result of the calculation
is the same as for emission with the understanding that
$k^+>0$. 

As we have already seen in the 34 case, the interference
term is negligible unless all components of $k$ are small so 
we can simplify that term by neglecting $k^+$ compared to the $p_i^+$.
For the future discussion we also write out separately the
outgoing and incoming gluon cases sending $k\to-k$ in the
outgoing case so that $k^+>0$ in both cases:
\bea
|A_{\rm Out}^\lor|^2+|A_{\rm Out}^\land|^2&\approx& 4g^2|A_{\rm Core}|^2
\bigg\{\left[{P_4^{+2}\over2(P_4^{+}+k^+)^2}
+{(P_4^{+}+k^+)^2\over2P_4^{+2}}\right]
{({\bfs k}-k^+{\bfs v}_4)^2\over[({\bfs k}-k^+{\bfs v}_4)^2
+\mu^2P_4^{+2}/(P_4^{+}+k^+)^2]^2}\nonumber\\
&&+\left[{P_1^{+2}\over2(P_1^{+}+k^+)^2}
+{(P_1^{+}+k^+)^2\over2P_1^{+2}}\right]
{({\bfs k}-k^+{\bfs v}_1)^2\over[({\bfs k}-k^+{\bfs v}_1)^2
+\mu^2P_1^{+2}/(P_1^{+}+k^+)^2]^2} \label{out14brem}\\
&&
-{2({\bfs k}-k^+{\bfs v}_1)\cdot({\bfs k}-k^+{\bfs v}_4)\over
[({\bfs k}-k^+{\bfs v}_1)^2
+\mu^2][({\bfs k}-k^+{\bfs v}_4)^2
+\mu^2]
}\bigg\}\nonumber\\
|A_{\rm In}^\lor|^2+|A_{\rm In}^\land|^2&=& 4g^2|A_{\rm Core}|^2
\bigg\{\left[{P_4^{+2}\over2(P_4^{+}-k^+)^2}
+{(P_4^{+}-k^+)^2\over2P_4^{+2}}\right]
{({\bfs k}-k^+{\bfs v}_4)^2\over[({\bfs k}-k^+{\bfs v}_4)^2
+\mu^2P_4^{+2}/(P_4^{+}-k^+)^2]^2}\nonumber\\
&&+\left[{P_1^{+2}\over2(P_1^{+}-k^+)^2}
+{(P_1^{+}-k^+)^2\over2P_1^{+2}}\right]
{({\bfs k}-k^+{\bfs v}_1)^2\over[({\bfs k}-k^+{\bfs v}_1)^2
+\mu^2P_1^{+2}/(P_1^{+}-k^+)^2]^2}\label{in14brem}\\
&&-
{2({\bfs k}-k^+{\bfs v}_1)\cdot({\bfs k}-k^+{\bfs v}_4)\over
[({\bfs k}-k^+{\bfs v}_1)^2
+\mu^2][({\bfs k}-k^+{\bfs v}_4)^2
+\mu^2]}\bigg\}\nonumber
\eea
Clearly simply adding in these two contributions can't be the
whole story since only one of them is needed to cancel IR
divergences from loops.

There is also a difficulty with the
contributions for a hard unobserved gluon.
A collinear divergence is present when an extra hard outgoing gluon 
is collinear with either leg 4 or leg 1. When collinear
with leg 4 it is simply part of the jet associated with that leg and
combines with the collinear gluon emission from the right of leg 4
to cancel the collinear divergence in the self energy correction
to leg 4. The divergence coming from an outgoing extra gluon collinear
with leg 1 has a very different meaning. Since we have stipulated that
none of the gluons in the core process are collinear, this hard extra gluon
is well separated from gluons 3 and 4 and therefore in principle detectable:
In this case the final state is unambiguously a three gluon state. A similar
situation applies when an extra hard gluon in the initial state is
collinear with leg 4. These non-jet-like collinear divergences have nothing
to do with self-energy corrections on external lines and must be canceled
by something else.

\begin{figure}[ht]
\psfrag{'+'}{{\large$+$}}
\psfrag{'.5'}{{\huge${1\over2}$}}
\psfrag{'triangle'}{\large Triangle-like}
\psfrag{'q'}{}
\psfrag{'4'}{$4$}
\psfrag{'1'}{$1$}
\psfrag{'2'}{$2$}
\psfrag{'3'}{$3$}
\begin{center}
\includegraphics[width=5.5in]{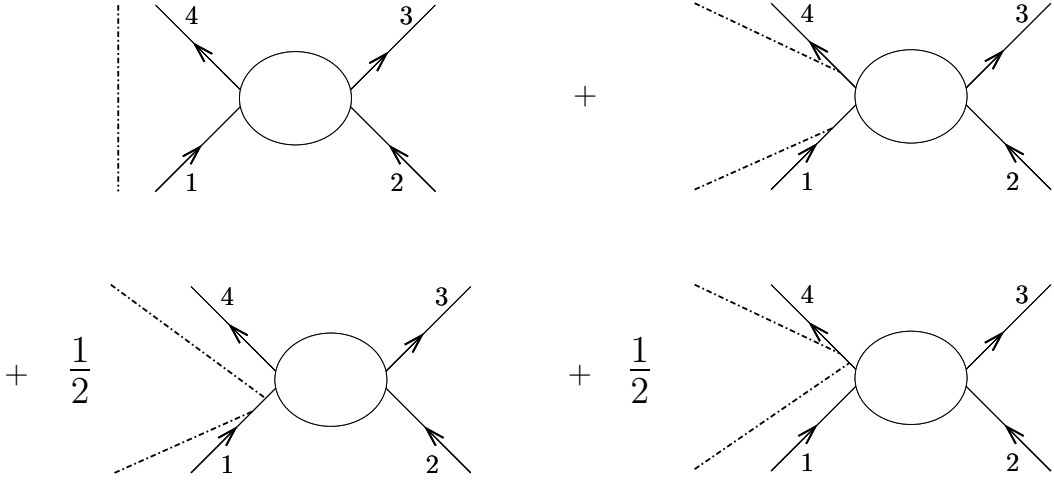}
\caption{Diagrams representing an unseen bremsstrahlung gluon in both
the initial and final state. The interference term, between the disconnected
diagram on the top left and the remaining  three diagrams, in the
square of the sum of the four diagrams is of order $O(g^6)$ and therefore
comparable to the bremsstrahlung probability for one unseen gluon in the
final state and none in the initial state or {\it vice versa}. 
To account for the factors of $1/2$ see Fig.~\ref{unitarity}.}
\label{inoutbrem}
\end{center}
\end{figure}

\begin{figure}[ht]
\psfrag{'+'}{{\large$+$}}
\psfrag{'.5'}{{\huge${1\over2}$}}
\psfrag{'triangle'}{\large Triangle-like}
\psfrag{'k'}{$k$}
\psfrag{'a'}{$A$}
\psfrag{'b'}{$B$}
\psfrag{'c'}{$C$}
\psfrag{'-k'}{$-k$}
\begin{center}
\includegraphics[width=5.5in]{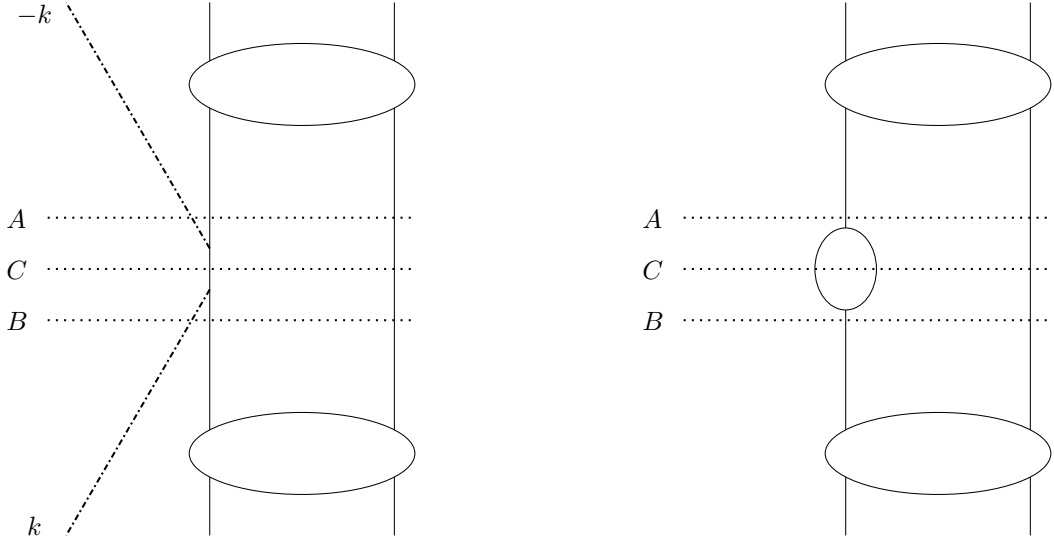}
\caption{One of the processes on the second line of Fig.~\ref{inoutbrem}
viewed as a unitarity cut of a larger diagram.
Because the unobserved gluon injects zero momentum into
the single line it scatters from, the unitarity cuts $A$ and $B$ are equivalent
and only one of them should be used in constructing the unitarity sum
of the squared amplitude.
This explains the factors of $1/2$ in Fig.~\ref{inoutbrem}. 
The situation is entirely analogous to the well-known procedure
of weighting self-energy bubbles on external lines with a factor of
$1/2$, as indicated on the right. In both cases the unitarity cut $C$
is distinct and unique.}
\label{unitarity}
\end{center}
\end{figure}

The mechanism \cite{leenauenberg} that takes care of the 
doubled soft bremsstrahlung and the
non-jet-like collinear divergences is shown in Fig.~\ref{inoutbrem}.
At first glance it seems that these diagrams wouldn't be relevant,
because they are either disconnected (the first diagram) or apparently
higher order (the diagrams involving two extra gluons). However, when
we square the sum of these diagrams, the cross terms between the
first term and the remaining three contribute as  connected structures
which are exactly the right order $O(g^6)$ to be comparable
to the one-loop and single gluon bremsstrahlung diagrams. The reason the
last two diagrams are multiplied by $1/2$ is 
explained in Fig.~\ref{unitarity}.
It is essentially the same reasoning as that for multiplying self-energy
bubbles on external lines by $1/2$. Otherwise the squared amplitude
would count equivalent contributions twice compared to
what is required by closure for the unitary evolution
operator $U(t_1,t_2)$ or, equivalently, by unitarity of the S-matrix.

We now discuss the evaluation of the cross terms in the square
of the sum of diagrams in Fig.~\ref{inoutbrem}. Since the gluon 
momentum can't simultaneously be collinear with both
leg 1 and leg 4, the cross term with the second diagram
on line 1 will only be singular
for a soft forward scattered gluon. With incoming gluon polarization
$\land$ this contribution is given by
\bea
-(2g)^2\left[{(p_4^+-k^+)(p_1^++k^+)K^\lor_{4,k}K^\land_{1,k}
\over k^{+2}p_1^+p_4^+(p_4-k)^2(p_1+k)^2}\quad +\quad {\rm c.c.}\right]
&=&-4g^2{(p_4^+-k^+)(p_1^++k^+){\bfs K}_{4,k}\cdot{\bfs K}_{1,k}
\over k^{+2}p_1^+p_4^+(p_4-k)^2(p_1+k)^2}\nonumber\\
&&\hskip-3in \approx\ 4g^2{({\bfs k}-k^+{\bfs v}_4)
\cdot({\bfs k}-k^+{\bfs v}_1)
\over [({\bfs k}-k^+{\bfs v}_4)^2+\mu^2]
[({\bfs k}-k^+{\bfs v}_1)^2+\mu^2]}
\to\ 4g^2{({\bfs k}-k^+{\bfs v}_4)
\cdot({\bfs k}-k^+{\bfs v}_1)
\over ({\bfs k}-k^+{\bfs v}_4)^2({\bfs k}-k^+{\bfs v}_1)^2}
\eea
where the first form on the second line neglects $k^+$ compared to
$p_1^+$ and $p_4^+$, and the second form also uses the fact that the
IR divergence of this contribution is insensitive to the temporary
gluon mass $\mu$. Although the $p^+$ dependent factors for the other
polarization of incoming gluon $\lor$ are slightly different,
this difference disappears for soft gluons. So adding the
two polarizations just multiplies this result by a factor 2.
Comparing this expression to (\ref{out14brem}), (\ref{in14brem}), 
we see that it will
cancel one of the two interference terms, so only one will be counted.

Next we turn to the diagrams on the second line of Fig.~\ref{inoutbrem}.
Since they are similar to each other, we only need do one
in detail, say the first. It will give a singular contribution 
not only when $k$
is soft, but also when it is hard and collinear with leg 1. In the
latter case there will be three well separated gluons in the
final state and therefore no confusion with the two gluon final state
we want to describe. Nonetheless the collinear singularity from this
contribution will cancel the one from a collinear brem gluon emission
from leg 1 with no extra gluon in the initial state. 

These diagrams with the forward scattering process entirely on an
external leg are formally singular because the propagator between the
last emission vertex and the rest of the diagram is on shell.
In this regard it is analogous to self energy corrections on external lines.
The most reliable way to handle such situations is to compute the
forward scattering process with $p_1$ off-shell, giving the
intermediate gluons a small mass $\mu$, and then go on shell 
by extracting the residue of the pole and factorizing it
which means keeping only half of the correction to the
residue (see Fig.~\ref{unitarity}). 
This process correctly
discards the double pole contribution which is properly interpreted
as an energy shift.\footnote{A rougher procedure is to attempt to work
on shell from the beginning and use the $i\epsilon$'s in the denominator
to keep things finite. Then we would obtain
\bea
A^{\land\lor}&=&{4g^2K_{k,1}^\lor K_{k,1}^\land
\over-i\epsilon[(p_1+k)^2+\mu^2-i\epsilon]}{(p_1^++k^+)^2\over k^{+2}p_1^{+2}}
A_{\rm Core}\\
A^{\lor\land}&=&{4g^2K_{k,1}^\land K_{k,1}^\lor
\over-i\epsilon[(p_1+k)^2+\mu^2-i\epsilon]}
\left[{p_1^{+2}\over k^{+2}(p_1^++k^+)^2}+{k^{+2}\over p_1^{+2}
(p_1^++k^+)^2}\right]A_{\rm Core}
\eea
When the cross term is constructed each of 
the above expressions will
be multiplied by $A_{\rm Core}^*$ and added to its complex conjugate, 
and finally multiplied by $1/2$:
\bea
{\rm Cross~Term}^{\land\lor}&=&{1\over2}{4g^2{\bfs K}_{k,1}^2 
\over-i\epsilon2}{(p_1^++k^+)^2\over k^{+2}p_1^{+2}}
|A_{\rm Core}|^2\left[{1\over(p_1+k)^2+\mu^2-i\epsilon}
-{1\over(p_1+k)^2+\mu^2+i\epsilon}
\right]\nonumber\\
&=&-{(p_1^++k^+)^2\over k^{+2}p_1^{+2}}
{4g^2{\bfs K}_{k,1}^2 |A_{\rm Core}|^2
\over2[(p_1+k)^2+\mu^2-i\epsilon][(p_1+k)^2+\mu^2+i\epsilon]}\\
{\rm Cross~Term}^{\lor\land}&=&-
\left[{p_1^{+2}\over k^{+2}(p_1^++k^+)^2}+{k^{+2}\over p_1^{+2}
(p_1^++k^+)^2}\right]{4g^2{\bfs K}_{k,1}^2 |A_{\rm Core}|^2
\over2[(p_1+k)^2+\mu^2-i\epsilon][(p_1+k)^2+\mu^2+i\epsilon]}
\eea
Compared to the treatment in the text, which defines jet
amplitudes as they would be extracted from larger diagrams with
unitarity cuts, this rough procedure
misses an overall factor $(p_1^++k^+)/p_1^+$. For soft $k$ this discrepancy
is negligible, but for hard collinear $k$ the treatment in the text
is the one that reflects the Lee-Nauenberg theorem.} 

Applying the Feynman rules to this diagram (without the factor of 
$1/2$) with $p_1$ off-shell and $k^-=({\bfs k}^2+\mu^2)/2k^+$ we find 
\bea
{A^{\land\lor}\over p_1^2}&=&{4g^2K_{k,1}^\lor K_{k,1}^\land
\over (p_1^2)^2
[(p_1+k)^2+\mu^2]}{(p_1^++k^+)^2\over k^{+2}p_1^{+2}}
A_{\rm Core}(p_1,p_2,p_3,p_4)\nonumber\\
&=&{4g^2(p_1^++k^+)^2\over k^{+2}p_1^{+2}(p_1^2)^2}
{-k^+p_1^+({\bfs k}-k^+{\bfs v}_1)^2
\over 
2[({\bfs k}-k^+{\bfs v}_1)^2+\mu^2-k^+(k^++p_1^+)p_1^2/p_1^{+2}]}
A_{\rm Core}(p_1,p_2,p_3,p_4)\nonumber\\
&\sim&-{2g^2(p_1^++k^+)^2\over k^{+}p_1^{+}(p_1^2)^2}
\left[{({\bfs k}-k^+{\bfs v}_1)^2
\over({\bfs k}-k^+{\bfs v}_1)^2+\mu^2}+{p_1^2\over p_1^{+2}}
{k^+(k^++p_1^+)({\bfs k}-k^+{\bfs v}_1)^2
\over[({\bfs k}-k^+{\bfs v}_1)^2+\mu^2]^2}\right]A_{\rm Core}
\eea
where the superscripts indicate the polarization of the unseen gluon.
The first term in square brackets is a double pole in $p_1^2$ with a
coefficient that will not have a collinear divergence (when smeared over
${\bfs k}$.) Its residue will be proportional to the
derivative of $A_{\rm Core}$ but will not contribute to the collinear 
divergence. The residue of only the second term in square brackets,
which is a single pole in $p_1^2$ will be divergent, and we
easily read off the
divergent contribution for this polarization
\bea
A^{\land\lor}&\sim&-{2g^2(p_1^++k^+)^2\over k^{+}p_1^{+}}
{k^+(k^++p_1^+)\over p_1^{+2}}
{({\bfs k}-k^+{\bfs v}_1)^2
\over[({\bfs k}-k^+{\bfs v}_1)^2+\mu^2]^2}{A_{\rm Core}}
\eea
The other polarization is given by 
\bea
A^{\lor\land}&=&-{2g^2}
{k^+(k^++p_1^+)\over p_1^{+2}}
\left[{p_1^{+3}\over k^{+}(p_1^++k^+)^2}+{k^{+3}\over p_1^{+}
(p_1^++k^+)^2}\right]{({\bfs k}-k^+{\bfs v}_1)^2
\over[({\bfs k}-k^+{\bfs v}_1)^2+\mu^2]^2}A_{\rm Core}
\eea
There are two contributions for $\lor\land$ polarization because
the gluon connecting the absorption and emission vertices can have
either polarization.

When the cross term is constructed each of 
the above expressions will
be multiplied by $A_{\rm Core}^*$ and added to its complex conjugate,
which doubles it, and finally multiplied by $1/2$ which undoubles it:
\bea
{\rm Cross~Term}^{\land\lor}&=&-{2g^2(p_1^++k^+)^3\over p_1^{+3}}
{({\bfs k}-k^+{\bfs v}_1)^2
\over[({\bfs k}-k^+{\bfs v}_1)^2+\mu^2]^2}|A_{\rm Core}|^2\\
{\rm Cross~Term}^{\lor\land}&=&-{2g^2}
\left[{p_1^{+}\over p_1^++k^+}+{k^{+4}\over p_1^{+3}
(p_1^++k^+)}\right]{({\bfs k}-k^+{\bfs v}_1)^2
\over[({\bfs k}-k^+{\bfs v}_1)^2+\mu^2]^2}|A_{\rm Core}|^2
\label{forward1}
\eea
Adding together the 
contribution for the two polarizations gives
\bea
{\rm Cross~Term}&=&-2g^2\left[{(p_1^++k^+)^3\over p_1^{+3}}
+{p_1^{+}\over p_1^++k^+}+{k^{+4}\over p_1^{+3}
(p_1^++k^+)}\right]
{({\bfs k}-k^+{\bfs v}_1)^2
\over[({\bfs k}-k^+{\bfs v}_1)^2+\mu^2]^2}|A_{\rm Core}|^2
\label{forward2}
\eea
The process described here is one where an outgoing extra gluon
is collinear with gluon 1 which is incoming. This gluon is
thus not in jet 3 or jet 4. It can therefore be
experimentally detected unless it is too soft. 
Comparing to (\ref{out14brem})
we see that the first two terms of (\ref{forward2}) almost
cancel the second term of (\ref{out14brem}). Integrating the difference
over a neighborhood of ${\bfs k}=k^+{\bfs v}_1$  
in the limit $\mu\to0$ involves
\bea
\int d{\bfs k}\left[{({\bfs k}-k^+{\bfs v}_1)^2\over[({\bfs k}-k^+{\bfs v}_1)^2
+\mu^2P_1^{+2}/(P_1^{+}+k^+)^2]^2}-{({\bfs k}-k^+{\bfs v}_1)^2
\over[({\bfs k}-k^+{\bfs v}_1)^2+\mu^2]^2}\right]\sim
\pi\ln{P_1^{+2}\over(P_1^{+}+k^+)^2}
\eea
This vanishes as $k^+\to0$, when the extra gluon is undetectable,
and when $k^+$ is finite it will be excluded from an outgoing jet
along $p_3$ or $p_4$. Thus it shouldn't be included in the
undetected bremsstrahlung associated with the core process.
The last diagram of Fig.~\ref{inoutbrem} removes the divergent
contribution of the first term of (\ref{in14brem}) in a similar way.
The last term of (\ref{forward2}) is negligible
for soft bremsstrahlung and, when $k^+=O(1)$ cancels a spurious collinear 
divergence from a brem gluon emitted from gluon 1 with the
opposite helicity to that considered here. 
The squared amplitude for that opposite helicity process is
\bea
|A_{\rm Brem}^{\prime\land}|^2&=&
2g^2{k^{+4}\over p_1^{+2}(p_1^+-k^+)^2}
{({\bfs k}-k^+{\bfs v}_1)^2|A_{\rm Core}(p_1-k,p_2,p_3,p_4)|^2\over
[({\bfs k}-k^+{\bfs v}_1)^2+\mu^2(p_1^+-k^+)^2/p_1^{+2}]^2}
\eea
Here the superscript denotes the polarization of the brem gluon, and
$-k$ is the incoming momentum of the brem gluon (so $k^+>0$). 
Notice that the first argument of $A_{\rm Core}$ is $P_1\equiv p_1-k$, not
$p_1$ as in (\ref{forward2}). 
To see all these cancellations, it is important to
recall that ${\bfs k}$ is to be smeared
in a narrow region about the collinear point ${\bfs k}=k^+{\bfs p_1}/p_1^+$.
Then ${\bfs P}_1\equiv{\bfs p}_1-{\bfs k}\approx (p_1^+-k^+){\bfs p}_1/p_1^+$
so the integration measure of the bremsstrahlung probability is
\bea
{d{\bfs k}\over2|k^+|}{d{\bfs p}_1\over2p_1^+}\approx
{d{\bfs k}\over2|k^+|}{d{\bfs P}_1\over2P_1^+}{P_1^++k^+\over P_1^+}
\eea 
In (\ref{forward2}) the first argument of $A_{\rm Core}$ is $p_1$
which is to be identified with $p$ here. So we should compare
(\ref{forward2}) to $(p^++k^+)/p^+$ times the appropriate term in
(\ref{out14brem}).

After these cancellations what remains of the bremsstrahlung
contributions (\ref{out14brem}), (\ref{in14brem}) is just
\bea
&& 4g^2|A_{\rm Core}|^2
\bigg\{\left[{P_4^{+2}\over2(P_4^{+}+k^+)^2}
+{(P_4^{+}+k^+)^2\over2P_4^{+2}}\right]
{({\bfs k}-k^+{\bfs v}_4)^2\over[({\bfs k}-k^+{\bfs v}_4)^2
+\mu^2P_4^{+2}/(P_4^{+}+k^+)^2]^2}\nonumber\\
&&
-{2({\bfs k}-k^+{\bfs v}_1)\cdot({\bfs k}-k^+{\bfs v}_4)\over
[({\bfs k}-k^+{\bfs v}_1)^2
+\mu^2][({\bfs k}-k^+{\bfs v}_4)^2
+\mu^2]
}\nonumber\\
&&+\left[{P_1^{+2}\over2(P_1^{+}-k^+)^2}
+{(P_1^{+}-k^+)^2\over2P_1^{+2}}\right]
{({\bfs k}-k^+{\bfs v}_1)^2\over[({\bfs k}-k^+{\bfs v}_1)^2
+\mu^2P_1^{+2}/(P_1^{+}-k^+)^2]^2}\bigg\}
\eea 
But now notice that these residual terms are the same as the
34 contribution if we identify $P_3$ with $-P_1$. Thus when
summed over $k^+$ the 14 total contribution can be obtained 
from the 34 contribution with the substitution $p_3\to -p_1$.
This is consistent because unlike $p_1+p_4$, which
is space-like, $p_4-p_1$ is time-like.

To summarize this section we collect together all the contributions
from hard collinear gluons, soft gluons and self energy 
corrections on external lines. The soft contributions boil down to
a contribution like (\ref{34brem}) for each pair of neighboring
lines:
\bea
&&\sum_{i=1}^4\sum_{|k^+|<|P_i^+|}{g^2|A_{\rm Core}|^2
\over 8\pi^2} 
\bigg({|P_i^+|\over|k^+(P_i^+-k^+)|}+{|P_i^+-k^+|^{3}
\over |k^+P_i^{+3}|}+{|k^+|^{3}
\over |(P_i^+-k^+)P_i^{+3}|}\bigg)
\ln{|k^+(P_i^+-k^+)|\Delta^2\delta e^\gamma\over|P_i^+|^2}\nonumber\\
&&+\sum_{i=1}^4{g^2|A_{\rm Core}|^2\over4\pi^2}
\bigg[\sum_{|k^+|<A}
{1\over|k^+|}\ln{k^{+2}{\bfs v}_{i,i+1}^4|P_i^+P_{i+1}^+|
\over \Delta^4}-\ln{\Delta^2\over A|P_i^+|v_{i,i+1}^2}\
\ln{\Delta^2\over A|P_{i+1}^+|v_{i,i+1}^2}\bigg]\nonumber
\eea 
Here $v_{ij}^2=({\bfs v}_i-{\bfs v}_j)^2=-{(p_i+p_j)^2/ P_i^+P_j^+}
={|(p_i+p_j)^2|/|P_i^+P_j^+|}$. Notice that the terms on the
first line correspond to contributions associated with each leg
of the diagram, whereas those on the second line involve
contributions associated with pairs of consecutive lines. The first 
category of terms includes the wave function renormalization
due to self energy bubbles on external lines
\bea
\sum_{i=1}^4(Z_i-1)|A_{\rm Core}|^2={g^2\over 4\pi^2}
\sum_{i=1}^4{|A_{\rm Core}|^2\over|p_i^+|}
\sum_{k^+}
\left(x_i(1-x_i)+{x_i\over 1-x_i}
+{1-x_i\over x_i}\right)\ln(\mu^2\delta e^{\gamma+1})
\eea
so that the collinear divergence as $\mu\to0$ cancels. 
So we see that the temporary cutoff $\mu$ can be removed
as soon as we combine everything together. We shall see that the
remaining divergences in these expressions cancel against
similar ones that come from the remaining one loop corrections
to the glue-glue scattering process. These include self energy insertions
on internal lines together with triangle and box diagrams.
\vskip14pt
\section{Cubic Vertex Function}
We shall not include calculational details for the one loop
corrections to the cubic vertex function. They can be found in
\cite{thornlcnotes}. Instead we present the final answers
for the vertex corrections with two on-shell gluons.
\begin{figure}[ht]
\psfrag{'q'}{$q$}
\psfrag{'k0'}{$k_0$}
\psfrag{'k1'}{$k_1$}
\psfrag{'k2'}{$k_2$}
\begin{center}
\includegraphics[width=8cm]{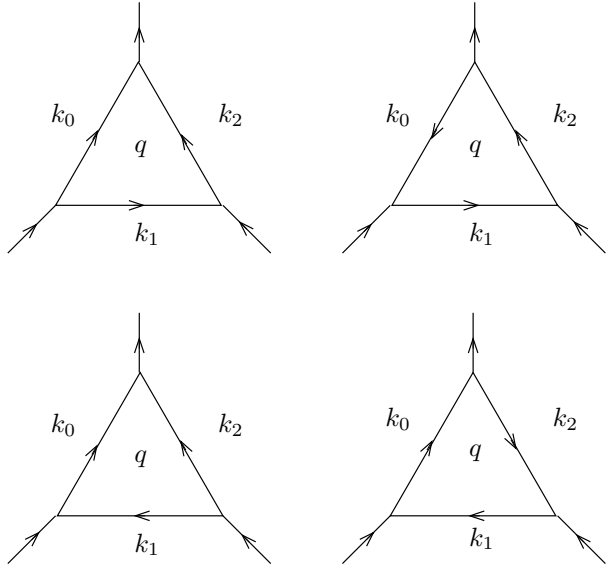}
\caption{The triangle diagrams contributing to $\Gamma^{\land\land\lor}$.
The labels $q,k_0,k_1,k_2$ are dual momenta. The actual momentum of any line
is the difference of the momenta of the regions it bounds. $p_1=k_1-k_0$,
etc.}
\label{uudgraphs}
\end{center}
\end{figure}
\begin{figure}[ht]
\psfrag{'q'}{$q$}
\psfrag{'k0'}{$k_0$}
\psfrag{'k1'}{$k_1$}
\psfrag{'k2'}{$k_2$}
\begin{center}
\includegraphics[width=14cm]{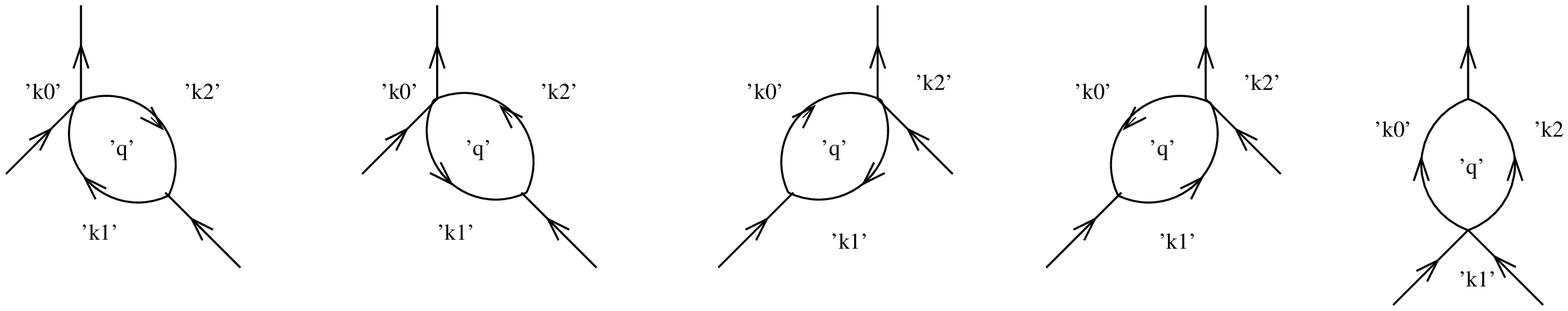}
\caption{The swordfish diagrams contributing to $\Gamma^{\land\land\lor}$.}
\label{uudsfgraphs}
\end{center}
\end{figure}
We put the 
combination of swordfish and triangle diagrams 
(see Figs.~\ref{uudgraphs},\ref{uudsfgraphs}) 
with two like-helicities
and two legs on-shell in the form
\bea
\Gamma_{\rm 1~loop}+\Gamma_{\rm C.T.}
=-{g^2\over8\pi^2}
\Gamma_{\rm tree}\left({70\over9}
-{11\over3}\ln(\delta p_o^2e^{\gamma})+S\right)
+\alpha{g^3\over12\pi^2}{K\over p_o^+}
\label{cubiccorrection}
\eea
where the vectors $k_i,K$ carry the polarization of the two like-
helicity gluons, $p_o$ is the four-momentum of the off-shell gluon,
$\alpha=1$ when the on-shell gluons have like-helicity,
and $\alpha=0$ otherwise. Finally $S$
is an infrared sensitive term that depends on the location of
the off-shell gluon, but not on any of the gluon helicities. 
In the
case $p_1^+, p_2^+>0$, we denote by $S_i^{q^+}(p_1,p_2)$ the
value of $S$ when leg $i$ is off-shell, and with loop momentum chosen
so that $q^+$  is the longitudinal momentum of the
internal line joining leg $1$ to leg $3$, satisfying $0<q^+<p_{12}^+$.
Then, \bea
S_1^{q^+}(p_1,p_2)&=&\sum_{q^+<p_1^+}  
\bigg\{
\bigg[{2\over q^+}+{1\over p_1^++p_2^+-q^+}+{1\over p_1^+-q^+}
\bigg]\left(\ln(\delta p_1^2e^{\gamma})
+\ln{q^+\over p_1^+}\right)\nonumber\\
&&+\bigg[{2\over p_1^+-q^+}-{1\over p_1^++p_2^+-q^+}+{1\over q^+}
\bigg]\ln{p_1^+-q^+\over p_1^+}\bigg\}\nonumber\\
&&+\sum_{q^+>p_1^+} \bigg\{
\bigg[{1\over q^+}+{2\over p_1^++p_2^+-q^+}+{1\over q^+-p_1^+}
\bigg]\left(\ln(\delta p_1^2e^{\gamma})
+\ln{p_1^++p_2^+-q^+\over p_2^+}\right)
\nonumber\\
&&+\sum_{q^+\neq p_1^+}\bigg[{1\over q^+}+{2\over p_1^++p_2^+-q^+}
+{1\over q^+-p_1^+}
\bigg]\ln{p_1^++p_2^+-q^+\over p_1^++p_2^+}\\
S_2^{q^+}(p_1,p_2)&=&\sum_{q^+\neq p_1^+}
\left[{2\over q^+}+{1\over p_1^++p_2^+-q^+}+{1\over p_1^+-q^+}\right]
\ln{q^+\over p_1^++p_2^+}\nonumber\\
&&+\sum_{q^+<p_1^+}  
\bigg\{\bigg[{2\over q^+}+{1\over p_1^++p_2^+-q^+}+{1\over p_1^+-q^+}
\bigg]\left(\ln(\delta p_2^2e^{\gamma})+\ln{q^+\over p_1^+}\right)
\bigg\}\nonumber\\
&&+\sum_{q^+>p_1^+}\bigg\{
\bigg[{1\over q^+}+{2\over p_1^++p_2^+-q^+}+{1\over q^+-p_1^+}
\bigg]\left(\ln(\delta p_2^2e^{\gamma})
+\ln{p_1^++p_2^+-q^+\over p_2^+}\right)
\nonumber\\
&&\qquad\qquad
+\bigg[{2\over q^+-p_1^+}+{1\over p_1^++p_2^+-q^+}-{1\over q^+}
\bigg]
\ln{q^+-p_1^+\over p_2^+}\bigg\}\\
S_3^{q^+}(p_1,p_2)&=&\sum_{q^+<p_1^+}  
\bigg\{
\bigg[{2\over q^+}+{1\over p_1^++p_2^+-q^+}+{1\over p_1^+-q^+}
\bigg]\left(\ln(\delta p_{12}^2e^{\gamma})
+\ln{q^+\over p_1^++p_2^+}\right)\nonumber\\
&&+\bigg[{1\over q^+}+{2\over p_1^++p_2^+-q^+}+{1\over q^+-p_1^+}
\bigg]\ln{p_1^++p_2^+-q^+\over p_1^++p_2^+}\nonumber\\
&&+\bigg[{2\over p_1^+-q^+}-{1\over p_1^++p_2^+-q^+}+{1\over q^+}
\bigg]\ln{p_1^+-q^+\over p_1^+}\bigg\}\nonumber\\
&&+\sum_{q^+>p_1^+} \bigg\{
\bigg[{1\over q^+}+{2\over p_1^++p_2^+-q^+}+{1\over q^+-p_1^+}
\bigg]\left(\ln(\delta p_{12}^2e^{\gamma})
+\ln{p_1^++p_2^+-q^+\over p_1^++p_2^+}\right)
\nonumber\\
&&+\bigg[{2\over q^+}+{1\over p_1^++p_2^+-q^+}+{1\over p_1^+-q^+}
\bigg]\ln{q^+\over p_1^++p_2^+}
\nonumber\\ &&
+\bigg[{2\over q^+-p_1^+}+{1\over p_1^++p_2^+-q^+}-{1\over q^+}
\bigg]\ln{q^+-p_1^+\over p_2^+}\bigg\}
\eea

\begin{figure}[ht]
\psfrag{'q'}{$q$}
\psfrag{'k0'}{$k_0$}
\psfrag{'k1'}{$k_1$}
\psfrag{'k2'}{$k_2$}
\begin{center}
\includegraphics[width=8cm]{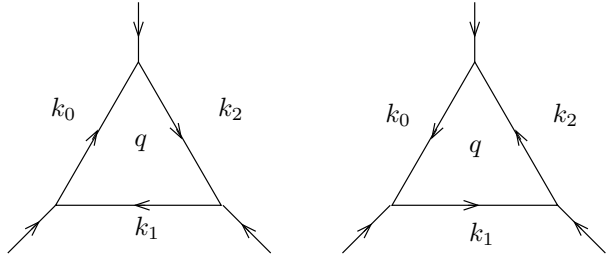}
\caption{The triangle diagrams contributing to $\Gamma^{\land\land\land}$.}
\label{uuugraphs}
\end{center}
\end{figure}

In addition to these corrections to the tree level cubic vertex, the
triangle diagram with three like-helicities (see Fig.~\ref{uuugraphs}) 
is non-zero, and it is given,
for the case of two on-shell legs, by
\bea
\Gamma_\triangle^{\land\land\land}&=&-{g^3\over6\pi^2}
{K^{\land3}
\over p_1^+p_2^+p_3^+ p_o^2}\\
\Gamma_\triangle^{\lor\lor\lor}&=&-{g^3\over6\pi^2}{K^{\lor3}
\over p_1^+p_2^+p_3^+ p_o^2}
\eea
where $p_o$ is the momentum of the off-shell gluon.

\section{Box Diagrams}
\begin{figure}[ht]
\psfrag{'='}{$=$}
\psfrag{'-'}{$-$}
\psfrag{'triangle'}{\large Triangle-like}
\psfrag{'+'}{$\hskip -.25in +$}
\psfrag{'q'}{}
\psfrag{'k0'}{}
\psfrag{'k1'}{}
\psfrag{'k2'}{}
\psfrag{'k3'}{}
\begin{center}
\includegraphics[width=5.5in]{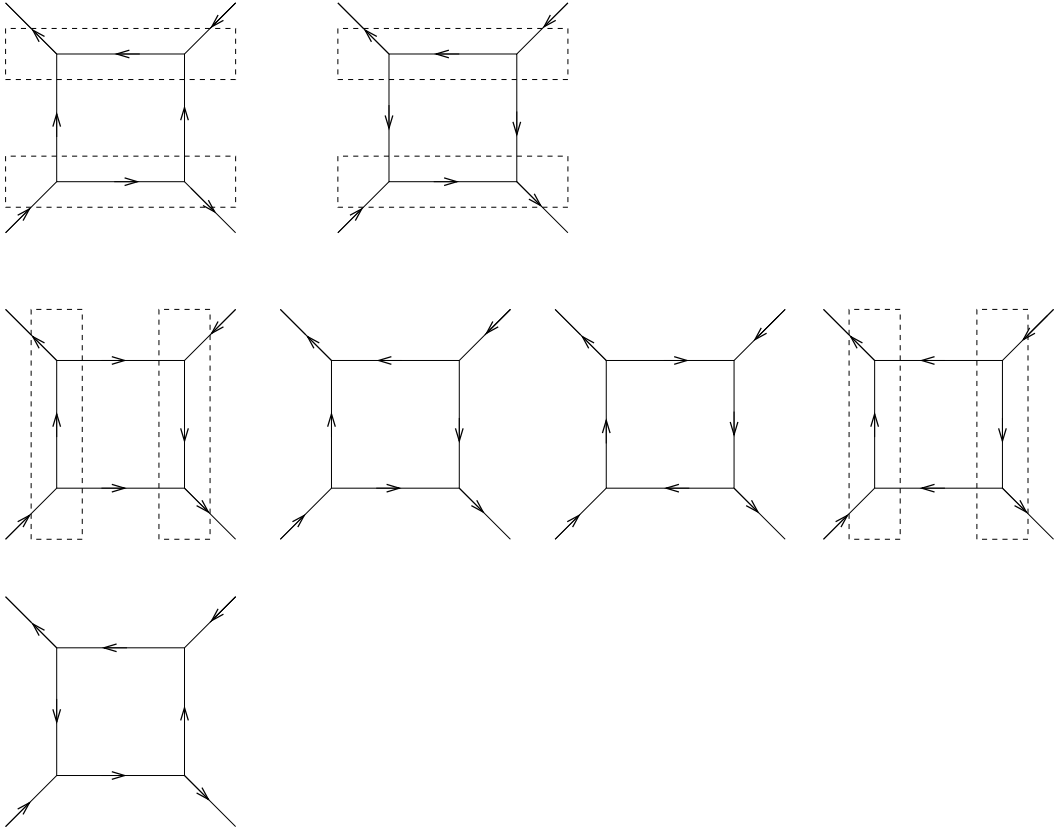}
\caption{The boxes for the $\land\lor\land\lor$ scattering process.
The dashed boxes enclose sub diagrams whose replacement would convert the
box to triangle-like loop integrals.}
\label{boxreduction3}
\end{center}
\end{figure}

\begin{figure}[ht]
\psfrag{'='}{$=$}
\psfrag{'-'}{$-$}
\psfrag{'triangle'}{\large Triangle-like}
\psfrag{'+'}{$\hskip -.25in +$}
\psfrag{'q'}{}
\psfrag{'k0'}{}
\psfrag{'k1'}{}
\psfrag{'k2'}{}
\psfrag{'k3'}{}
\begin{center}
\includegraphics[width=5.5in]{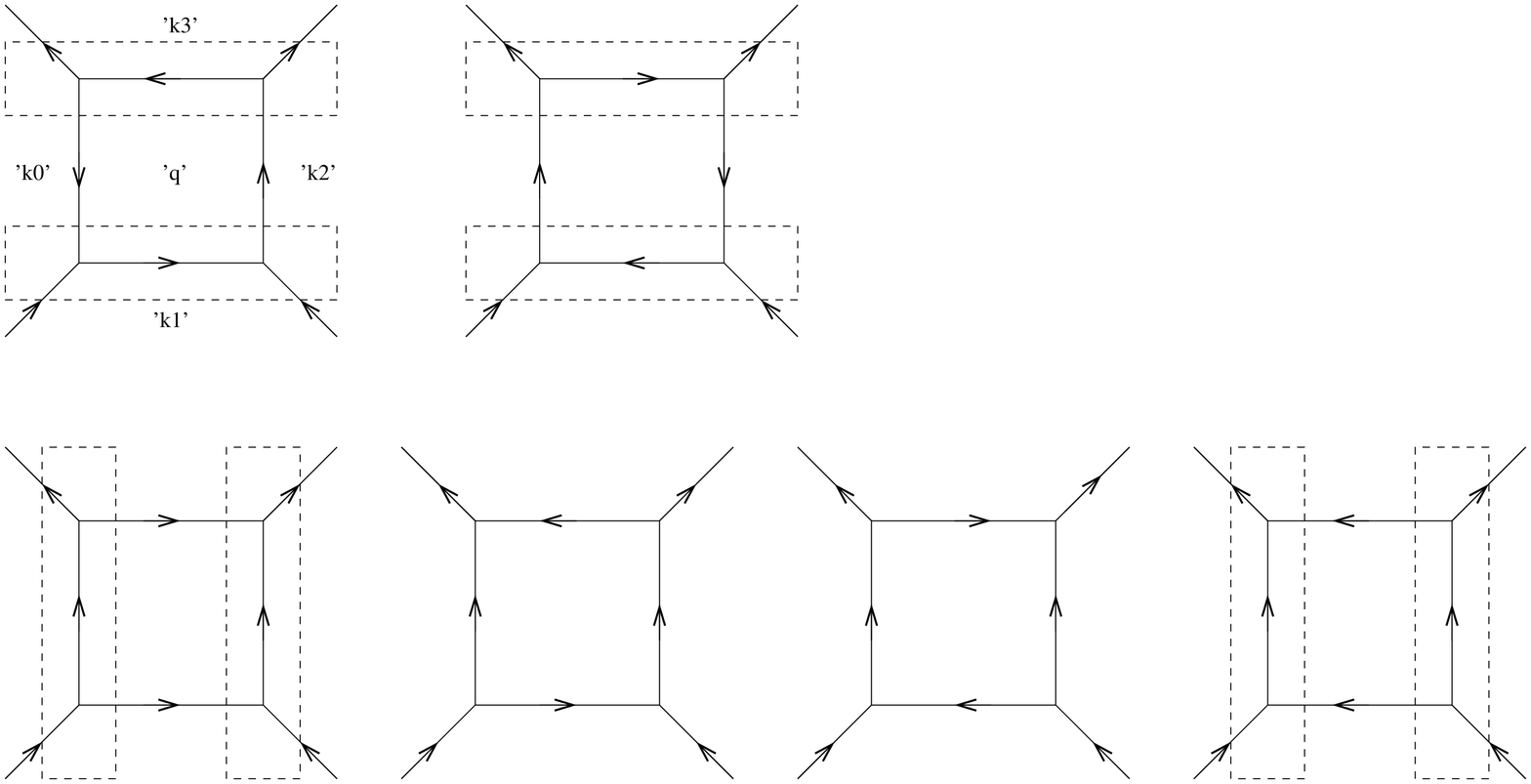}
\caption{The boxes for the $\land\land\lor\lor$ scattering process.
The dashed boxes enclose sub-diagrams whose replacement would convert the
box to triangle-like loop integrals.}
\label{boxreduction2}
\end{center}
\end{figure}

For box diagrams the presence of $q^+$ pole and double pole singularities
in the integrand makes a direct evaluation of the integrals
horrendous. Fortunately, it is possible to manipulate these integrands
so that all of these problematic singularities reside in triangle-like
diagrams. This is because on-shell tree
amplitudes do not possess these singularities.
We can identify 
tree amplitudes as sub-diagrams of one loop diagrams,
but some of the legs of these sub-diagrams will be off-shell, so it
would seem that features of the on-shell limit
can't be exploited. However, if one leaves the denominators of the
off-shell tree subdiagrams alone, then the numerators can always be
written as the on-shell expression (with no $q^+$ singularities)
plus terms each of
which contain at least one factor of the virtuality $q_i^2$ of one
of the off-shell legs. In a box diagram such terms will
cancel a propagator reducing
the required loop integrand to one with the structure of a
triangle diagram. Since triangle integrals with
$q^+$ singularities are considerably easier to
analyze than such box integrals, the resulting simplification is very useful.
In the following  subsection, we apply this technique to all of the 
helicity conserving box diagrams. (The helicity violating case was
done in \cite{chakrabartiqt}, where  each box integrand was
completely reduced to a sum of triangle-like integrands.) In the
remaining subsections we complete the evaluation of the box diagrams.   
\begin{subsection}
{Box Reduction} 
\end{subsection}
The thirteen box diagrams, seven 
for the helicity patterns $\land\lor\land\lor$,
and six for $\land\land\lor\lor$, are shown in Figure \ref{boxreduction3}
and Figure \ref{boxreduction2} respectively.
The integrand of any of these box diagrams has the structure
\bea
{1\over(2\pi)^4}
{R{\cal N}
\over p_1^+p_2^+p_3^+p_4^+
(q-k_0)^2(q-k_1)^2(q-k_2)^2(q-k_3)^2}
\eea
where ${\cal N}$ is a quartic monomial of $K_{ij}$'s carrying the
gluon polarization information, and $R$ is a rational function of
$q^+, p_i^+$. 
\noindent There are only six possible ${\cal N}$'s:
\begin{enumerate}
\item $K_{61}^{\lor} K_{25}^{\lor}K_{35}^{\land} K_{64}^{\land}$:
First diagram of Fig. \ref{boxreduction3}.
\item $K_{61}^{\land} K_{25}^{\land}K_{35}^{\lor} K_{64}^{\lor}$:
Second diagram of Fig. \ref{boxreduction3}; first two diagrams of
Fig. \ref{boxreduction2}.
\item $K_{61}^{\lor} K_{25}^{\land}K_{35}^{\land} K_{64}^{\lor}$:
Third diagram of Fig. \ref{boxreduction3}; third diagram of 
Fig. \ref{boxreduction2}.
\item $K_{61}^{\land} K_{25}^{\lor}K_{35}^{\lor} K_{64}^{\land}$:
Sixth diagram of Fig. \ref{boxreduction3}; sixth diagram of
Fig. \ref{boxreduction2}.
\item $K_{61}^{\lor} K_{25}^{\land}K_{35}^{\lor} K_{64}^{\land}$:
Fourth diagram of Fig. \ref{boxreduction3}; fourth diagram of 
Fig. \ref{boxreduction2}.
\item $K_{61}^{\land} K_{25}^{\lor}K_{35}^{\land} K_{64}^{\lor}$:
Fifth and seventh diagrams of Fig. \ref{boxreduction3}; fifth 
diagram of Fig. \ref{boxreduction2}.
\end{enumerate}
Since these structures come in complex conjugate pairs there are
really only three essentially different structures.
The forms of the rational functions $R$ change from diagram to diagram
and we just list them in order:
\bea
&&\hskip-.25in{p_1^{+2}p_4^{+2}\over q^{+2}(q^+-p_{12}^+)^2},\quad
{p_2^{+2}p_3^{+2}\over q^{+2}(q^+-p_{12}^+)^2},\quad
{p_1^{+2}p_2^{+2}\over (q^{+}-p_1^+)^2(q^++p_4^+)^2},\quad
{p_1^{+2}p_2^{+2}p_3^{+2}p_4^{+2}\over q^{+2}
(q^{+}-p_1^+)^2(q^++p_4^+)^2(q^+-p_{12}^+)^2},\nonumber\\
&&\hskip-.25in{q^{+2}(q^+-p_{12}^+)^2\over(q^{+}-p_1^+)^2(q^++p_4^+)^2},\quad
{p_3^{+2}p_3^{+2}\over (q^{+}-p_1^+)^2(q^++p_4^+)^2},\quad
{(q^{+}-p_1^+)^2(q^++p_4^+)^2\over q^{+2}(q^+-p_{12}^+)^2},\nonumber\\
&&\hskip-.25in
{(q^+-p_{12}^+)^2\over q^{+2}},\quad {q^{+2}\over(q^+-p_{12}^+)^2},\quad
{p_1^{+2}p_3^{+2}\over (q^{+}-p_1^+)^2(q^++p_4^+)^2},\nonumber\\
&&\hskip-.25in{p_1^{+2}p_4^{+2}(q^+-p_{12}^+)^2\over 
q^{+2}(q^{+}-p_1^+)^2(q^++p_4^+)^2},
\quad
{p_2^{+2}p_3^{+2}q^{+2}\over (q^{+}-p_1^+)^2(q^++p_4^+)^2(q^+-p_{12}^+)^2},
\quad
{p_2^{+2}p_4^{+2}\over (q^{+}-p_1^+)^2(q^++p_4^+)^2}
\eea
We can expand each of these thirteen rational functions in
partial fractions and each will then be expressed as a sum of
pure double poles, pure single poles, and a constant.
\bea
R=C+\sum_i\left[{A_i\over(q^+-k_i^+)^2}+{B_i\over q^+-k_i^+}\right]
\eea
where $k_i^+$ is one of the four values $0, p_1^+, p_{12}^+, -p_4^+$.
Except for the fourth diagram of Fig. \ref{boxreduction3}, one
or more of the pole terms will be absent. Also
$C=1$ for the fifth and seventh diagrams of
Fig. \ref{boxreduction3} and for the first and second diagrams of
Fig. \ref{boxreduction2}; and $C=0$ otherwise.

The eight box diagrams with a
helicity violating subdiagram (enclosed by dashed boxes in the
figures) can be completely reduced to triangle-like
diagrams without collinear divergences. 
Each of these completely reducible diagrams has one of the
first four polarization structures in our list, i.e.
two neighboring $K$'s have the same polarization. 
Then for a like-polarization pair, one can use an identity like
\bea
{K_{35}^{\land} K_{64}^{\land}\over q_3^2}+{K_{34}^\land K_{65}^\land
\over p_{12}^2}={K_{34}^\land K_{35}^\land q_0^2+K_{64}^\land K_{34}^\land
q_2^2\over p_{12}^2q_{3}^2}
\label{92}
\eea
which underlies the on-shell vanishing of the three like-helicity
amplitude to convert the integrand to a triangle-like one
which is free of collinear divergences. The integrals over ${\bfs q},q^-$
can be evaluated as in \cite{chakrabartiqt}. The sum over $q^+$ can
be converted to an integral and carried out for the terms with no
poles in $q^+$. The sums with $q^+$ poles are left undone as what we call
the infrared divergent contribution, and will eventually be canceled
against real gluon  bremsstrahlung in their contribution to cross sections. 
For more detail on the calculation of these eight diagrams, see
Appendix E.

The remaining five diagrams cannot
be completely reduced to triangle-like diagrams. But we can
manipulate the integrand so that all $q^+$ divergences are located
in triangle-like diagrams. Then the remaining box integral can be
straightforwardly evaluated.
Since the procedure works in essentially the
same way for each of these five diagrams, we shall illustrate
the method by picking one of the two polarization structures,
the last in our list, which
appears in the fifth and seventh box diagram of Fig. \ref{boxreduction3}
and in the fifth diagram of Fig. \ref{boxreduction2}.
As already mentioned the constant term in $R$ only appears
in these two (the fifth and seventh of Fig. \ref{boxreduction3}) 
of the five ``difficult'' diagrams. Its
contribution is the same for both diagrams, the integrand being
\bea
{1\over(2\pi)^4}
{K_{61}^{\land} K_{25}^{\lor}K_{35}^{\land} K_{64}^{\lor}
\over p_1^+p_2^+p_3^+p_4^+
(q-k_0)^2(q-k_1)^2(q-k_2)^2(q-k_3)^2}
\label{ceeterm}
\eea 
and the integral of (\ref{ceeterm}) is evaluated in Appendix D.

Of the pole terms in $R$ it is sufficient for our illustrative
evaluation to single out only 
one of the $q^+$ pole locations, say that
at $q^+=p_1^+$. Thus we wish to integrate the following
integrand
\bea
{1\over(2\pi)^4}\left({A\over (q^+-p_{1}^+)^2}+{B\over q^+-p_{1}^+}\right)
{K_{61}^{\land} K_{25}^{\lor}K_{35}^{\land} K_{64}^{\lor}
\over p_1^+p_2^+p_3^+p_4^+q_0^2q_1^2q_2^2q_3^2
}
\eea
where we have introduced the shorthand notation $q_i\equiv q-k_i$,
with $k_i$ the dual momenta related to the actual momenta
by $p_i=k_{i+1}-k_i$. We usually take $k_0^{\pm}=0$ but leave
${\bfs k}_0$ arbitrary.
Next we list eight identities, the last two of which 
enable the desired manipulation of this model integrand.
\bea
K_{61}^{\land} K_{64}^{\lor}+K_{64}^{\land} K_{61}^{\lor}
&=&K_{61}\cdot K_{64}={q^{+2}p_{14}^2\over2}+{q_0^2\over2}(p_1^+(q^++p_4^+)
+p_4^+(p_1^+-q^+))\nonumber\\
&&\hskip1in+{q^+p_4^+q_1^2\over2}-{q^+p_1^+q_3^2\over2}\label{softenid1}\\
K_{61}^{\land} K_{64}^{\lor}-K_{64}^{\land} K_{61}^{\lor}
&=&{q^+\over p_{14}^+}(K_{14}^\land K_{61}^\lor- K_{14}^\lor K_{61}^\land
+K_{14}^\land K_{64}^\lor- K_{14}^\lor K_{64}^\land)\label{softenid1a}\\
K_{35}^{\land} K_{25}^{\lor}+K_{25}^{\land} K_{35}^{\lor}
&=&K_{35}\cdot K_{25}={(p_{12}^+-q^{+})^2p_{14}^2\over2}
+{q_2^2\over2}(-p_2^+(q^++p_4^+)
-p_3^+(p_1^+-q^+))\nonumber\\
&&\hskip1in+{(q^+-p_{12}^+)p_2^+q_3^2\over2}-
{(q^+-p_{12}^+)p_3^+q_1^2\over2}\\
K_{35}^{\land} K_{25}^{\lor}-K_{25}^{\land} K_{35}^{\lor}
&=&{p_{12}^+-q^+\over p_{23}^+}(K_{32}^\land K_{52}^\lor
- K_{32}^\lor K_{52}^\land
+ K_{32}^\land K_{53}^\lor- K_{32}^\lor K_{53}^\land) \\
K_{35}^{\land} K_{64}^{\lor}+K_{64}^{\land} K_{35}^{\lor}
&=&K_{35}\cdot K_{64}={(p_{4}^++q^{+})^2p_{12}^2\over2}
-{q_3^2\over2}(p_3^+q^+
+p_4^+(p_{12}^+-q^+))\nonumber\\
&&\hskip1in+{(q^++p_{4}^+)p_3^+q_0^2\over2}-
{(q^++p_{4}^+)p_4^+q_2^2\over2}\\
K_{35}^{\land} K_{64}^{\lor}-K_{64}^{\land} K_{35}^{\lor}
&=&{p_{4}^++q^{+}\over p_{34}^+}(K_{34}^\land K_{65}^\lor
-K_{34}^\lor K_{65}^\land)\\
K_{61}^{\land} K_{25}^{\lor}+K_{25}^{\land} K_{61}^{\lor}
&=&K_{61}\cdot K_{25}={(p_{1}^+-q^{+})^2p_{12}^2\over2}
+{q_1^2\over2}(p_2^+q^+
+p_1^+(p_{12}^+-q^+))\nonumber\\
&&\hskip1in+{(p_{1}^+-q^+)p_2^+q_0^2\over2}-
{(p_{1}^+-q^+)p_1^+q_2^2\over2}\\
K_{61}^{\land} K_{25}^{\lor}-K_{25}^{\land} K_{61}^{\lor}
&=&{q^{+}-p_{1}^+\over p_{12}^+}(K_{21}^\land K_{65}^\lor
-K_{21}^\lor K_{65}^\land)
\label{softenid4}
\eea
The first six identities are needed for other box integrands.
All but the first terms of any of the right sides contain a factor of
$q_i^2$ which would cancel one of the propagators converting the
box to a triangle-like diagram. Depending on which term in the partial fraction
expansion we consider, we can use one of these identities to switch $\land$
and $\lor$ on a pair of $K$ factors. For our model case we use the
last two identities to write the first two $K$ factors in two different ways
\bea
K_{61}^{\land} K_{25}^{\lor}&=&-K_{61}^{\lor}K_{25}^{\land} 
+{(p_{1}^+-q^{+})^2p_{12}^2\over2}
+{q_1^2\over2}(p_2^+q^+
+p_1^+(p_{12}^+-q^+))
+{(p_{1}^+-q^+)p_2^+q_0^2\over2}-
{(q^+-p_{1}^+)p_1^+q_2^2\over2}\nonumber\\
K_{61}^{\land} K_{25}^{\lor}&=&K_{61}^{\lor}K_{25}^{\land} 
+{q^+-p_{1}^+\over p_{12}^+}(K_{21}^\land K_{65}^\lor
-K_{21}^\lor K_{65}^\land)
\eea
We use the first rewrite for the double pole and the second
for the single pole. The first terms of either convert the
polarization structure to one which completely reduces to
triangle-like diagrams free of collinear divergences just as
with the eight box diagrams with helicity violating subdiagrams.
Their calculation follows the models given in Appendix E.
The second terms of either have an explicit factor that 
cancels the singular $q^+$ denominators, leaving a box integrand
free of $q^+$ divergences. However, only the box from the
second line is free of collinear divergences; its
evaluation is therefore straightforward and is given in 
Appendix F.
There are no more terms for the second rewrite, and
the remaining terms of the
first each have a virtuality factor $q_0^2,q_1^2,q_2^2$
which converts the integrand to a triangle-like one.
Unfortunately, both these last triangle-like diagrams and
the box integrand left after the first rewrite individually have
collinear divergences. They must, of course, cancel among themselves,
but to achieve clean results we discuss in the next subsection
a way to rearrange the integrands to finesse this difficulty.

Summarizing this subsection, we have 
rewritten our model integrand in the form
\bea
&&\hskip-.5in{1\over(2\pi)^4}\left({A\over (q^{+}-p_1^+)^2}
+{B\over q^+-p_1^+}\right)
{K_{61}^{\land} K_{25}^{\lor}
K_{35}^{\land} K_{64}^{\lor}
\over p_1^+p_2^+p_3^+p_4^+
q_0^2q_1^2q_2^2q_3^2}\ =\ 
{A\over(2\pi)^4}{p_{12}^2K_{35}^{\land}K_{64}^{\lor}
\over 2 p_1^+p_2^+p_3^+p_4^+
q_0^2q_1^2q_2^2q_3^2}\nonumber\\
&&\hskip2.25in +{B\over(2\pi)^4}{(K_{21}^\land K_{65}^\lor
- K_{21}^\lor K_{65}^\land)
K_{35}^{\land}K_{64}^{\lor}
\over p_{12}^+p_1^+p_2^+p_3^+p_4^+
q_0^2q_1^2q_2^2q_3^2} +{\rm Triangle-like}
\label{104}
\eea
In the next subsection we show how to deal with the 
collinear divergence in the first term of the right side and in 
the triangle-like diagrams associated with it:
\bea
{A\over(2\pi)^4}{K_{35}^{\land}K_{64}^{\lor}
\over 2(p_1^+-q^+)^2 p_1^+p_2^+p_3^+p_4^+}
\left[{p_2^+q^+
+p_1^+(p_{12}^+-q^+)\over q_0^2q_2^2q_3^2}
+{(p_{1}^+-q^+)p_2^+\over q_1^2q_2^2q_3^2}-
{(p_{1}^+-q^+)p_1^+\over q_0^2q_1^2q_3^2 }\right]
\label{collineartri}
\eea
Actually the prefactor kills the collinear divergence in the 
first term in the square brackets so we only need deal with the
last two terms.

\begin{subsection}
{Subtracting Collinear Divergences}
\end{subsection}
The box reduction we have so far accomplished has the undesirable feature that
the new box integrands have collinear divergences that were not present in the
original box integrands. This means that
there must be canceling collinear divergences among the
triangle-like diagrams that we generated in the reduction
procedure. The terms with four $K$'s in the numerator will not
have this problem but all the terms with only 2 $K$'s in the numerator do.
In order to deal with this problem we must add some triangle-like diagrams
to these problematic box integrands in such a way as to regulate 
these divergences.
To begin, we note that the terms linear in loop momentum dependent
$K$'s can be made
IR finite by a simple subtraction of two triangle-like terms. We find
that each of the combinations
\bea
{K_{25}^\lor\over q_0^2q_1^2q_2^2q_3^2}
-{K_{12}^\lor\over q_0^2q_1^2p_{12}^2q_3^2}
-{K_{23}^\lor\over q_0^2q_2^2q_3^2p_{14}^2}&=&
{K_{25}^\lor p_{12}^2p_{14}^2-K_{12}^\lor p_{14}^2q_2^2
-K_{23}^\lor p_{12}^2q_1^2\over 
q_0^2q_1^2q_2^2q_3^2p_{12}^2p_{14}^2}\nonumber\\
{K_{61}^\land\over q_0^2q_1^2q_2^2q_3^2}
-{K_{41}^\land\over q_0^2q_2^2p_{14}^2q_3^2}
-{K_{12}^\land\over q_1^2q_2^2q_3^2p_{12}^2}&=&
{K_{61}^\land p_{12}^2p_{14}^2-K_{41}^\land p_{12}^2q_1^2
-K_{12}^\land p_{14}^2q_0^2\over 
q_0^2q_1^2q_2^2q_3^2p_{12}^2p_{14}^2}\nonumber\\
{K_{35}^\land\over q_0^2q_1^2q_2^2q_3^2}
-{K_{34}^\land\over q_0^2q_1^2p_{12}^2q_3^2}
-{K_{23}^\land\over q_0^2q_1^2q_2^2p_{14}^2}&=&
{K_{35}^\land p_{12}^2p_{14}^2-K_{34}^\land p_{14}^2q_2^2
-K_{23}^\land p_{12}^2q_3^2\over 
q_0^2q_1^2q_2^2q_3^2p_{12}^2p_{14}^2}\nonumber\\
{K_{64}^\lor\over q_0^2q_1^2q_2^2q_3^2}
-{K_{41}^\lor\over q_0^2q_1^2p_{14}^2q_2^2}
-{K_{34}^\lor\over q_1^2q_2^2q_3^2p_{12}^2}&=&
{K_{64}^\lor p_{12}^2p_{14}^2-K_{41}^\lor p_{12}^2q_3^2
-K_{34}^\lor p_{14}^2q_0^2\over 
q_0^2q_1^2q_2^2q_3^2p_{12}^2p_{14}^2}
\label{irreg1}
\eea
is finite integrated in the infra-red, and the triangle-like subtractions
are quadratically convergent in the ultra-violet. This nice IR
behavior is not spoiled by multiplying each expression by further
factors of loop momentum dependent $K$'s, and up to two such factors could
be applied keeping the UV behavior no worse than logarithmic. 
Thus we can satisfactorily regulate the terms quadratic 
in loop momenta by simply multiplying one of these expressions by
the appropriate $K$. There are several choices for regulating each
term, so we arbitrarily choose one of them.
Of course, as already mentioned the term
quartic in the loop momenta is IR convergent by itself and needs no
subtractions.

When we pass to the Schwinger parameterization (with the notation of
Appendix C) the numerator factors
in (\ref{irreg1}) enjoy a nice simplification after an appropriate
shift in $q$:
\bea
K_{25}^\lor p_{12}^2p_{14}^2-K_{12}^\lor p_{14}^2q_2^2
-K_{23}^\lor p_{12}^2q_1^2&\to&(p_2^+q^\lor-q^+p_2^\lor)p_{12}^2p_{14}^2
-2q\cdot(K_0-p_1-p_2)K_{12}^\lor p_{14}^2\nonumber\\
&&\hskip-0.25in-2q\cdot(K_0-p_1)K_{23}^\lor p_{12}^2
+(K_{23}^\lor p_{12}^2+K_{12}^\lor
p_{14}^2)(x_1x_3p_{12}^2+x_2x_4p_{14}^2-q^2)\nonumber\\
K_{61}^\land p_{12}^2p_{14}^2-K_{41}^\land p_{12}^2q_1^2
-K_{12}^\land p_{14}^2q_0^2&\to&
(p_1^+q^\land-q^+p_1^\land) p_{12}^2p_{14}^2
-2q\cdot K_0 K_{12}^\land p_{14}^2
-2q\cdot(K_0-p_1)K_{41}^\land p_{12}^2\nonumber\\
&&\hskip-.25in+(K_{41}^\land p_{12}^2+K_{12}^\land
p_{14}^2)(x_1x_3p_{12}^2+x_2x_4p_{14}^2-q^2)\nonumber\\
K_{35}^\land p_{12}^2p_{14}^2-K_{34}^\land p_{14}^2q_2^2
-K_{23}^\land p_{12}^2q_3^2&\to&(p_3^+q^\land-q^+p_3^\land)p_{12}^2p_{14}^2
-2q\cdot(K_0-p_1-p_2)K_{34}^\land p_{14}^2\nonumber\\
&&\hskip-.25in-2q\cdot(K_0+p_4)K_{23}^\land p_{12}^2
+(K_{23}^\land p_{12}^2+K_{34}^\land
p_{14}^2)(x_1x_3p_{12}^2+x_2x_4p_{14}^2-q^2)\nonumber\\
K_{64}^\lor p_{12}^2p_{14}^2-K_{41}^\lor p_{12}^2q_3^2
-K_{34}^\lor p_{14}^2q_0^2&\to&(p_4^+q^\lor-q^+p_4^\lor)p_{12}^2p_{14}^2
-2q\cdot K_0 K_{34}^\lor p_{14}^2-2q\cdot(K_0+p_4)K_{41}^\lor p_{12}^2
\nonumber\\
&&\hskip-.25in+(K_{41}^\lor p_{12}^2+K_{34}^\lor
p_{14}^2)(x_1x_3p_{12}^2+x_2x_4p_{14}^2-q^2)
\label{irregnum1}
\eea
Here $K_0=x_2p_1+x_3(p_1+p_2)-x_4p_4$ is the $\delta\to0$ limit of
$K-k_0$ where $K$ has been defined in Appendix A.

Now consider the terms quadratic in loop momentum dependent $K$'s. 
These terms involve one of the pairs $(K_{25},K_{35})$, $(K_{35},K_{64})$,
 $(K_{64},K_{61})$, $(K_{61},K_{25})$, each pair is associated with
a neighboring pair of external lines of the box
diagram. The two members of each pair have opposite
polarization, with both possibilities occurring.
It is easy to confirm that the terms involving
each pair can be obtained from one another by cyclic symmetry. 
Therefore we need only analyze one class of terms,say,
those involving the
pair occurring in the model integrand of the previous
subsection, $K_{35}^\land, K_{64}^\lor$. 
These terms can be regulated in the infra-red
either by multiplying the last line of (\ref{irreg1}) by $K_{35}^\land$ or
the third line by $K_{64}^\lor$. We choose the former and for the other pairs
make the choice dictated by cyclic symmetry. 
The terms in the numerator that will survive integration over $q$
are
\bea
&&\hskip-.35in 
K_{35}^\land(K_{64}^\lor p_{12}^2p_{14}^2-K_{41}^\lor p_{12}^2q_3^2
-K_{34}^\lor p_{14}^2q_0^2)\to    
-2q^\land q^\lor(x_1K^\land_{34}+x_2K^\land_{23}) (K_{34}^\lor p_{14}^2+
K_{41}^\lor p_{12}^2)
\nonumber\\
&&+(x_2K_{23}^\land+x_1K_{34}^\land)(K_{41}^\lor p_{12}^2+K_{34}^\lor
p_{14}^2)(x_1x_3p_{12}^2+x_2x_4p_{14}^2-q^2)
\eea
In addition to dropping terms that directly integrate to 0,
we also used $-q^+q^-\to q^\land q^\lor$ valid under $q$
integration.
As shown in Appendix C, the effect of $q^2$ in the numerator is its 
replacement by a factor $2H\equiv 2(x_1x_3p_{12}^2+x_2x_4p_{14}^2)$
and the effect of $q^\land q^\lor$ is its replacement by $H/2$.
Thus we can replace
\bea
K_{35}^\land\left(K_{64}^\lor -{K_{34}^\lor q_0^2\over
p_{12}^2}
-{K_{41}^\lor q_3^2\over p_{14}^2}\right)&\to&
-2(x_1x_3p_{12}^2+x_2x_4p_{14}^2)
(x_2K_{23}^\land+x_1K_{34}^\land)\left({K_{41}^\lor\over p_{14}^2}
+{K_{34}^\lor\over
p_{12}^2}\right)\\
&=&-4(x_1x_3p_{12}^2+x_2x_4p_{14}^2)
(x_2K_{23}^\land+x_1K_{34}^\land){K_{13}^\land K_{41}^\lor 
K_{34}^\lor\over p_1^+p_3^+p_{12}^2 p_{14}^2}
\eea
The loop integral involving this pair of $K$'s can be done (see Appendix C)
\bea
&&\hskip-1in\int{d^4q\over16\pi^4}{p_{12}^2K_{35}^\land K_{64}^\lor\over
2p_{1}^{+}p_2^+p_3^+p_4^+q_0^2q_1^2q_2^2q_3^2}\to 
I_{34}\equiv\int{d^4q\over16\pi^4}
{p_{12}^2K_{35}^\land \left(K_{64}^\lor -{K_{34}^\lor q_0^2/
p_{12}^2}
-{K_{41}^\lor q_3^2/ p_{14}^2}\right)\over
2p_{1}^{+}p_2^+p_3^+p_4^+q_0^2q_1^2q_2^2q_3^2}\nonumber\\
I_{34}&=&{K_{13}^\land K_{42}^\lor\over16\pi^2p_1^{+2}p_2^{+2}p_3^+p_4^+}
\left\{
{K_{12}^\land K_{41}^\lor \ln(p_{14}^2/p_{12}^2)
\over p_{14}^2(p_{12}^2+p_{14}^2)}
+{p_{12}^2K_{23}^\land K_{41}^\lor[\pi^2+ \ln^2(p_{12}^2/p_{14}^2)]
\over 2p_{14}^2(p_{12}^2+p_{14}^2)^2}\right\}
\eea
The box integrands involving other pairs of $K$'s can be 
regulated and evaluated in an exactly parallel fashion.
We shall quote the results of combining all contributions
in our results Section 9.
\vskip14pt
\begin{subsection}
{Calculation of the Triangle Diagrams with Collinear Divergences}
\end{subsection}
We now turn to the triangle-like diagrams
which contain collinear divergences, the last
two terms of  (\ref{collineartri}). These divergences must be canceled
when we add back in the triangle-like diagrams we subtracted from
the box to cancel its collinear divergences.
\bea
{A\over(2\pi)^4}{K_{35}^{\land}K_{64}^{\lor}
\over 2(p_1^+-q^+) p_1^+p_2^+p_3^+p_4^+}
\left[{p_2^+\over q_1^2q_2^2q_3^2}-{p_1^+\over q_0^2q_1^2q_3^2 }\right]
+{A\over(2\pi)^4}{K_{35}^{\land}
\over 2 p_1^+p_2^+p_3^+p_4^+}
\left[{p_{12}^2K_{41}^\lor\over p_{14}^2 q_0^2q_1^2q_2^2}+
{K_{34}^\lor\over q_1^2q_2^2q_3^2 }\right]\ = \ \nonumber\\
{AK_{35}^{\land}\over(2\pi)^42 p_1^+p_2^+p_3^+p_4^+}
\left[{p_2^+K_{64}^{\lor}+(p_1^+-q^+)K_{34}^\lor
\over (p_1^+-q^+)q_1^2q_2^2q_3^2}
-{p_1^+K_{64}^{\lor}\over (p_1^+-q^+)q_0^2q_1^2q_3^2 }
+{p_{12}^2K_{41}^\lor\over p_{14}^2 q_0^2q_1^2q_2^2}\right]
\label{collineartria}
\eea
The last two terms in the square brackets on the
right have a collinear divergence due to the vanishing of
$q_0^2$ and $q_1^2$ when $q$ is collinear with $p_1$,
specifically at $q=q^+p_1/p_1^+$. In this limit we have
\bea
-{p_1^+K_{35}^{\land}K_{64}^{\lor}\over (p_1^+-q^+)q_0^2q_1^2q_3^2 }\to
-{[(p_1^+-q^+)K_{34}^\land+q^+K_{23}^\land]
K_{41}^{\lor}\over (p_1^+-q^+)q_0^2q_1^2p_{14}^2},\quad
{p_{12}^2K_{35}^{\land}K_{41}^\lor\over p_{14}^2 q_0^2q_1^2q_2^2}\to
+{[(p_1^+-q^+)K_{34}^\land
+q^+K_{23}^\land]K_{41}^{\lor}\over (p_1^+-q^+)q_0^2q_1^2p_{14}^2}
\label{113}
\eea
We see that this collinear divergence cancels.
Similarly the first and
last terms have a collinear divergence due to the vanishing of
$q_1^2$ and $q_2^2$ when $q_1$ is collinear with $p_2$,
specifically at $q=p_1+(q^+-p_1^+)p_2/p_2^+$. In this limit
\bea
{K_{35}^{\land}[p_2^+K_{64}^{\lor}+(p_1^+-q^+)K_{34}^\lor]
\over (p_1^+-q^+)q_1^2q_2^2q_3^2}\to
{(p_{12}^+-q^+)K_{23}^\land K_{41}^{\lor}\over 
(p_1^+-q^+)q_1^2q_2^2p_{14}^2},\qquad
{p_{12}^2K_{35}^{\land}K_{41}^\lor\over p_{14}^2 q_0^2q_1^2q_2^2}\to
-{(p_{12}^+-q^+)K_{23}^\land K_{41}^{\lor}\over (p_1^+-q^+)q_1^2q_2^2p_{14}^2}
\label{114}
\eea
and they also cancel.
The other possible collinear divergences in these expressions are killed
by the polarization factors $K_{64}^\lor$ and $K_{35}^\land$.

To calculate these triangle-like contributions, we remove from each
triangle structure its collinear limit. We work out one case,
the first one in (\ref{113}), explicitly.
We integrate $dq^-$ first\footnote{All our formulas have treated $d^4q$
as Euclidean, but to integrate over $q^-$ we convert to Minkowski
measure $d^4q\equiv -id^4q_M$, and use the usual prescription for
propagator denominators $q_i^2\to q_i^2-i\epsilon$. Then the 
integral over $q^-$ just gives $2\pi i$ times the sum of residues
and the $i$'s cancel.}. The $q^-$ integrals of all six types of bubble
integrands and all four types
of triangle-like integrands are listed in Appendix B.  
Note that, under the assumption that
$p_1^+<-p_4^+$, for $0<q^+<p_1^+$ only the $q_0$ pole contributes,
for $p_1^+<q^+<-p_4^+$ only the $q_3$ pole contributes, and the
integral gives 0 otherwise:
\bea
&&\hskip-.25in -i\int {dq^-\over2\pi}\left[
-{p_1^+K_{35}^{\land}K_{64}^{\lor}\over (p_1^+-q^+)q_0^2q_1^2q_3^2 }
+{[(p_1^+-q^+)K_{34}^\land+q^+K_{23}^\land]
K_{41}^{\lor}\over (p_1^+-q^+)q_0^2q_1^2p_{14}^2}\right]\\
&&\hskip-.25in =\ {K_{61}^{\land}K_{64}^{\lor}K_{34}^\land K_{41}^{\lor}(q^++p_4^+)
\over p_1^+p_4^{+2}(p_1^+-q^+)p_{14}^2({\bfs q}-q^+{\bfs p}_1/p_1^+)^2
({\bfs q}-q^+{\bfs p}_4/p_4^+)^2}\hskip2in{\rm for}\quad 0<q^+<p_1^+\nonumber\\
&&\hskip-.25in =\ {K_{35}^\land K_{64}^\lor q^+p_1^+|q^++p_4^+|/(q^+-p_1^+)
\over p_4^+({\bfs q}-q^+{\bfs p}_4/p_4^+)^2
[p_4^+(q^+-p_1^+)({\bfs q}-q^+{\bfs p}_4/p_4^+)^2+p_1^+(q^++p_4^+)
({\bfs q}-q^+{\bfs p}_1/p_1^+)^2]}\quad{\rm for}\ p_1^+<q^+<-p_4^+\nonumber\\
&&\hskip-.25in =\ {K_{35}^\land K_{64}^\lor p_1^+|q^++p_4^+|/(q^+-p_1^+)
\over p_{4}^{+2}({\bfs q}-q^+{\bfs p}_4/p_4^+)^2
[({\bfs q}+{\bfs p}_4-(q^++p_4^+){\bfs p}_{14}/p_{14}^+)^2
-(q^+-p_1^+)(q^++p_4^+)p_{14}^2/p_{14}^{+2}
]}\quad{\rm for}\ p_1^+<q^+<-p_4^+\nonumber
\eea
We see indeed that no collinear divergences are encountered when these
expressions are integrated over ${\bfs q}$.

The other cases are similar. The fact that the $012$ integrand
has two collinear divergences that require cancellation is not a
problem, because, as seen in appendix B, after the integral
over $q^-$, the two divergences occur in disjoint regions
of $q^+$, so one can simply remove them additively:
\bea
&&\hskip-.25in -i\int {dq^-\over2\pi}\left[
{p_{12}^2K_{35}^{\land}K_{41}^{\lor}\over p_{14}^2q_0^2q_1^2q_2^2 }
-{[(p_{1}^+-q^+)K_{34}^\land+q^+K_{23}^\land]
K_{41}^{\lor}\over (p_1^+-q^+)q_0^2q_1^2p_{14}^2}
+{(p_{12}^+-q^+)K_{23}^\land K_{41}^{\lor}
\over (p_1^+-q^+)q_1^2q_2^2p_{14}^2}\right]
\eea
is thus free of collinear divergences as is
\bea
&&\hskip-.25in -i\int {dq^-\over2\pi}\left[
{K_{35}^{\land}[p_2^+K_{64}^{\lor}+(p_1^+-q^+)K_{34}^{\lor}
\over(p_1^+-q^+)q_1^2q_2^2q_3^2 }
-{[(p_{1}^+-q^+)K_{34}^\land+q^+K_{23}^\land]
K_{41}^{\lor}\over (p_1^+-q^+)q_0^2q_1^2p_{14}^2}
\right]
\eea
Another approach to the collinear divergence
problem is to calculate the individual diagrams
with a mass regulator
$\mu^2$, and send $\mu\to0$ only after combining the terms
that together are free of collinear divergences. In fact,
for the purposes of automating our calculations, we found
this latter regulator method more efficient, and used it to
obtain the combined final results of all these
integrations quoted at length in Appendix G. 
\vskip14pt
\section{Non-box One Loop Corrections to Scattering of Glue by Glue}
\begin{figure}[ht]
\begin{center}
\includegraphics[width=4.5in]{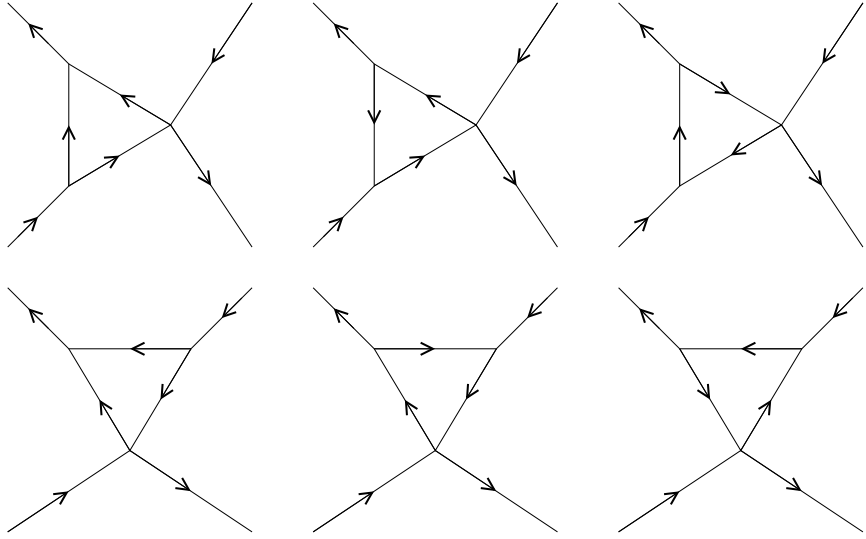}
\caption{Half of the quartic triangle diagrams for the 
$\land\lor\land\lor$ scattering process. The six others
are similar but with the quartic vertex at the left or top. 
Particle labels $1234$ are applied counter-clockwise
starting at the lower left of each diagram.}
\label{quarttripol}
\end{center}
\end{figure}

\begin{figure}[ht]
\begin{center}
\includegraphics[width=4.5in]{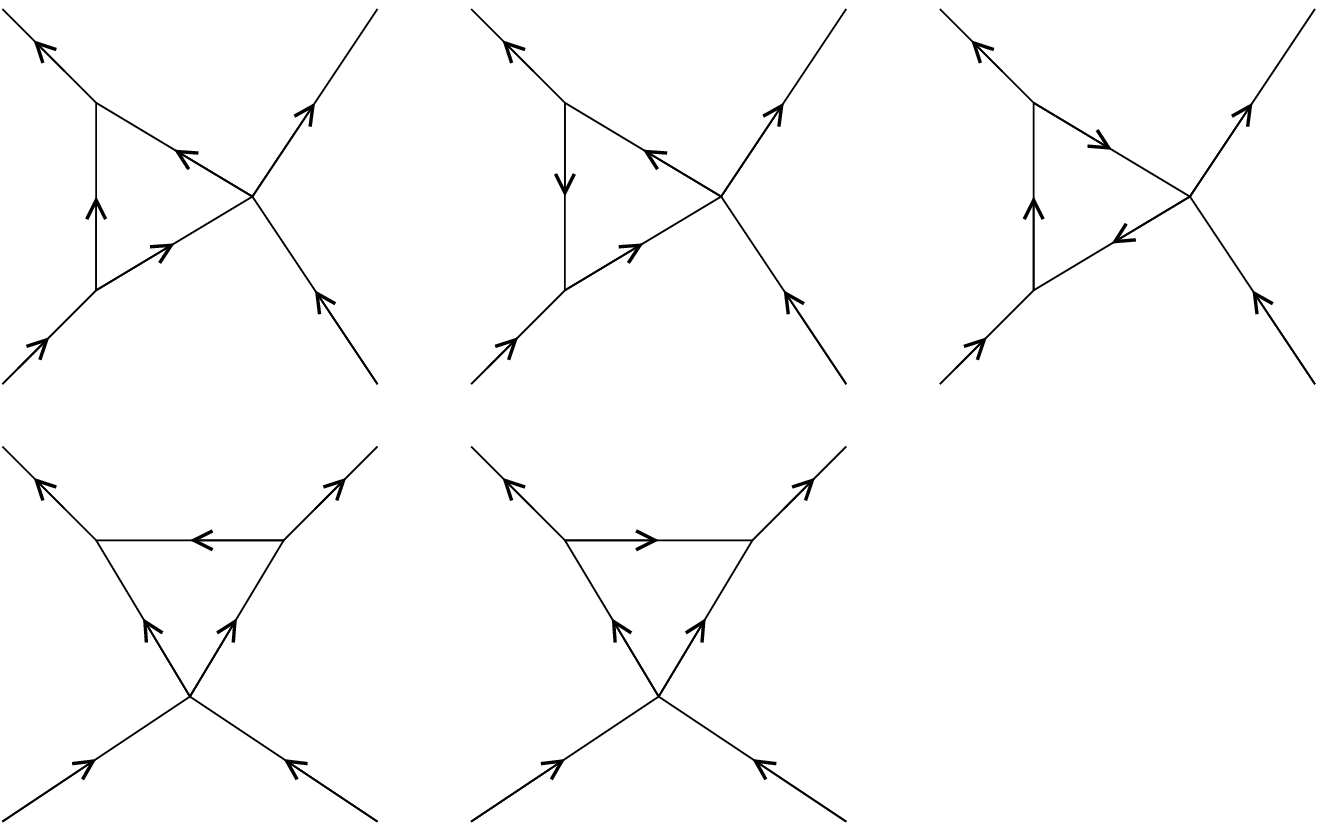}
\caption{Half of the quartic triangle diagrams for the 
$\land\land\lor\lor$ scattering process. The five others
are similar but with the quartic vertex at the left or top. 
Particle labels $1234$ are applied counter-clockwise
starting at the lower left of each diagram.}
\label{quarttripol1}
\end{center}
\end{figure}

\begin{subsection}
{Cubic Vertex Corrections and Self Energy Insertions on Internal
Lines.}
\end{subsection}
We quote here the contribution of triangle corrections to 
glue-glue scattering combined with the self energy bubbles on internal
lines (see \cite{thornlcnotes,chakrabartiqt}):
\bea
\Gamma^{\land\land\lor\lor}_{\triangle+\rm SE}
&=& -{p_{12}^{+2}K_{34}^\lor K_{12}^\land
\over32\pi^2p_1^+p_2^+p_3^+p_4^+p_{12}^2}
\bigg[{11\over3}\ln(p_{12}^2e^\gamma\delta)-{73\over9}+{p_1^+p_2^++p_3^+p_4^+
\over 3p_{12}^{+2}}\nn\\
&&
\hskip1.5in-S_3(p_1,p_2)-S_3(-p_4,-p_3)+{\cal A}(p_{12}^2,p_{12}^+)\bigg]\nonumber\\
&&-{1\over32\pi^2}
\left({p_1^+p_3^+\over p_2^+p_4^+}{K_{23}^\land K_{41}^\lor\over p_{14}^{+2}
p_{14}^2}
+{p_2^+p_4^+\over p_1^+p_3^+}{K_{23}^\lor K_{41}^\land\over p_{14}^{+2}
p_{14}^2}\right)
\bigg[{11\over3}\ln(p_{14}^2e^\gamma\delta)-{73\over9}\nn\\
&&
\hskip1.5in-S_2(-p_4,-p_{23})-S_1(p_{14},p_2)+{\cal A}(p_{14}^2,p_{14}^+)\bigg]
\eea
\bea
\Gamma^{\land\lor\land\lor}_{\triangle+{\rm SE}}
&=&-{1\over32\pi^2}
\left({p_1^+p_4^+\over p_2^+p_3^+}{K_{34}^\land K_{12}^\lor\over p_{12}^{+2}
p_{12}^2}
+{p_2^+p_3^+\over p_1^+p_4^+}{K_{34}^\lor K_{12}^\land\over p_{12}^{+2}
p_{12}^2}\right)
\bigg[{11\over3}\ln(p_{12}^2e^\gamma\delta)-{73\over9}\nn\\&&\hskip1.5in
-S_3(p_1,p_2)-S_3(-p_4,-p_3)+{\cal A}(p_{12}^2,p_{12}^+)\bigg] \nonumber\\
&&-{1\over32\pi^2}
\left({p_1^+p_2^+\over p_3^+p_4^+}{K_{23}^\land K_{41}^\lor\over p_{14}^{+2}
p_{14}^2}
+{p_3^+p_4^+\over p_1^+p_2^+}{K_{23}^\lor K_{41}^\land\over p_{14}^{+2}
p_{14}^2}\right)
\bigg[{11\over3}\ln(p_{14}^2e^\gamma\delta)-{73\over9}\nn\\&&\hskip1.5in
-S_2(-p_4,-p_{23})-S_1(p_{14},p_2)+{\cal A}(p_{14}^2,p_{14}^+)\bigg]
\eea
where
\bea
{\cal A}(p^2,p^+)&=&\sum_{q^+}\left[{1\over q^+}+{1\over p^+-q^+}\right]
\ln\left\{{q^+\over p^+}\left(1-{q^+\over p^+}\right)p^2e^\gamma\delta
\right\}
\eea
\begin{subsection}
{Quartic Triangle Diagrams}
\end{subsection}
\begin{figure}[ht]
\begin{center}
\includegraphics[width=5in]{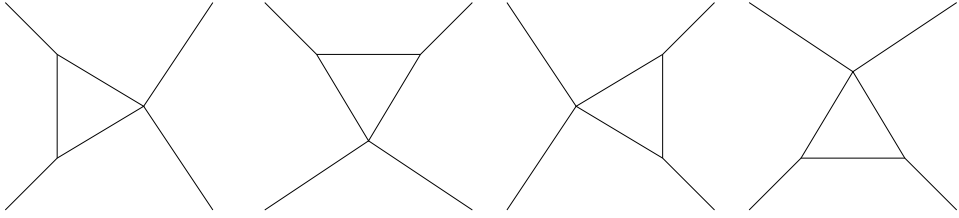}
\caption{The quartic triangle diagrams shown generically, 
without arrows indicating
spin flow. Particle labels $1234$ are applied counter-clockwise
starting at the lower left of each diagram.}
\label{quarttri}
\end{center}
\end{figure}
There are four distinct quartic triangle structures (see Fig.\ref{quarttri}), 
which we label by the two legs entering the quartic vertex. Half of the
diagrams for each polarization configuration are given in
Figs.\ref{quarttripol},\ref{quarttripol1}.
The integrand of each diagram has three of the four possible
propagator factors $1/q_i^2$ for $i=0,1,2,3$, a numerator
consisting of one of the eight polarization structures
$K_{61}^\lor K_{64}^\land$, $K_{61}^\land K_{64}^\lor$,
$K_{35}^\land K_{64}^\lor$, $K_{35}^\lor K_{64}^\land$,
$K_{35}^\land K_{25}^\lor$, $K_{35}^\lor K_{25}^\land$,
$K_{61}^\lor K_{25}^\land$, $K_{61}^\land K_{25}^\lor$,
times a rational function of $q^+$ and the $p_i^+$.
The $q^-$ and ${\bfs q}$ integrations are virtually identical from
one diagram to the other.

A model integrand for a quartic triangle is: \bea -K_{6,4}^{\lor}
K_{3,5}^\land
\frac{1}{q_0^2}\frac{1}{q_3^2}\frac{1}{q_2^2}\nn\eea
\bea &&\pref g^4 \Nc\times\nn\\
\Pt<\Kp<\Pth&&\frac{1}{4}
[\log{H_s}+\log{(\delta e^{\gamma})}]\frac{({\Pt}-{\Pth})({\Pf}-\Kp)}{({\Pt}-{\Pf})}-\frac{1}{4} [\log{H_d}+\log{(\delta e^{\gamma})}]({\Pth}-\Kp)\nn\\
\nn\\
\Pth<\Kp<\Po&&\frac{1}{4}
[\log{H_s}+\log{(\delta e^{\gamma})}]\frac{(-{\Pth}+{\Pf})({\Pt}-\Kp)}{({\Pt}-{\Pf})}+\frac{1}{4}[\log{H_d}+\log{(\delta e^{\gamma})}]({\Pth}-\Kp)\nn\\
\nn\\
\Po<\Kp<\Pf&&\frac{1}{4} [\log{H_s}+\log{(\delta
 e^{\gamma})}]\frac{(-{\Pth}+{\Pf})({\Pt}-\Kp)}{({\Pt}-{\Pf})}+\frac{1}{4}
[\log{H_d}+\log{(\delta e^{\gamma})}]({\Pth}-\Kp)\nn
\eea 
The various H's are defined: \bea \rleft\nn\\
&&H_s=\frac
{{{p_{12}^2}}(\Kp-\Pt)({\Pf}-\Kp)}{({\Pt}-{\Pf})^{2}}\hspace{.5in}
H_d=-{\frac
{({\Pth}-\Kp)({\Pt}-\Kp){{p_{12}^2}}}{({\Pt}-{\Pth})({\Pt}-{\Pf})}}\nn\\
&&H_u={\frac
{{{p_{12}^2}}({\Pt}-\Kp)(-\Kp+{\Po})}{({\Po}-{\Pt})({\Pt}-{\Pf})}}\hspace{.5in} H_r=-{\frac {({\Pt}-\Kp)^{2}{{p_{14}^2}}}{({\Po}-{\Pt})({\Pt}-{\Pth})}}\nn\\
\rmid\nn\\
&&H_s=\frac
{{{p_{12}^2}}(\Kp-\Pt)({\Pf}-\Kp)}{({\Pt}-{\Pf})^{2}}\hspace{.5in}H_t=-{\frac {{{p_{14}^2}}(-\Kp+{\Po})({\Pth}-\Kp)}{({\Po}-{\Pth})^{2}}}\nn\\
&&H_d={\frac
{{{p_{12}^2}}({\Pth}-\Kp)({\Pf}-\Kp)}{({\Pt}-{\Pf})(-{\Pth}+{\Pf})}}\hspace{.5in}H_l=-{\frac {{{p_{14}^2}}({\Pf}-\Kp)({\Pth}-\Kp)}{(-{\Pth}+{\Pf})({\Po}-{\Pth})}}\nn\\
&&H_u={\frac
{{{p_{12}^2}}({\Pt}-\Kp)(-\Kp+{\Po})}{({\Po}-{\Pt})({\Pt}-{\Pf})}}\hspace{.5in}H_r=-{\frac {({\Pt}-\Kp)(-\Kp+{\Po}){{p_{14}^2}}}{({\Po}-{\Pt})({\Po}-{\Pth})}}\nn\\
\rright\nn\\
&&H_s=\frac
{{{p_{12}^2}}(\Kp-\Pt)({\Pf}-\Kp)}{({\Pt}-{\Pf})^{2}}\hspace{.5in}H_d={\frac {{{p_{12}^2}}({\Pth}-\Kp)({\Pf}-\Kp)}{(-{\Pth}+{\Pf})({\Pt}-{\Pf})}}\nn\\
&&H_l=-{\frac
{{{p_{14}^2}}({\Pf}-\Kp)^{2}}{(-{\Pf}+{\Po})(-{\Pth}+{\Pf})}}\hspace{.5in}H_u=-{\frac
{{{p_{12}^2}}(-\Kp+{\Po})({\Pf}-\Kp)}{(-{\Pf}+{\Po})({\Pt}-{\Pf})}}\nn\eea
Assume
that the coefficient of this diagram is
$A+B(\Kp-\Pth)+\mathrm{single\;pole}+\mathrm{double\;pole}$.
Multiply the above result by the prefactor, partial fraction the
coefficient of the logarithm into pole terms and polynomials. The
polynomials can be integrated to give:: \bea&&{\frac {1}{72}}
({\Pth}-{\Pt})({\Pth}-{\Pf})(-4 B{\Pth}+2 B{\Pf}+9 A+2
B{\Pt})\nn\\&&-\frac{1}{24} (B{\Pt}-2 B{\Pth}+3
A+B{\Pf})({\Pth}-{\Pf})({\Pth}-{\Pt})\log{(p_{12}^2\delta
 e^{\gamma})}\nn\eea It's quite interesting that although in
principle, a single pole term in the prefactor can still
contribute, their net effect is zero. After the polynomial terms
are integrated out, the remaining terms will be the infrared
terms, they either cancel or combine with infrared terms from
other diagrams into complete trees.

Similarly, the quartic triangle integrand
 \bea -K_{1,6}^{\lor}
K_{5,2}^\land
\frac{1}{q_1^2}\frac{1}{q_0^2}\frac{1}{q_2^2}\nn\eea gives:
\bea\Pt<\Kp<\Pth&+&\frac{1}{4}[\log{H_s}+\log{(\delta e^{\gamma})}]\frac{({\Pf}-\Kp)({\Po}-{\Pt})}{({\Pf}-{\Pt})}-\frac{1}{4}[\log{H_u}+\log{(\delta e^{\gamma})}]({\Po}-\Kp)\nn\\
\nn\\
\Pth<\Kp<\Po&+&\frac{1}{4}[\log{H_s}+\log{(\delta e^{\gamma})}]\frac{({\Pf}-\Kp)({\Po}-{\Pt})}{({\Pf}-{\Pt})}-\frac{1}{4}[\log{H_u}+\log{(\delta e^{\gamma})}]({\Po}-\Kp)\nn\\
\nn\\
\Po<\Kp<\Pf&-&\frac{1}{4}[\log{H_s}+\log{(\delta
 e^{\gamma})}]\frac{({\Po}-{\Pf})(-{\Pt}+\Kp)}{({\Pf}-{\Pt})}+\frac{1}{4}
[\log{H_u}+\log{(\delta e^{\gamma})}]({\Po}-\Kp)\nn\eea 
Assume that the prefactor of this integrand is
$A+B(\Kp-\Po)+\mathrm{single\;pole}+\mathrm{double\;pole}$, a
similar procedure gives for the $A$, $B$ terms:
\bea&&{\frac {1}{72}}
(-{\Pt}+{\Po})(-{\Pf}+{\Po})(-4 B{\Po}+2 B{\Pf}+2 B{\Pt}+9
A)\nn\\&&+\frac{1}{24}({\Po}-{\Pf})({\Pt}-{\Po})(B{\Pt}-2
B{\Po}+B{\Pf}+3 A)\log((p_{12}^2\delta e^{\gamma}))\nn\eea
Next, the integrand
\bea -K_{6,1}^{\lor} K_{6,4}^\land
\frac{1}{q_1^2}\frac{1}{q_0^2}\frac{1}{q_3^2}\nn\eea
gives: \bea\Pt<\Kp<\Pth&+&\frac{1}{4}[\log{H_r}+\log{(\delta e^{\gamma})}]({\Pt}-\Kp)\nn\\
\nn\\
\Pth<\Kp<\Po&-&\frac{1}{4}
[\log{H_t}+\log{(\delta e^{\gamma})}]\frac{({\Pth}-\Kp)({\Po}-{\Pt})}{({\Po}-{\Pth})}+\frac{1}{4} [\log{H_r}+\log{(\delta e^{\gamma})}]({\Pt}-\Kp)\nn\\
\nn\\
\Po<\Kp<\Pf& &0\nn\eea Assume the prefactor of this model
integrand is:
$A+B(\Kp-\Pt)+\mathrm{single\;pole}+\mathrm{double\;pole}$. Then the $A$, $B$
terms can be integrated to give:
\bea&&{\frac {1}{72}}
({\Pth}-{\Pt})(-{\Pt}+{\Po})(2 B{\Po}+2 B{\Pth}+9 A-4
B{\Pt})\nn\\&&+\frac{1}{24} ({\Pt}-{\Po})({\Pth}-{\Pt})(B{\Pth}-2
B{\Pt}+3 A+B{\Po})\log{(p_{14}^2\delta e^{\gamma})}\nn\eea
Finally, the integrand
\bea -K_{3,5}^{\lor} K_{2,5}^\land
\frac{1}{q_1^2}\frac{1}{q_3^2}\frac{1}{q_2^2}\nn\eea
gives \bea\Pt<\Kp<\Pth& &0\nn\\
\nn\\
\Pth<\Kp<\Po&+&\frac{1}{4}
[\log{H_t}+\log{(\delta e^{\gamma})}]\frac{(-{\Pth}+{\Pf})(-\Kp+{\Po})}{\Po-\Pth}-\frac{1}{4}[\log{H_l}+\log{(\delta e^{\gamma})}]({\Pf}-\Kp)\nn\\
\nn\\
\Po<\Kp<\Pf &-&\frac{1}{4}[\log{H_l}+\log{(\delta
 e^{\gamma})}]({\Pf}-\Kp)\nn\eea 
Assume the prefactor of this
model integrand is:
$A+B(\Kp-\Pf)+\mathrm{single\;pole}+\mathrm{double\;pole}$. And the $A$,$B$
terms can be integrated to give: 
\bea&&{\frac {1}{72}}
({\Pth}-{\Pf})({\Po}-{\Pf})(2 B{\Po}+2 B{\Pth}-4 {\Pf}
B+9A)\nn\\&&-\frac{1}{24}
({\Po}-{\Pf})({\Pth}-{\Pf})(B{\Pth}+B{\Po}+3 A-2 {\Pf}
B)\log{(p_{14}^2\delta e^{\gamma})}\nn\eea
\vskip14pt
\begin{subsection}
{Double Quartic Diagrams}
\end{subsection}
\begin{figure}[ht]
\begin{center}
\includegraphics[width=5in]{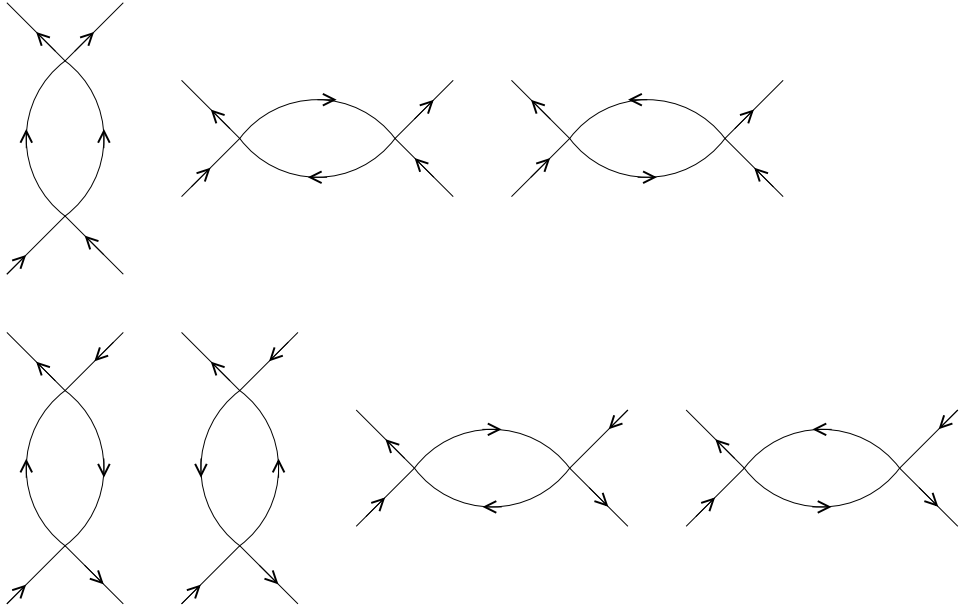}
\caption{The double quartic diagrams for the two possible
polarization patterns. Particle labels $1234$ are applied counter-clockwise
starting at the lower left of each diagram.}
\label{doublequart}
\end{center}
\end{figure}
A typical double quartic integrand is:
$\frac{1}{q_1^2}\frac{1}{q_3^2}$
This gives: \bea-\pref
g^4\Nc\int_{\Pth}^{\Po}\frac{d\Kp}{\Po-\Pth}\frac{1}{2}\log{\left[\frac{(\Kp-\Pth)(\Po-\Kp)}{(\Po-\Pth)^2}p_{14}^2\delta
e^{\gamma}\right]}\nn\eea

We can also have a diagram like $\frac{1}{q_0^2}\frac{1}{q_2^2}$.
Its contribution is similar, but spans over all three regions.
\bea-\pref
g^4\Nc\int_{\Pt}^{\Pf}\frac{d\Kp}{\Pf-\Pt}\frac{1}{2}\log{\left[\frac{(\Kp-\Pt)(\Pf-\Kp)}{(\Pf-\Pt)^2}p_{12}^2\delta
e^{\gamma}\right]}\nn\eea

Assuming that the prefactors of the double quartic diagrams are
$A+\mathrm{pole\;terms}$. We can integrating out the A term,
leaving the rest as infrared terms. Thus we have: \bea \pref
\frac{-1}{2}\left[-2+\log{(p_{14}^2\delta
e^{\gamma})}\right]\nn\eea for the first case and \bea \pref
\frac{-1}{2}\left[-2+\log{(p_{12}^2\delta
e^{\gamma})}\right]\nn\eea for the second case.
\vskip14pt
\section{Final Results}
In the previous sections we have described our calculational methods
by choosing a single example of each distinct type, and analyzing it in 
detail (relegating the more tedious parts to appendices). These
examples are chosen to illustrate every type of technical
complication we encountered. However, along with each such
example there are quite a few others involving essentially identical
calculations. In fact, there are so many that we chose to automate
their calculation using Matlab and Mathematica. After the results
of all these many calculations are combined, there ensues a stunning
simplification that allows us to 
present the complete elastic glue-glue scattering amplitude in the
first subsection, and in the second subsection, to obtain the complete answer
for probabilities including unseen gluons in the initial and
final states. This last result is compact, infrared
finite, manifestly Lorentz invariant, and displays the
ultraviolet behavior dictated by asymptotic freedom.

\begin{subsection}
{One Loop Corrections to Elastic Scattering of Glue by Glue}
\end{subsection}
We quote here the {\it amputated} four gluon amplitudes, 
which do not include any external leg corrections. 
\bea \Anei^{1-loop}&=&-{g^2\over8\pi^2}
\left[-(\log^2{\frac{p_{12}^2}{p_{14}^2}}+\pi^2)-\frac{11}{3}
\log[{p_{14}^2}\delta e^\gamma]+\frac{73}{9}\right]\Anei^{tree}\nn\\
&&+\pref g^2 \Nc\left[\frac{1}{3}\textbf{(4 pt
vertex)}+g^2\frac{2}{3}\right]+\left[\mathrm{IR\;terms}\right]\nn\\
\nn\\
\Aalt^{1-loop}&=&-{g^2\over8\pi^2}\bigg[
-\frac{(p_{12}^4+p_{12}^2p_{14}^2+p_{14}^4)^2}{(p_{14}^2+p_{12}^2)^4}
\left(\log^2{\frac{p_{12}^2}{p_{14}^2}}+\pi^2\right)
+\frac{p_{12}^2}{3}\frac{(14p_{14}^4+19p_{12}^2p_{14}^2+11p_{12}^4)}
{(p_{14}^2+p_{12}^2)^3}\log{\frac{p_{12}^2}{p_{14}^2}}\nn\\
&&-\frac{11}{3}\log[{p_{12}^2\delta e^\gamma}]+\frac{p_{14}^2
p_{12}^2}{(p_{14}^2+p_{12}^2)^2}+\frac{73}{9}\bigg]
\Aalt^{tree}\nn\\
&&+{g^2\over8\pi^2}\Nc\left[\frac{1}{3}\textbf{(4 pt
vertex)}+g^2\frac{2}{3}\right]+\left[\mathrm{IR\;terms}\right]\nn\eea
The infrared terms for both helicity 
arrangements are the same multiples of the corresponding trees:
\bea &&\Gamma_{IR}=-\pref g^2\Nc A^{tree}\times\nn\\
\rleft\nn\\
& &\left[\frac{1}{\Kp-\Pth}+\frac{1}{\Kp-\Po}\right]\log{\frac
{({\Pf}-\Kp)(-{\Pt}+\Kp)p_{12}^2\delta e^\gamma}{({\Pt}-{\Pf})^{2}}}
\nn\\
&-&\left[\frac{2}{\Kp-\Po}\right]\log{\frac
{({\Po}-\Kp)(\Kp-{\Pt})p_{12}^2\delta e^\gamma}{({\Po}-{\Pt})({\Pf}-{\Pt})}}\nn\\
&-&\left[\frac{2}{\Kp-\Pth}\right]\log{\frac
{({\Pth}-\Kp)(-{\Pt}+\Kp)p_{12}^2\delta e^\gamma}{({\Pth}-{\Pt})({\Pf}-{\Pt})}}\nn\\
&+&\left[\frac{2}{\Kp-\Pt}\right]\log{\frac
{(-{\Pt}+\Kp)^{2}p_{14}^2\delta e^\gamma}{({\Po}-{\Pt})({\Pth}-{\Pt})}}\nn\\
\nn\\
\rmid\nn\\
&-&\left[\frac{1}{\Kp-\Pth}-\frac{1}{\Kp-\Po}\right]
\log{\frac{({\Pf}-\Kp)(-{\Pt}+\Kp)p_{12}^2\delta e^\gamma}{({\Pt}-{\Pf})^{2}}}\nn\\
&-&\left[\frac{1}{\Kp-\Pt}-\frac{1}{\Kp-\Pf}\right]
\log{\frac{({\Pth}-\Kp)(\Kp-{\Po})p_{14}^2\delta e^\gamma}{(-{\Po}+{\Pth})^{2}}}\nn\\
&-&\left[\frac{2}{\Kp-\Po}\right]\log{\frac
{({\Pt}-\Kp)(\Kp-{\Po})p_{12}^2\delta e^\gamma}{(-{\Po}+{\Pt})({\Pt}-{\Pf})}}\nn\\
&+&\left[\frac{2}{\Kp-\Pth}\right]\log{\frac
{(\Kp-\Pth)({\Pf}-\Kp)p_{12}^2\delta e^\gamma}{({\Pf}-{\Pt})({\Pf}-{\Pth})}}\nn\\
&-&\left[\frac{2}{\Kp-\Pf}\right]\log{\frac
{(-{\Pf}+\Kp)({\Pth}-\Kp)p_{14}^2\delta e^\gamma}{({\Pth}-{\Pf})(-{\Po}+{\Pth})}}\nn\\
&+&\left[\frac{2}{\Kp-\Pt}\right]\log{\frac
{({\Pt}-\Kp)(\Kp-{\Po})p_{14}^2\delta e^\gamma}{(-{\Po}+{\Pt})(-{\Po}+{\Pth})}}\nn\\
\nn\\
\rright\nn\\
&-&\left[\frac{1}{\Kp-\Pth}+\frac{1}{\Kp-\Po}\right]
\log{\frac{({\Pf}-\Kp)(-{\Pt}+\Kp)p_{12}^2\delta e^\gamma}{({\Pt}-{\Pf})^{2}}}\nn\\
&+&\left[\frac{2}{\Kp-\Po}\right]\log{\frac
{(\Kp-{\Po})(-{\Pf}+\Kp)p_{12}^2\delta e^\gamma}{({\Pf}-{\Po})({\Pt}-{\Pf})}}\nn\\
&+&\left[\frac{2}{\Kp-\Pth}\right]\log{\frac
{({\Kp}-\Pth)({\Pf}-\Kp)p_{12}^2\delta e^\gamma}{({\Pf}-{\Pt})({\Pf}-{\Pth})}}\nn\\
&-&\left[\frac{2}{\Kp-\Pf}\right]
\log{\frac{(-{\Pf}+\Kp)^{2}p_{14}^2\delta e^\gamma}{({\Pf}-{\Po})({\Pf}-{\Pth})}}\nn\eea
The reader will note that the infrared sensitive terms depend on the
ultraviolet cutoff $\delta$. This entangling of infrared and ultraviolet
divergences is a familiar consequence of the way we have
cut off $p^+=0$ singularities. These entangled
divergences are precisely cancelled by similar divergences in the self-energy
corrections to external lines which contribute to the $\prod \sqrt{Z_i}$ factors
that convert the amputated amplitudes to properly normalized scattering
amplitudes. When these factors are included (as they will be in the
next subsection on physical probabilities), the net coefficient of
$\ln\delta$ becomes $-11g^2/24\pi^2$ in precise agreement with asymptotic
freedom \cite{thornfreedom,perryfreedom}.

The terms in these amplitudes that are not multiplied by trees
are Lorentz violating anomalies that must be removed by counterterms.
In Section 10 we show how this can be done locally in
target space and described locally on the worldsheet. As we shall 
see, after these counterterms are taken into account, 
the only change in the rest of the
formula is a change $73/9\to 67/9$ in the constant terms
multiplying the respective trees. We assume these changes have
been done in our discussion of unseen gluons in the following
subsection.

The expressions for the loop amplitudes are real in the unphysical
region for scattering in the 12 channel where $p_{12}^2=-s$ and
$p_{14}^2=-t$ are both positive. The physical region for this
process is $s>0, t<0$, which we can obtain by analytic continuation.
Since the physical $s$ is above the cut on the positive real axis,
we obtain the physical amplitudes by substituting $p_{12}^2=e^{-i\pi}s$
in the above formulas. In this way we see that in the
physical region the amplitudes acquire an imaginary part. In the
next subsection we use these physical region amplitudes in the 
calculation of gluon detection probabilities.

\begin{subsection}
{Probabilities Including Bremsstrahlung and Unseen Initial Gluons}
\end{subsection}
In this section, we will focus on the case when the extra gluon(we
can break the Bose symmetry by defining it to be the softest one)
is between particle 3 and 4. Recalling the results of section 5,
there is a total of three terms that contribute to the infrared
and collinear singularity. 
\bea
M_{coll}&=&\frac{g^2}{8\pi^2}\sum_{i=3,4}
\left|A_{core}^i\right|^2\int_0^{|P_i^+|}dk^+\nn\\
&&\left(\frac{|P_i^+|}{k^+(|P_i^+|-k^+)}
+\frac{(|P_i^+|-k^+)^3}{|k^+P_i^{+3}|}
+\frac{k^{+3}}{(|P_i^+|-k^+)|P_i^+|^{3}|}\right)
\log{\frac{k^+(|P_i^+|-k^+)\Delta^2}{P_i^{+2}}}\nn
\eea

\bea M_{soft\
brem}&=&\frac{g^2}{4\pi^2}\left|A_{core}\right|^2\left[\int_{|k^+|<A}\frac{dk^+}{|k^+|}\log{\frac{k^{+2}s^2}{|P_3^+P_4^+|\Delta^4}}-\log{\frac{\Delta^2|P_3^+|}{As}}\log{\frac{\Delta^2|P_4^+|}{As}}\right]\nn\\
&=&\frac{g^2}{4\pi^2}\left|A_{core}\right|^2\bigg[\int_{|k^+|<A}\frac{dk^+}{|k^+|}\log{\frac{k^{+4}s^2}{|P_3^+P_4^+|^2}}+\int_{|k_+|<A}\frac{dk^+}{|k^+|}\log{\frac{|P_3^+P_4^+|}{k^{+2}\Delta^4}}\nn\\
&&-\left(\log{\frac{\Delta^2|P_3^+|}{s}}-\log{A}\right)\left(\log{\frac{\Delta^2|P_4^+|}{s}}-\log{A}\right)\bigg]\nn\eea

While the infrared terms from loop calculation in region 34 is:
\bea
M_{loop}&=&\frac{-g^2|A_{core}|^2}{4\pi^2}\bigg[\int_{0}^{|P_4^+|}\frac{dk^+}{k^+}\log{\frac{k^{+2}s}{|P_4^+||P_3^+|}}+\int_{0}^{|P_3^+|}\frac{dk^+}{k^+}\log{\frac{k^{+2}s}{|P_3^+||P_4^+|}}\nn\\
&+&\int_{-|P_4^+|}^{|P_3^+|}\frac{dk^+}{k^+}\log{\frac{(|P_3^+|-k^+)|P_4^+|}{(|P_4^+|+k^+)|P_3^+|}}\bigg]\nn\eea

First, notice that $M_{coll}$ is organized according to which leg
the collinear emission is attached, but for the sake of the
current discussion, we need to break them up into different
regions. For example, rewrite the term (take $i=3$)
\bea\frac{|P_3^+|}{k^+(|P_3^+|-k^+)}+\frac{(|P_3^+|-k^+)^3}{k^+|P_3^+|^{3}}+\frac{k^{+3}}{(|P_3^+|-k^+)|P_3^+|^{3}}\nn\eea
as
\bea\frac{1}{k^+}+\frac{(|P_3^+|-k^+)^3}{|k^+P_3^{+3}|}+\frac{1}{|P_3^+|-k^+}+\frac{k^{+3}}{(|P_3^+|-k^+)|P_3^{+3}|}\nn\eea
The first two term are related to the last two by substituting
$k^+\to\,(P_3^+-k^+)$. The first two will be divergent
when the extra gluon becomes soft. The last two will be divergent
when the extra gluon becomes dominating over gluon 3, so, by the
definition of 'the extra gluon' given above, gluon 3 becomes the
'extra one'. Hence we assign the last two terms to region 23.

Second combine $M_{loop}$ with $M_{soft\ brem}$ (leaving out a
common factor $\frac{g^2}{8\pi^2}\left|A_{core}\right|^2$) \bea
M_{loop}+M_{soft\
brem}&=&-2\bigg[\int_{A}^{|P_4^+|}\frac{dk^+}{k^+}\log{\frac{k^{+2}s}{|P_4^+||P_3^+|}}+\int_{A}^{|P_3^+|}\frac{dk^+}{k^+}\log{\frac{k^{+2}s}{|P_3^+||P_4^+|}}+\mathcal{I}_{34}\bigg]\nn\\
&+&2\int_{|k_+|<A}\frac{dk^+}{|k^+|}\log{\frac{|P_3^+P_4^+|}{k^{+2}\Delta^4}}-2\left(\log{\frac{\Delta^2|P_3^+|}{s}}-\log{A}\right)\left(\log{\frac{\Delta^2|P_4^+|}{s}}-\log{A}\right)\nn\eea

where $\mathcal{I}_{34}$ is \bea
\int_{-|P_4^+|}^{|P_3^+|}\frac{dk^+}{k^+}\log{\frac{(|P_3^+|-k^+)|P_4^+|}{(|P_4^+|+k^+)|P_3^+|}}\nn\eea
We can see that the first two terms in $M_{soft\ brem}$ cancel the
divergence in the first two terms of $M_{loop}$. Performing the
first two integrals: \bea
&-&2\left[\log^2{|P_3^+|}+\log^2{|P_4^+|}-2\log^2{A}+\log{\frac{|P_3^+P_4^+|}{A^2}}\log{\frac{s}{|P_3^+P_4^+|}}+\mathcal{I}_{34}\right]\nn\\
&+&2\int_{0<k^+<A}\frac{dk^+}{k^+}\log{\frac{|P_3^+P_4^+|}{k^{+2}\Delta^4}}-2\left(\log{\frac{\Delta^2|P_3^+|}{s}}-\log{A}\right)\left(\log{\frac{\Delta^2|P_4^+|}{s}}-\log{A}\right)\nn\eea

Rewrite the divergent integral as\bea
&&2\int_{0<k^+<A}\frac{dk^+}{k^+}\log{\frac{|P_3^+P_4^+|}{k^{+2}\Delta^4}}\nn\\
&&=\int_{0}^{|P_3^+|}\frac{dk^+}{k^+}\log{\frac{|P_3^+P_4^+|}{k^{+2}\Delta^4}}+\int_{0}^{|P_4^+|}\frac{dk^+}{k^+}\log{\frac{|P_3^+P_4^+|}{k^{+2}\Delta^4}}\nn\\
&&-\bigg[-\log^2{|P_3^+|}-\log^2{|P_4^+|}+2\log^2{A}+\log{\frac{|P_3^+P_4^+|}{A^2}}\log{\frac{|P_3^+P_4^+|}{\Delta^4}}\bigg]\nn\eea
The first two integrals will be cancelled by $M_{coll}$ later. We
get: \bea M_{loop}+M_{soft\
brem}&=&-2\left[\log^2{|P_3^+|}+\log^2{|P_4^+|}+\log{|P_3^+P_4^+|}\log{\frac{s}{|P_3^+P_4^+|}}+\mathcal{I}_{34}\right]\nn\\
&+&\int_{0}^{|P_3^+|}\frac{dk^+}{k^+}\log{\frac{|P_3^+P_4^+|}{k^{+2}\Delta^4}}+\int_{0}^{|P_4^+|}\frac{dk^+}{k^+}\log{\frac{|P_3^+P_4^+|}{k^{+2}\Delta^4}}\nn\\
&+&\bigg[\log^2{|P_3^+|}+\log^2{|P_4^+|}-\log{|P_3^+P_4^+|}\log{\frac{|P_3^+P_4^+|}{\Delta^4}}\bigg]-2\log\frac{\Delta^2|P_3^+|}{s}\log\frac{\Delta^2|P_4^+|}{s}\nn\\
&=&-2
\mathcal{I}_{34}+\int_{0}^{|P_3^+|}\frac{dk^+}{k^+}\log{\frac{|P_3^+P_4^+|}{k^{+2}\Delta^4}}+\int_{0}^{|P_4^+|}\frac{dk^+}{k^+}\log{\frac{|P_3^+P_4^+|}{k^{+2}\Delta^4}}-2\log^2{\frac{\Delta^2}{s}}\nn\eea

\bea
M_{coll}&=&\int_0^{|P_3^+|}\frac{dk^+}{k^+}\left(1+\frac{(|P_3^+|-k^+)^3}{|P_3^{+3}|}\right)\log{\frac{k^+(|P_3^+|-k^+)\Delta^2}{P_3^{+2}}}+3\,\leftrightarrow\,4\nn\\
&=&\int_0^{|P_3^+|}\frac{dk^+}{k^+}\left(2+\frac{(|P_3^+|-k^+)^3-|P_3^{+3}|}{|P_3^{+3}|}\right)\log{\frac{k^+(|P_3^+|-k^+)\Delta^2}{P_3^{+2}}}+3\,\leftrightarrow\,4\nn\eea

Putting everything together, we get: \bea &
&-2\mathcal{I}_{34}-2\log^2{\frac{\Delta^2}{s}}+\left\{\int_{0}^{|P_3^+|}\frac{dk^+}{k^+}\log{\left[\frac{|P_3^+P_4^+|}{k^{+2}\Delta^4}\frac{k^{+2}(|P_3^+|-k^+)^2\Delta^4}{P_3^{+4}}\right]}+3 \leftrightarrow 4\right\}\nn\\
&&+\left\{\int_0^{|P_3^+|}\frac{dk^+}{k^+}\frac{(|P_3^+|-k^+)^3-|P_3^{+3}|}{|P_3^{+3}|}\log{\frac{k^+(|P_3^+|-k^+)\Delta^2}{P_3^{+2}}}+3\leftrightarrow4\right\}\nn\\
&&=-2\mathcal{I}_{34}-2\log^2{\frac{\Delta^2}{s}}+\left\{\int_{0}^{|P_3^+|}\frac{dk^+}{k^+}\left[\log{\frac{|P_4^+|}{|P_3^+|}}+\log{\frac{(|P_3^+|-k^+)^2}{P_3^{+2}}}\right]+3 \leftrightarrow 4\right\}+2\left[\frac{67}{18}-\frac{11}{6}\log{\Delta^2}\right]\nn\\
&&=-2\mathcal{I}_{34}-2\log^2{\frac{\Delta^2}{s}}+(-\frac{2}{3}\pi^2)-\log^2{\frac{P_3^+}{P_4^+}}+2\left[\frac{67}{18}-\frac{11}{6}\log{\Delta^2}\right]\nn\\
&&=-2\left(-\frac{1}{2}\pi^2-\frac{1}{2}\log^2{\frac{P_3^+}{P_4^+}}\right)-2\log^2{\frac{\Delta^2}{s}}+(-\frac{2}{3}\pi^2)-\log^2{\frac{|P_3^+|}{|P_4^+|}}+2\left[\frac{67}{18}-\frac{11}{6}\log{\Delta^2}\right]\nn\\
&&=-2\log^2{\frac{\Delta^2}{s}}+\frac{1}{3}\pi^2+\left[\frac{67}{9}-\frac{11}{3}\log{\Delta^2}\right]\nn
\eea
Thus, after including the (12), (23), and (41) cases,
the total probabilities are:
\bea
P_{\land\land\lor\lor}&=&|A_{\land\land\lor\lor}|^2
\bigg[1+\frac{g^2}{4\pi^2}\bigg[-2\log^2{\frac{\Delta^2}{s}}
-2\log^2{\frac{\Delta^2}{|t|}}+2\cdot\frac{\pi^2}{3}+\frac{67}{9}\nonumber\\
&&\hskip2in-\frac{11}{3}\left[\log{(\Delta^2\delta e^{\gamma})}
+\log{\frac{\Delta^2}{|t|}}\right]+\log^2{\frac{s}{|t|}}\bigg]\bigg]\nn\\
&&\phantom{or}\label{jetprobfinal}\\
P_{\land\lor\land\lor}&=&|A_{\land\lor\land\lor}|^2
\bigg[1+\frac{g^2}{4\pi^2}\bigg[-2\log^2{\frac{\Delta^2}{s}}
-2\log^2{\frac{\Delta^2}{|t|}}+2\cdot\frac{\pi^2}{3}
+\frac{67}{9}-\frac{11}{3}[\log{(\Delta^2\delta e^{\gamma})}
+\frac{1}{2}\log{\frac{\Delta^4}{s|t|}}]\nn\\
&&\hskip.5in+\frac{(s^2+st+t^2)^2}{(t+s)^4}\log^2{\frac{s}{|t|}}
+\frac{(5st^2-5s^2t+11t^3-11s^3)}{6(t+s)^3}\cdot\log{\frac{s}{|t|}}
-\frac{ts}{(t+s)^2}\bigg]\bigg]
\label{jetprobfinal2}
\eea
We see that all IR divergences have cancelled, that the UV divergences 
are exactly those dictated by asymptotic freedom, and Lorentz invariance
is manifest.
\vskip14pt
\section{Worldsheet Description of Counterterms}
As we have seen, counterterms that are polynomials in
the dual momenta must be added to the two, three and four
point functions in order achieve the correct Lorentz covariant results.
Specifically in \cite{chakrabartiqt} we required the following
counterterms:
\bea
\Pi^{\land\land}_{\rm C.T.}&=&-{g^2\over12\pi^2}[k^{\land2}_0
+k^{\land}_0k^{\land}_1+k^{\land2}_1]\\
\Pi^{\land\lor}_{\rm C.T.}&=&-{g^2\over24m\delta}p^++{g^2\over4\pi^2\delta}\\
\Gamma^{\land\land\lor}_{\rm C.T.}&=&{g^3\over12\pi^2}[k^\land_0+k^\land_1
+k^\land_2]\label{cubicct1}
\eea
which are all polynomials in the dual momenta.

Quartic counterterms must also be included
in the list. At first sight the terms in $\Gamma$ that need to be canceled
seem to be rational functions of the $p^+_i$ 
(which would be nonlocal in $x^-$):
\bea
\Gamma^{\land\land\lor\lor}_{\rm anom}&=&{g^2\over24\pi^2}\left[
-2{g^2}{p^+_1p^+_3+p^+_2p^+_4\over(p^+_1+p^+_4)^2}
+2g^2\right]\\
\Gamma^{\land\lor\land\lor}_{\rm anom}&=&{g^2\over24\pi^2}\left[
2g^2{p^+_1p^+_2+p^+_3p^+_4\over(p^+_1+p^+_4)^2}
+2g^2{p^+_1p^+_4+p^+_2p^+_3\over(p^+_1+p^+_2)^2}
+2g^2\right]
\eea
Local counterterms must be polynomials in the momenta.
However, we note that the nonpolynomial parts of these anomalies
are proportional to the quartic vertex contributions to the corresponding
tree amplitudes. The addition of a term to $\Pi^{\land\lor}_{\rm C.T.}$
proportional to $p^2$, which is an allowed 
counterterm by power counting, would
contribute a term proportional to 
the part of the same trees built from pairs of cubic vertices. 
By tuning the coefficient
of $p^2$ we can convert these nonpolynomial anomalies to complete
trees, obviating the need to cancel them with a counterterm. 
They just correspond to a 
perfectly allowed {\it finite} coupling renormalization. The
change in $\Pi^{\land\lor}_{\rm C.T.}$ which accomplishes this is just
\bea
\Pi^{\land\lor}_{\rm C.T.}&\to&
-{g^2\over24m\delta}p^++{g^2\over4\pi^2\delta}+{g^2\over24\pi^2}p^2
\eea
After this rearrangement we see
that the required quartic counterterms are simply constants:
\bea
\Gamma^{\land\land\lor\lor}_{\rm C.T.}&=& -{g^4\over12\pi^2},\qquad\qquad
\Gamma^{\land\lor\land\lor}_{\rm C.T.}\ =\ -{g^4\over12\pi^2}
\eea
It is noteworthy that these quartic counterterms are spin independent. 
This is consistent with the
interpretation of the anomalies as ultraviolet artifacts of box diagrams: 
The large momentum behavior of box integrands must of 
necessity be of the form $q^{\mu_1}q^{\mu_2}
q^{\mu_3}q^{\mu_4}/q^8$ which integrates to completely 
symmetrized Kronecker deltas.

This is all quite satisfactory from the field theoretic point of
view, but the worldsheet description makes more stringent
requirements on the counterterms: They must be generated by
purely local changes to the worldsheet action. Worldsheet locality
is quite independent of field theoretic (target space) locality,
and we must still show how it can be preserved. We start with
$\Pi^{\land\land}_{\rm C.T.}$. Interpreted as a contribution to the
worldsheet path integral representation of a gluon
propagator, it should be multiplied by $T/2p^+$ \cite{chakrabartiqt}.
\bea
{T\Pi^{\land\land}_{\rm C.T.}\over2p^+}&=&-{g^2\over16\pi^2}
\int d\tau{q^{\land2}(0)+q^{\land2}(p^+)\over p^+}
+{g^2\over48\pi^2}
\int d\tau d\sigma \left({\partial q^\land\over\partial\sigma}\right)^2
\eea
When exponentiated (through higher loop corrections) this expression
can be interpreted as adding new terms to $-S$, where $S$ is the worldsheet
action. The first term modifies the treatment of the worldsheet
boundary in a similar way to the 
description of a mass term \cite{thornscalar}, 
and the second term
is a new bulk term. Both terms violate helicity in the right
way to cancel the helicity violation implied by the nonvanishing
of $\Pi^{\land\land}$. These new terms
also produce new effects from contact contributions arising
when the bulk term sits on the same time slice as an interaction
vertex, similarly to the generation of quartic vertices
from two cubics \cite{thornsheet}. Using the generating function
obtained in that reference we find, for correlators on the
same time slice,
\bea
\left\langle\left({q_{i+1}-q_i\over m}\right)^\lor
\left({q_{j+1}-q_j\over m}\right)^{\land 2}\right\rangle
&=&{p^{\land2}p^{\lor}\over p^{+3}}
+{2\over a}{p^{\lor}\over p^{+2}}
\left[{\delta_{ij}\over m}-{1\over p^+}\right]\\
am\sum_j\left\langle\left({q_{i+1}-q_i\over m}\right)^\lor
\left({q_{j+1}-q_j\over m}\right)^{\land 2}\right\rangle
&=&a{p^{\land2}p^{\lor}\over p^{+2}}\\
\left\langle\left({q_{i+1}-q_i\over m}\right)^\lor
\left({q_{j+1}-q_j\over m}\right)^{\land 2}
\left({q_{l+1}-q_l\over m}\right)^\lor\right\rangle
&=&{p^{\land2}p^{\lor2}\over p^{+4}}
+{2\over a}{p^{\land}p^{\lor}\over p^{+2}}
\left[{\delta_{ij}+\delta_{lj}\over m}-{2\over p^+}\right]\nonumber\\
&&+{2\over a^2}\left[{\delta_{ij}\over m}-{1\over p^+}\right]
\left[{\delta_{lj}\over m}-{1\over p^+}\right]\\
am\sum_j\left\langle\left({q_{i+1}-q_i\over m}\right)^\lor
\left({q_{j+1}-q_j\over m}\right)^{\land 2}
\left({q_{l+1}-q_l\over m}\right)^\lor\right\rangle
&=&a{p^{\land2}p^{\lor2}\over p^{+3}}
+{2\over a}\left[{\delta_{il}\over m}-{1\over p^+}\right]
\eea
Here $a,m$ are the discretization units of $\tau,\sigma$ respectively.
The terms proportional to $a$ will be negligible in the continuum limit,
so we see that the only contact contribution that survives is
the last term on the last line. Its effect is to produce new
quartic vertices similar to the quartic anomalies already
discussed. Indeed the
usual quartic vertices arise from the second term of the correlator
\bea
\left\langle\left({q_{i+1}-q_i\over m}\right)^\lor
\left({q_{j+1}-q_j\over m}\right)^{\land }\right\rangle
&=&{p^{\land}p^{\lor}\over p^{+2}}
+{1\over a}\left[{\delta_{ij}\over m}-{1\over p^+}\right]
\eea
which has an exactly analogous $1/a$ contribution. By a parallel
calculation we easily find that the contraction terms produced
by $\Pi^{\land\land}_{\rm C.T.}$  give the following 
quartic vertices:
\bea
\Gamma^{\land\land\lor\lor}_\Pi&=&-{g^4\over12\pi^2}\left[
{p_1^+p_3^++p_2^+p_4^+\over p_{14}^{+2}}+1\right]\\
\Gamma^{\land\lor\land\lor}_\Pi&=&{g^4\over12\pi^2}\left[
{p_1^+p_2^++p_3^+p_4^+\over p_{14}^{+2}}
+{p_1^+p_4^++p_2^+p_3^+\over p_{12}^{+2}}+2\right]
\eea
These new terms add to the anomalous 
contributions previously discussed:
\bea
\Gamma^{\land\land\lor\lor}_{\rm anom}&\to&
\Gamma^{\land\land\lor\lor}_{\rm anom~ws}
\ =\ {g^2\over24\pi^2}\left[
-4{g^2}{p^+_1p^+_3+p^+_2p^+_4\over(p^+_1+p^+_4)^2}\right]\\
\Gamma^{\land\lor\land\lor}_{\rm anom}&\to&
\Gamma^{\land\lor\land\lor}_{\rm anom~ws}\ =\ 
{g^2\over24\pi^2}\left[
4g^2{p^+_1p^+_2+p^+_3p^+_4\over(p^+_1+p^+_4)^2}
+4g^2{p^+_1p^+_4+p^+_2p^+_3\over(p^+_1+p^+_2)^2}
+6g^2\right]
\eea
The nonlocal terms in these expressions can be handled as
before by retuning
the $p^2$ term in $\Pi^{\land\lor}$ a little differently:
\bea
\Pi^{\land\lor}_{\rm C.T.}&\to&\Pi^{\land\lor}_{\rm C.T.ws}
\ =\ -{g^2\over24m\delta}p^++{g^2\over4\pi^2\delta}+{g^2\over12\pi^2}p^2
\eea
 The worldsheet description of the first two terms in
$\Pi^{\land\lor}_{\rm C.T.ws}$ has been explained in
\cite{chakrabartiqt}: the first term can be absorbed in 
a worldsheet boundary cosmological constant, and the second
term, which has the interpretation as a shift in the
gluon (mass)${}^2$, can be absorbed in the worldsheet description
of a mass counterterm \cite{thornscalar}.
The way to put the last term in the worldsheet description is
a little more subtle. We first note that its effect is
simply a finite contribution to wave function renormalization $Z$ which
multiplies the cubic vertex by $Z^{3/2}$ and the quartic vertex by $Z^2$.
It is convenient to, at the same time, make a finite renormalization of the
coupling constant $g\to g/Z$ to reduce the effect to multiplying
the cubic vertex by $Z^{1/2}$ leaving the quartic vertex untouched.
Then the net effect, to be described by the
worldsheet formalism, is to modify the constant
part of the correction
multiplying the part of the tree involving gluon exchange
from $73/9\to 67/9$ without touching the correction multiplying the
quartic vertex part of the tree, which gets adjusted to $67/9$
by the nonlocal part of the quartic anomaly. One's first
thought is to simply change the coefficient of the cubic
vertex appropriately. But then the worldsheet contact contributions
would make a corresponding modification to the quartic
vertex contribution and the change would only
amount to a finite renormalization of $g$. Fortunately it is possible to
prevent the $\partial q/\partial\sigma$ worldsheet insertions
from generating quartic contributions by altering the ghost 
worldsheet action near the interaction point.

\begin{figure}[ht]
\psfrag{'='}{$=$}
\psfrag{'-'}{$-$}
\psfrag{'triangle'}{\large Triangle-like}
\psfrag{'+'}{$\hskip -.25in +$}
\psfrag{'q'}{}
\psfrag{'k0'}{}
\psfrag{'k1'}{}
\psfrag{'k2'}{}
\psfrag{'k3'}{}
\begin{center}
\includegraphics[width=2in]{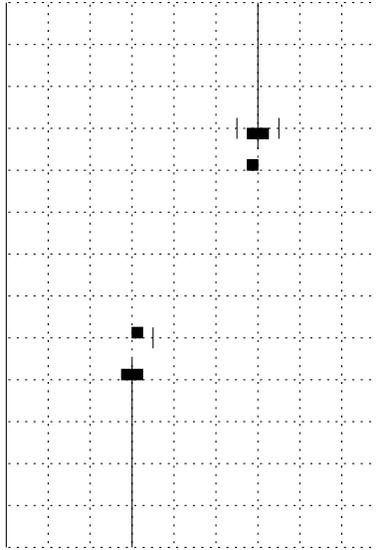}
\caption{Discretized worldsheet for a four gluon tree. The solid squares
indicate where $\partial q/\partial\sigma$ insertions can be located.
The short vertical lines indicate the links to be deleted in the
worldsheet ghost action. All indicated ghost link
deletions are present regardless of the insertion location.}
\label{treeghosts}
\end{center}
\end{figure}

\begin{figure}[htp]
\psfrag{'='}{$=$}
\psfrag{'-'}{$-$}
\psfrag{'triangle'}{\large Triangle-like}
\psfrag{'+'}{$\hskip -.25in +$}
\psfrag{'q'}{}
\psfrag{'k0'}{}
\psfrag{'k1'}{}
\psfrag{'k2'}{}
\psfrag{'k3'}{}
\begin{center}
\includegraphics[width=5.5in]{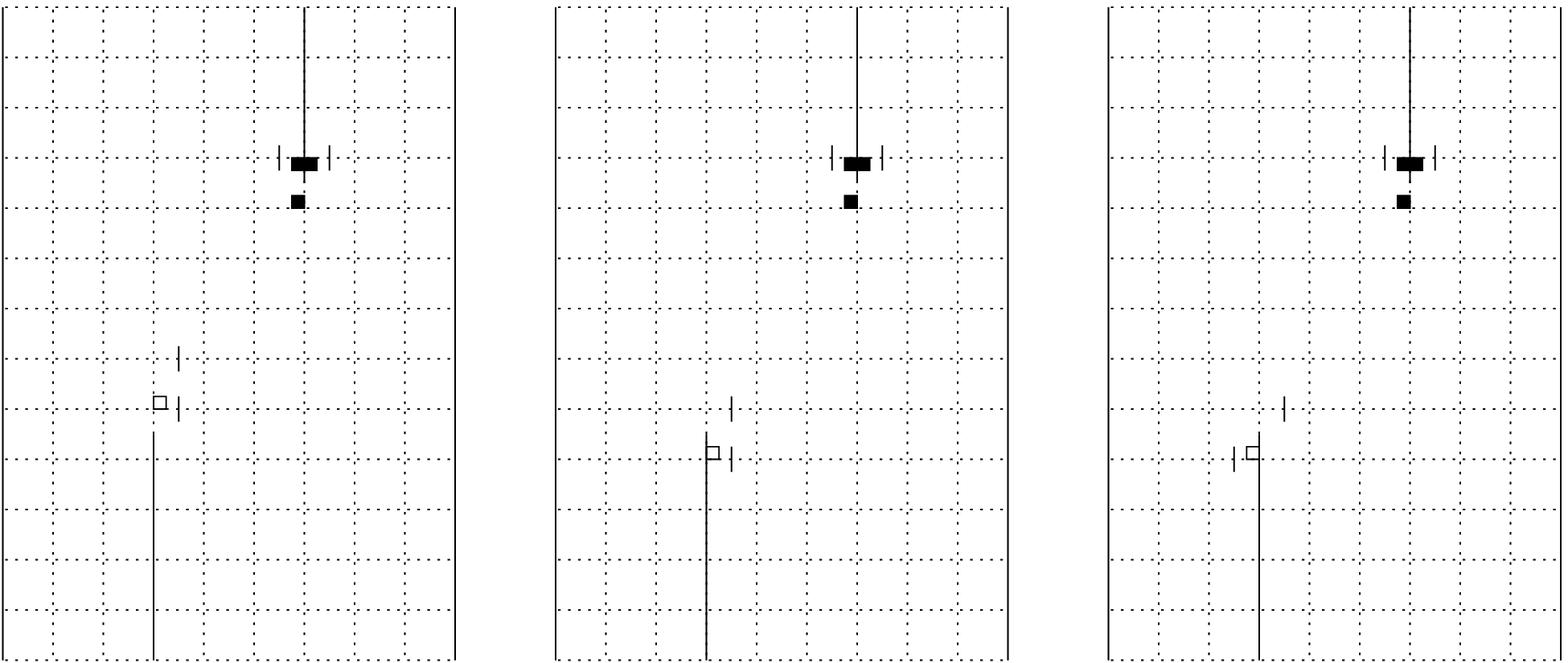}\\
\vskip14pt
\includegraphics[width=5.5in]{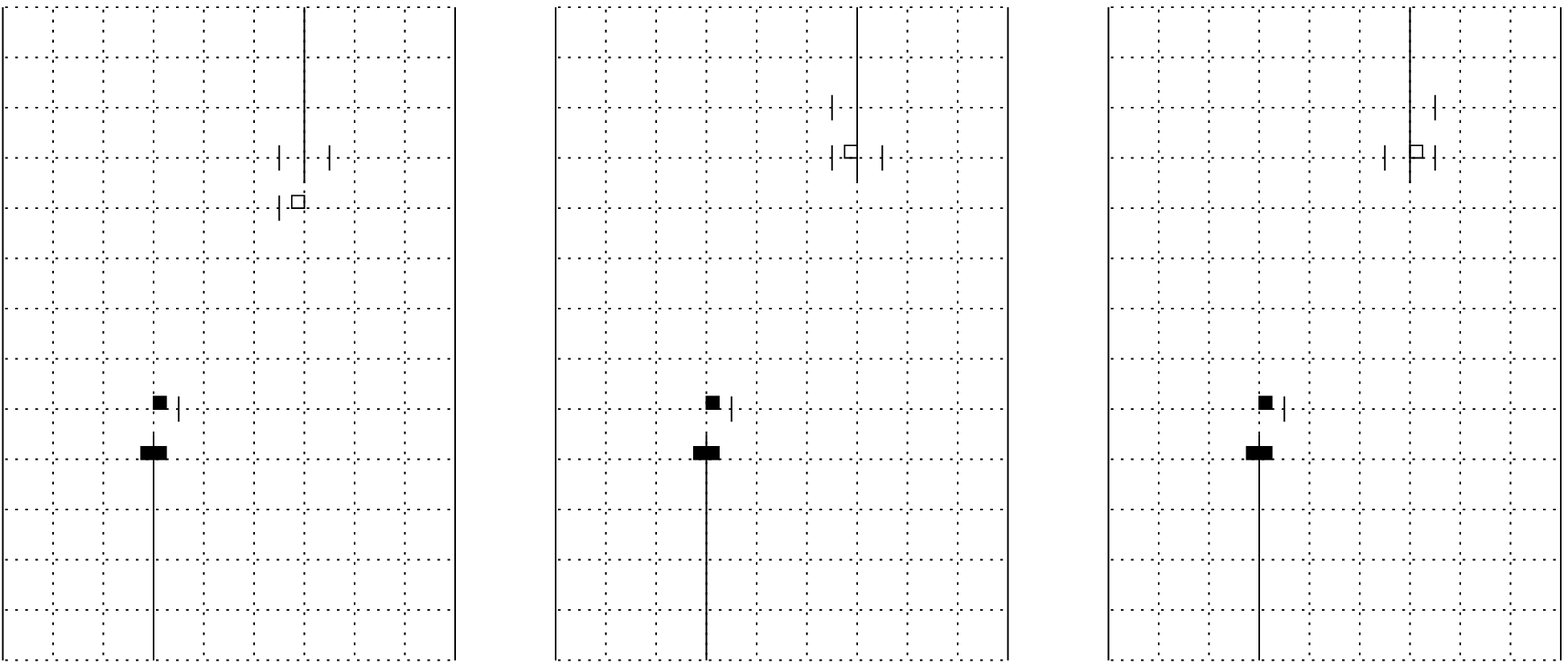}
\caption{Discretized worldsheet for a four gluon amplitude with
one tree cubic vertex (solid squares) and one cubic
counterterm vertex (open squares). The squares
indicate where $\partial q/\partial\sigma$ insertions can be located.
The short vertical lines indicate the links to be deleted in the
worldsheet ghost action. Notice that each counterterm insertion is
accompanied by an extra ghost link deletion. Inspection shows
that whenever two insertions are on the same time slice of the
same gluon propagator, there are {\it two} deletions and hence the
contact contribution is suppressed.}
\label{cubiccts}
\end{center}
\end{figure}

The crucial feature of the ghost path integral that
permits this was explained in
\cite{bardakcit}. The discretized ($p^+=Mm$)  
ghost action on a fixed time slice is
\bea
\sum_{i=0}^{M-1}(b_{i+1}-b_i)(c_{i+1}-c_i)
\eea
where $b_0,b_M,c_0,c_M=0$. The effect of the worldsheet integral of the
exponential of this expression is to supply a factor of $M$
which cancels a $1/M$ from the coordinate part of the path integral.
If a single link in the sum is deleted, the result of integration
is down by $1/M$ in other words it is 1. This was the mechanism
we used to generate needed $1/p^+$ factors in the vertex functions.
The location of these ghost deletions is indicated by short
vertical lines in Fig.~\ref{treeghosts}.
If {\it two} links are deleted on the same time slice
of the same gluon propagator, the worldsheet integral gives zero.
Thus, when the insertions are on the same time slice there will be
{\it two} deleted links and the contribution will be suppressed.
Since the deleted links produce unwanted $1/p^+$ factors, 
one must also include dummy ghost insertions (defined in \cite{thornsheet})
to put back corresponding factors of $p^+$.
Thus we can suppress contact contributions from being
produced by the cubic counterterms by accompanying each
$\partial q/\partial\sigma$ insertion with an extra deleted link.
We show in Fig.~\ref{cubiccts} which
ghost link deletion is made for each of the six possible 
$\partial q/\partial\sigma$ insertions
(three for the fusion vertex and three for the fission vertex).
This ghost deletion scheme allows us in effect to change the
cubic vertex by a multiple of itself without affecting the quartic vertex,
and this in turn allows the conversion of the nonlocal parts of the quartic
counterterms to complete trees.

\begin{figure}[htp]
\psfrag{'A'}{$A$}
\psfrag{'B'}{$B$}
\psfrag{'p1+'}{$p_1^+$}
\psfrag{'p2+'}{$p_2^+$}
\psfrag{'q'}{}
\psfrag{'k0'}{$k_0$}
\psfrag{'k1'}{$k_1$}
\psfrag{'k2'}{$k_2$}
\psfrag{'k3'}{}
\begin{center}
\includegraphics[width=2in]{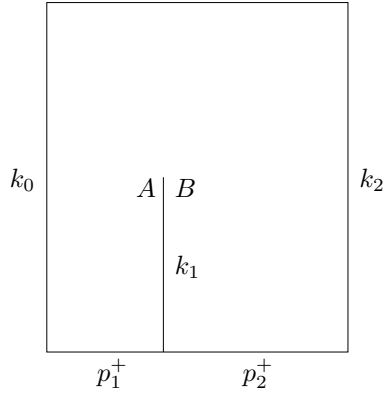}
\caption{Worldsheet for cubic fusion vertex.}
\label{cubicws}
\end{center}
\end{figure}

The same scheme is very useful in translating the cubic counterterm 
(\ref{cubicct1}) to the worldsheet. 
\bea
\Gamma^{\land\land\lor}_{\rm C.T.}&=&{g^3\over12\pi^2}[k^\land_0+k^\land_1
+k^\land_2]={g^3\over12\pi^2}[k^\land_0-k^\land_1
+k^\land_2-k^\land_1+3k^\land_1]\nonumber\\
&=&{g^3\over12\pi^2}[p^\land_2-p^\land_1+3k^\land_1]
={g^3\over12\pi^2}\left\langle p^+_2{\partial q^\land\over\partial\sigma}(B)
-p^+_1{\partial q^\land\over\partial\sigma}(A)+3q^\land(A)\right\rangle
\eea
Here $A$ and $B$ label worldsheet points just to the left and right
of the internal boundary separating the two gluon propagators that fuse to
or fission from the third gluon propagator (see Fig.~\ref{cubicws}). 
We can then use the
ghost deletion scheme just described to guarantee that these insertions
produce no modification of the quartic counterterms.

We have now shown how the self-energy and the cubic counterterms,
together with the nonlocal parts (in target space)  
of the quartic counterterms can all be given 
a local worldsheet description. It remains to find
a local worldsheet description of the purely
constant parts of the quartic counterterms 
\bea
\Gamma^{\land\land\lor\lor}_{\rm C.T.ws}&=& 0,\qquad\qquad
\Gamma^{\land\lor\land\lor}_{\rm C.T.ws}\ =\ -{g^4\over4\pi^2}
\label{constantsfinal}
\eea
We cannot simply postulate a direct quartic interaction vertex
in the worldsheet formalism without destroying worldsheet
locality. We therefore search for a cubic vertex whose contact
contributions generate constant quartic vertices. Consider the simple
ansatz
\bea
C^{\land\land\lor}=g^3\xi(p_2^\land-p_1^\land)\to
g^3\xi\left\langle p_2^+{\partial q^\land\over\partial\sigma}(2)
-p_1^+{\partial q^\land\over\partial\sigma}(1)\right\rangle
\eea
where the legs of the vertex are labeled $1,2,3$ counterclockwise
and $1,2$ have like helicity. Then it is not hard to see that
the four gluon trees built from one tree cubic and one of these
vertices, generate the contact contributions
\bea
C^{\land\land\lor\lor}&=& -g^4\xi,\qquad C^{\land\lor\land\lor}
\ =\ +2g^4\xi
\eea
That is the ratio of the two polarization structures is the same as that
coming from the $[A_\mu,A_\nu]^2$ term, in the field theoretic Lagrangian.
To make this work we need to suppress the new cubic couplings while retaining
the contact contributions. To do this we can write
\bea 
0=C^{\land\land\lor}-C^{\land\land\lor}
\eea
 and apply ghost link deletions on the second term so that it will not produce
contact contributions. Then the exchange graphs will cancel leaving
only the contact quartic vertex!

We have found no simple variation
of this scheme that provides us with exactly the counterterm
(\ref{constantsfinal}). Instead, our proposal is
to increase the flexibility of the worldsheet
formalism by increasing the dimensionality of the worldsheet
fields ${\bfs q}(\sigma,\tau)$. This is not unprecedented.
Recall that in the $AdS/CFT$ correspondence \cite{maldacena}, the string theory
is formulated in ten space-time dimensions whereas the supersymmetric
gauge theory is formulated in only four space-time dimensions.
Similarly,
in developing the worldsheet description of ${\cal N}=4$ supersymmetric
gauge theories \cite{gudmundssontt}, we found it particularly
convenient to add six extra dimensions, that is the index of $q^i$
took the values $i=1,2,\ldots,8$. The boundary conditions
on the six new $q$'s were strict Dirichlet conditions $q^i=0$
on all boundaries, internal or external.
At the same time we added three new sets of $b,c$ ghosts, which
like the original set have strict Dirichlet boundary conditions.
Since the extra $q$'s and ghosts share identical boundary conditions,
their contributions to the path integral exactly cancel: they are
just dummy integration variables. 

\begin{figure}[htp]
\psfrag{'a'}{$ag^2$}
\psfrag{'b'}{$bg^2$}
\psfrag{'c'}{$cg^2$}
\psfrag{'d'}{$dg^2$}
\psfrag{'p1+'}{$p_1^+$}
\psfrag{'p2+'}{$p_2^+$}
\psfrag{'q'}{}
\psfrag{'k0'}{$k_0$}
\psfrag{'k1'}{$k_1$}
\psfrag{'k2'}{$k_2$}
\psfrag{'k3'}{}
\begin{center}
\includegraphics[width=5in]{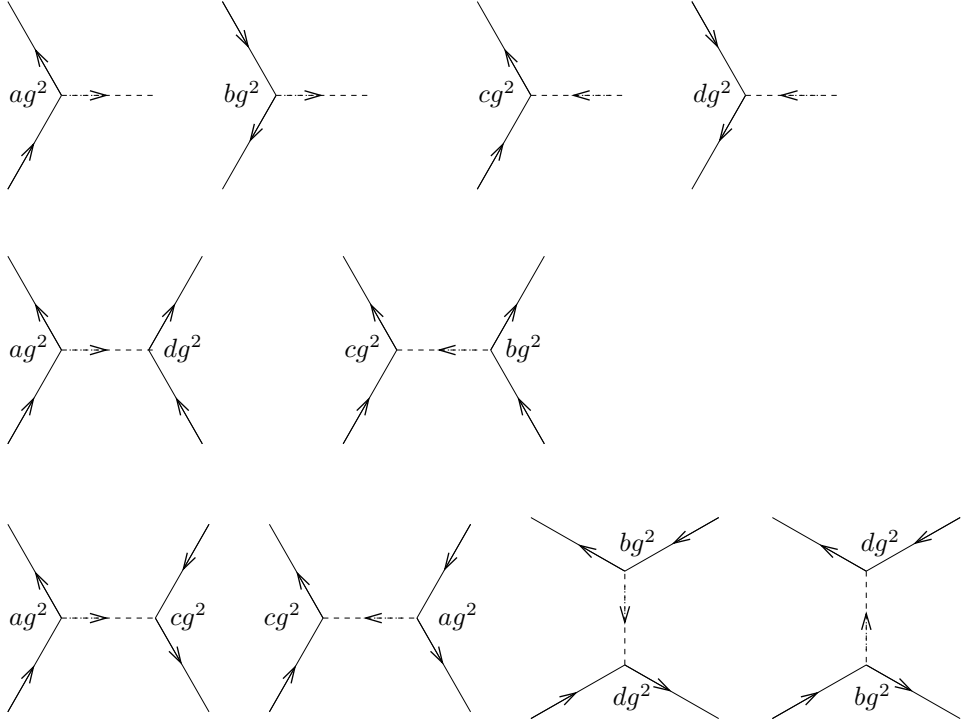}
\caption{New cubic counterterms, with a spurion (dashed line) 
coupling to two gluons.}
\label{extradcubics}
\end{center}
\end{figure}

For our purposes, to locally
produce the necessary quartic counterterms in pure gauge theories,
two extra dimensions and one extra set of $b,c$ ghosts suffice.
We thus have four transverse dual momenta corresponding to six 
dimensional space-time.
Let us call the new dimensions $r^k$, $k=1,2$ and we can, if we wish,
use a complex basis $r^\land, r^\lor$. But here $\land,\lor$ 
do {\it not} represent helicity, but rather an analogous charge in the extra 
dimensions. Next we allow spurions with values $\pm1$ of this charge to
couple to two gluons as indicated in the top line Fig.~\ref{extradcubics}.
In order to guarantee that the spurion decouples, we insert a factor
$p^+_r\partial r/\partial\sigma$ on the spurion propagator near
the interaction point. Because $r^k=0$ on all worldsheet
boundaries, the average of this factor over worldsheet
fields vanishes, except when there is another such factor on the
same time slice. In other words all the four gluon diagrams
exchanging the spurion with
internal propagators vanish, leaving only the contact contributions.
By inspecting the coupling assignments shown in Fig.~\ref{extradcubics},
we see that 
\bea
\Gamma^{\land\land\lor\lor}_{spur}&=&-(ad+bc)g^4\\
\Gamma^{\land\lor\land\lor}_{spur}&=&-2(ac+bd)g^4
\eea 
Altogether we have five adjustable parameters to produce two
independent counterterms:
\bea
\Gamma^{\land\land\lor\lor}_{\rm C.T.}&=&C^{\land\land\lor\lor}
+\Gamma^{\land\land\lor\lor}_{spur}\ =\ -(ad+bc+\xi)g^4\\
\Gamma^{\land\lor\land\lor}_{\rm C.T.}&=&C^{\land\lor\land\lor}
+\Gamma^{\land\lor\land\lor}_{spur}\ =\ -2(ac+bd-\xi)g^4
\eea
Since $\xi,a,b,c,d$ are arbitrary, 
there is more than enough flexibility to produce the 
necessary counterterms, and more generally to adjust them appropriately
at each order in perturbation theory. It is perhaps most economical to
employ the extra dimensions only to cancel the part of the anomaly
due to UV artifacts, which would demand spin independence
for this part: $ad+bc=2(ac+bd)$. With this choice we then determine
$\xi=-(ad+bc)=-1/12\pi^2$.

\section{Conclusion}
In this paper we have completed the lightcone gauge 
calculations of the scattering of glue by glue through one loop.
Our results completely agree with those obtained using covariant 
methods \cite{kunsztst}.
In addition to obtaining the elastic amplitudes through one loop
which are divergent in the infrared,
we have also calculated their contributions to probabilities
and have shown that infrared divergences cancel against
contributions from extra gluons in the initial and final
states. This is all in accord with the Lee-Nauenberg theorem.

The expressions for the final infrared finite probabilities, including the
bremsstrahlung gluons, are extremely compact, and are manifestly
Lorentz invariant. Infrared divergences have been traded for a
resolution parameter $\Delta$ characterizing unseen gluons and jets.

All calculations were done in 4 space-time dimensions without the
benefit of dimensional regularization, which means that (local)
counterterms beyond wave function and coupling renormalization
must be included. In spite of artificial
$p^+=0$ divergences (which raise the ugly possibility of requiring
nonpolynomial counterterms), all necessary counterterms are
polynomials in the external momenta of the degree dictated
by power counting. We would like to underline here the fact that
we did nothing sophisticated with $p^+$ zero modes. We simply discretized
the $p^+$ integrals and omitted the zero modes \cite{gilest,thornfishnets}. 
Apart from collinear divergences, which are only a problem for self
energy insertions on external lines, this discretization provides
an apt infrared regulator for lightcone calculations. 
Our calculation shows that in all
infrared safe calculations, including on-shell gluon scattering with
due care taken with jets and gluon bremsstrahlung, the continuum
limit of this discretization is finite and yields all previously
known results obtained in covariant gauges  
\cite{ellissexton,bernk4gluons,kunsztst}. 
{\it A sophisticated treatment of $p^+$ zero modes is not required}.

Finally we have discussed how to incorporate all of the counterterms
we require in the lightcone worldsheet formalism. A particularly convenient
way to do this is to interpret the QCD ``string'' dynamics as
occurring in 6 dimensional space-time. 
We stress that the extra two dimensions are
holographically generated on the ``string'' side of
Field/String duality and are not present at all in the
field theoretic description of the ``field'' side of the duality.
Significantly, this can all
be done while preserving a local worldsheet dynamics. 
 
For perturbative QCD, the next step is to prove that the
lightcone gauge calculational procedure we have adopted carries through 
to all orders in perturbation theory. We believe it will because
we have only needed to introduce strictly local counterterms,
consistent with the concept that gauge violating artifacts are
entirely associated with the ultraviolet part of the dynamics.
The fact that we needed some counterterms that were not in the
input classical Lagrangian is completely standard with the use
of a gauge noninvariant regulation and should not obstruct
the usual renormalization program. This is because the new counterterms
introduced obey the power counting rules of renormalizable theories.

If the renormalization program goes through as we expect, then
the modifications we have made in the ``bare'' worldsheet description to
accommodate the counterterms should suffice to all orders in
perturbation theory. This would fulfill our ambitious goal
of establishing a string theory dual of the large $N_c$ limit
of QCD, working entirely from the field theory side of the
duality. Then the exciting prospect before us would be to
use this duality to deepen our insight into nonperturbative
aspects of the strong interactions. 
\vskip14pt
\noindent\underline{ Acknowledgments}: 
We would like to thank Zvi Bern for valuable discussions. 
We also thank Lisa Everett for critical comments on the manuscript.
This research was supported in part by the Department
of Energy under Grant No. DE-FG02-97ER-41029.

\appendix
\section{Bremsstrahlung Integrals}
In the evaluation of soft and collinear bremsstrahlung cross
sections we need to do several integrations over phase
space. For the hard collinear case the integral was 
simple enough to treat in the text. Here we sketch the evaluation
of the integrals for soft radiation which are more complex.

At $N_c=\infty$ each amplitude is the sum of gluon emissions
from two neighboring lines, so the squared amplitude has two direct terms
and a cross term. We first consider the transverse integration over
the resolution of a direct term, which has the structure
\bea
I_{\rm Direct}({\bfs v})&=&
{1\over2}\int d{\bfs k} {({\bfs k}+{\bfs v})^2
\over [({\bfs k}+{\bfs v})^2+M^2]^2}
\nonumber\\
&=&{1\over2}\int_0^{\Delta^\prime} kdk\int_0^{2\pi}d\phi
{k^2+v^2+2kv\cos\phi\over
(k^2+v^2+M^2+2kv\cos\phi)^2}
\eea
The $\phi$ integral is easily done by transforming to a contour
integral over $z=e^{i\phi}$ and evaluating residues. Changing variables from
$k$ to $t=k^2$ then leaves us with
\bea
I_{\rm Direct}({\bfs v})&=&{\pi\over2}
\int_0^{\Delta^{\prime2}} dt{(t-v^2)^2+M^2(t+v^2)\over
[(t-v^2)^2+2M^2(t+v^2)+M^4]^{3/2}}\\
&=&{-{\pi\over4}{(\Delta^{\prime2}-v^2)-M^2\over\sqrt{
(\Delta^{\prime2}-v^2)^2+M^4+2\Delta^{\prime2}M^2+2M^2v^2}}}
-{\pi\over4}\nonumber\\
&&+{\pi\over2}{{\ln{(M^2-v^2+\Delta^{\prime2}
+\sqrt{(\Delta^{\prime2}-v^2)^2+M^4
+2\Delta^{\prime2}M^2+2M^2v^2})\over2M^2}}}\\
&\sim&{\pi\over2}\ln{\Delta^{\prime2}-v^2\over M^2e}\quad{\rm for}~
\Delta^{\prime2}>v^2;\qquad{\pi\over2}\ln{v^2\over
v^2-\Delta^{\prime2}} \quad{\rm for}~
\Delta^{\prime2}<v^2
\eea
where the last line applies as $M\to0$.

The transverse 
momentum integral of a cross term has the structure
\bea
I_{\rm Cross}&=&
\int d{\bfs k} {({\bfs k}+{\bfs v})\cdot({\bfs k}+{\bfs w})\over 
[({\bfs k}+{\bfs v})^2+M^2][({\bfs k}+{\bfs w})^2+M^2]}
\nonumber\\
&=&\int_0^{\Delta^\prime} kdk\int_0^{2\pi}d\phi{k^2+vw\cos\alpha+kv\cos\phi
+kw\cos(\phi-\alpha)\over
(k^2+v^2+M^2+2kv\cos\phi)(k^2+w^2+M^2+2kw\cos(\phi-\alpha))}
\eea
Again the $\phi$ integral is easily done by converting to a contour
integral in $z=e^{i\phi}$. In this case we can simplify life by taking 
$M=0$ from the beginning, in which case the poles of the
integrand are at $z=-v/k,-k/v,-we^{i\alpha}/k,-ke^{i\alpha}/w$.
The contour at the unit circle will enclose precisely two of these poles:
$z=-{\rm min}(v/k,k/v),-e^{i\alpha}{\rm min}(w/k,k/w)$, with four possibilities
depending on the relative size of $k,v,w$. For definiteness let's assume
that $v>w$. Then when $w<k<v$ it turns out that the two pole contributions
exactly cancel. Then the angular integral is
\bea
\pi\left[{1\over k^2-vwe^{-i\alpha}}+{1\over k^2-vwe^{i\alpha}}\right],
\qquad {\rm for}~k>v>w\nonumber\\
-\pi\left[{1\over k^2-vwe^{-i\alpha}}+{1\over k^2-vwe^{i\alpha}}\right],
\qquad {\rm for}~v>w>k\nonumber\\
0,\qquad {\rm for}~v>k>w
\eea
and we get with $u=k^2$, assuming $\Delta^{\prime2}>v^2,w^2$,
\bea
I_{\rm Cross}&=&
-{\pi\over2}\int_0^{w^2} du\left[{1\over u-vwe^{-i\alpha}}
+{1\over u-vwe^{i\alpha}}\right]+{\pi\over2}\int_{v^2}^{\Delta^{\prime2}} du
\left[{1\over u-vwe^{-i\alpha}}+{1\over u-vwe^{i\alpha}}\right]\nonumber\\
&=&{\pi\over2}\ln{\Delta^{\prime4}-2\Delta^{\prime2}vw\cos\alpha+v^2w^2
\over (v^2+w^2-2vw\cos\alpha)^2}\nonumber\\
&=& {\pi\over2}\ln{\Delta^{\prime4}
-2\Delta^{\prime2}{\bfs v}\cdot{\bfs w}+{\bfs v}^2{\bfs w}^2
\over ({\bfs v}-{\bfs w})^4},\qquad{\rm for}~w^2<v^2<\Delta^{\prime2};
\nonumber\\
&&-{\pi\over2}\ln{({\bfs v}-{\bfs w})^2\over {\bfs v}^2},
\hskip1.1in{\rm for}~w^2<\Delta^{\prime2}<v^2;\nonumber\\
&&-{\pi\over2}\ln{\Delta^{\prime4}
-2\Delta^{\prime2}{\bfs v}\cdot{\bfs w}+{\bfs v}^2{\bfs w}^2
\over {\bfs v}^2{\bfs w}^2},\quad{\rm for}~\Delta^{\prime2}<w^2<v^2\nonumber
\eea
If $v<w$, the same result follows.
The complete transverse integral for soft gluon radiation
is the combination
\bea
I_{\rm Total}&=&I_{\rm Direct}({\bfs v})+I_{\rm Direct}({\bfs w})-I_{\rm Cross}
\nonumber\\
&=&{\pi\over2}\ln{(\Delta^{\prime2}-{\bfs v}^2)(\Delta^{\prime2}-{\bfs w}^2)
\over
\Delta^{\prime4}
-2\Delta^{\prime2}{\bfs v}\cdot{\bfs w}+{\bfs v}^2{\bfs w}^2 }
+\pi\ln{({\bfs v}-{\bfs w})^2\over M^2e},
\qquad{\rm for}~w^2<v^2<\Delta^{\prime2};\nonumber\\
&=&{\pi\over2}\ln{(\Delta^{\prime2}-{\bfs w}^2)
\over {\bfs v}^2-\Delta^{\prime2}}
+{\pi\over2}\ln{({\bfs v}-{\bfs w})^2\over M^2e},
\qquad{\rm for}~w^2<\Delta^{\prime2}<v^2;\nonumber\\
&=&-{\pi\over2}\ln{(\Delta^{\prime2}-{\bfs v}^2)(\Delta^{\prime2}-{\bfs w}^2)
\over
\Delta^{\prime4}
-2\Delta^{\prime2}{\bfs v}\cdot{\bfs w}+{\bfs v}^2{\bfs w}^2 },
\qquad{\rm for}~\Delta^{\prime2}<w^2<v^2 \nonumber
\eea
For $k^+\to0$ both $v/\Delta^{\prime},w/\Delta^{\prime}\to0$ 
so we see that the small $k^+$
region is insensitive to the resolution $\Delta^\prime$.

\section{Evaluation of Bubble and Triangle Integrals}\label{triangleintegral}
We list here the $q^-$ integrations of bubble and triangle 
integrands that are useful in analyzing collinear divergences. 
First the six bubble integrands:
\bea
-i\int {dq^-\over2\pi} {1\over q_0^2q_1^2}&=&
{1\over2p_1^+({\bfs q}_0-q_0^+{\bfs p}_1/p_1^+)^2},\qquad{\rm for}~0<q_0^+<p_1^+
\\
-i\int {dq^-\over2\pi} {1\over q_1^2q_2^2}&=&
{1\over2p_2^+({\bfs q}_1-q_1^+{\bfs p}_2/p_2^+)^2},
\qquad{\rm for}~0<q_1^+<p_{2}^+
\\
-i\int {dq^-\over2\pi} {1\over q_2^2q_3^2}&=&
{-1\over2p_3^+({\bfs q}_3-q_3^+{\bfs p}_3/p_3^+)^2},
\qquad{\rm for}~0<q_3^+<-p_{3}^+
\\
-i\int {dq^-\over2\pi} {1\over q_0^2q_3^2}&=&
{-1\over2p_4^+({\bfs q}_0-q_0^+{\bfs p}_4/p_4^+)^2},
\qquad{\rm for}~0<q^+<-p_4^+
\\
-i\int {dq^-\over2\pi} {1\over q_0^2q_2^2}&=&
{1\over2p_{12}^+\left[({\bfs q}_0-q_0^+{\bfs p}_{12}/p_{12}^+)^2
+q_0^+(p_{12}^+-q_0^+)p_{12}^2/p_{12}^{+2}\right]},
\qquad{\rm for}~0<q_0^+<p_{12}^+
\\
-i\int {dq^-\over2\pi} {1\over q_1^2q_3^2}&=&
{1\over2|p_{14}^+|[({\bfs q}_1-q_1^+{\bfs p}_{14}/p_{14}^+)^2
-q_1^+(q_1^++p_{14}^+)p_{14}^2/p_{14}^{+2}]},
\qquad{\rm for}~0<|q_1^+|<|p_{14}^+|
\eea
Here we recall the notation $q_i\equiv q-k_i$, where $k_i$ are
the dual momenta, related to the gluon momenta by $p_i=k_i-k_{i-1}$,
with $k_4\equiv k_0$. We normally take $k_0^\pm=0$.
Next we list the triangle integrals:
\bea
&&\hskip-.3in-i\int {dq^-\over2\pi} {1\over q_0^2q_1^2q_2^2}\nonumber\\
&=&{q_0^+\over2p_1^+p_{12}^+({\bfs q}_0-q_0^+{\bfs p}_1/p_1^+)^2
[({\bfs q}_0-q_0^+{\bfs p}_{12}/p_{12}^+)^2
+q_0^+(p_{12}^+-q_0^+)p_{12}^2/p_{12}^{+2}]
},\quad {\rm for}~0<q_0^+<p_1^+\nonumber\\
&=&{p_{12}^+-q_0^+\over2p_2^+p_{12}^+({\bfs q}_0-{\bfs p}_{12}
-(q_0^+-p_{12}^{+}){\bfs p}_2/p_2^+)^2
[({\bfs q}_0-q_0^+{\bfs p}_{12}/p_{12}^+)^2+q_0^+(p_{12}^+-q_0^+)p_{12}^2/p_{12}^{+2}]
},~ {\rm for}~p_1^+<q_0^+<p_{12}^+\nonumber\\
&&\hskip-.3in-i\int {dq^-\over2\pi} {1\over q_0^2q_1^2q_3^2}\nonumber\\
&=&{-q_0^+\over2p_1^+p_{4}^+({\bfs q}_0-q_0^+{\bfs p}_1/p_1^+)^2
({\bfs q}_0-q_0^+{\bfs p}_{4}/p_{4}^+)^2},
\quad {\rm for}~0<q_0^+<p_1^+\nonumber\\
&=&{-q_0^+-p_4^+\over2p_4^+p_{14}^+({\bfs q}_0
-q_0^+{\bfs p}_4/p_4^+)^2
[({\bfs q}_0+{\bfs p}_4-(q_0^++p_4^+){\bfs p}_{14}/p_{14}^+)^2-(q_0^++p_4^+)
(q_0^+-p_{1}^+)p_{14}^2/p_{14}^{+2}]
},\nonumber\\
&&\hskip4.5in\quad {\rm for}~p_1^+<q_0^+<-p_4^+\nonumber\\
&&\hskip-.3in-i\int {dq^-\over2\pi} {1\over q_1^2q_2^2q_3^2}\nonumber\\
&=&{-q_1^+\over2p_2^+p_{14}^+({\bfs q}_1
-q_1^+{\bfs p}_2/p_2^+)^2
[({\bfs q}_1-q_1^+{\bfs p}_{14}/p_{14}^+)^2-(q_1^++p_{14}^+)
q_1^+p_{14}^2/p_{14}^{+2}]},\nonumber\\
&&\hskip4.5in
\quad {\rm for}~0<q_1^+<-p_{14}^+\nonumber\\
&=&{q_1^+-p_{2}^+\over2p_2^+p_{3}^+({\bfs q}_1
-q_1^+{\bfs p}_2/p_2^+)^2({\bfs q}_1-{\bfs p}_{2}
-(q_1^+-p_{2}^{+}){\bfs p}_3/p_3^+)^2
},\quad {\rm for}~-p_{14}^+<q_1^+<p_{2}^+\nonumber\\
&&\hskip-.3in-i\int {dq^-\over2\pi} {1\over q_0^2q_2^2q_3^2}\nonumber\\
&=&{-q_0^+\over2p_{12}^+p_{4}^+[({\bfs q}_0-q^+{\bfs p}_{12}/p_{12}^+)^2
+q_0^+(p_{12}^+-q_0^+)p_{12}^2/p_{12}^{+2}]
({\bfs q}_0-q^+{\bfs p}_{4}/p_{4}^+)^2},
\quad {\rm for}~0<q_0^+<-p_4^+\nonumber\\
&=&{q_0^+-p_{12}^+\over2p_3^+p_{12}^+({\bfs q}_0-{\bfs p}_{12}-(q_0^+-p_{12}^+)
{\bfs p}_3/p_3^+)^2
[({\bfs q}_0-q^+{\bfs p}_{12}/p_{12}^+)^2+q_0^+
(p_{12}^+-q_0^+)p_{12}^2/p_{12}^{+2}]
},\nonumber\\
&&\hskip4.5in\quad {\rm for}~-p_4^+<q_0^+<p_{12}^+\nonumber
\eea
Once the $q^-$ integrals have been performed, the transverse ${\bfs q}$
integrals can be done using one or two Schwinger parameters
to exponentiate the one or two denominators. 

\section{Evaluation of Box Integrals}\label{boxintegral}
The box integrals we encounter can be most easily handled through the
introduction of Schwinger parameters $T_1,T_2,T_3,T_4$ for the
internal line propagators 
$(q-k_0)^{-2}, (q-k_1)^{-2},(q-k_2)^{-2},(q-k_3)^{-2}$ respectively.
Since some of them are divergent in the ultra-violet, we also retain
the worldsheet UV cutoff factors $e^{-\delta{\bfs q}^2}$. The integration over
$q$ is then a Gaussian that is easily done by completing the
square and shifting ${q}\to{q}+{K}$, with
\bea
{\bfs K}={{\bfs k}_0 T_1 +{\bfs k}_1 T_2 +{\bfs k}_2 T_3 +{\bfs k}_3 T_4 \over
T_{14}+\delta},\qquad K^{\pm}
={({k}_0 T_1 +{k}_1 T_2 +{k}_2 T_3 +{k}_3 T_4)^\pm \over
T_{14}}
\eea 
where we use the shorthand $T_{14}=T_1+T_2+T_3+T_4$. One then finds, 
using the Feynman parameters $x_i\equiv T_i/T_{14}$ that
\bea
{\bfs K}_{16}&\to& -p_1^+{\bfs q}+q^+{\bfs p}_1
-x_3{\bfs K}_{12}-x_4{\bfs K}_{41}
+p_1^+{\delta{\bfs K}\over T_{14}}\\
{\bfs K}_{52}&\to& -p_2^+{\bfs q}+q^+{\bfs p}_2
-x_4{\bfs K}_{23}-x_1{\bfs K}_{12}
+p_2^+{\delta{\bfs K}\over T_{14}}\\
{\bfs K}_{35}&\to& p_3^+{\bfs q}-q^+{\bfs p}_3
+x_2{\bfs K}_{23}+x_1{\bfs K}_{34}
-p_3^+{\delta{\bfs K}\over T_{14}}\\
{\bfs K}_{64}&\to& p_4^+{\bfs q}-q^+{\bfs p}_4
+x_3{\bfs K}_{34}+x_2{\bfs K}_{41}
-p_4^+{\delta{\bfs K}\over T_{14}}
\eea
and the Gaussian factor left over is just $\exp\{-T_{14}(x_1x_3p_{12}^2
+x_2x_4p_{14}^2)+O(\delta)\}$. The $O(\delta)$ term in the exponent
is negligible in the box integrals because the log divergence is insufficient 
to overwhelm it. Also when the ${\bfs K}_{ij}$ occur in the numerator
of the box integrand the terms $ p_i^+{\delta{\bfs K}/T_{14}}$
are negligible since they are $O(1)$ only when all $T_i=O(\delta)$
and they occur only in integrals convergent in this region. The Gaussian
integration over $q$ involves up to four powers of $q$ as prefactors.
\bea
\int d^4q\ e^{-T_{14}q^2-\delta{\bfs q}^2}&=&
{\pi^2\over T_{14}(T_{14}+\delta)}, \qquad \int d^4q\ q^\land q^\lor
e^{-T_{14}q^2-\delta{\bfs q}^2}=
{\pi^2\over 2T_{14}(T_{14}+\delta)^2}\nonumber\\
\int d^4q\ q^2e^{-T_{14}q^2-\delta{\bfs q}^2}&=&
{\pi^2(2T_{14}+\delta)\over T_{14}^2(T_{14}+\delta)^2},
\qquad \int d^4q\ (q^\land q^\lor)^2
e^{-T_{14}q^2-\delta{\bfs q}^2}={\pi^2\over 2T_{14}(T_{14}+\delta)^3}
\nonumber\\
\int d^4q\ q^\land q^\lor q^2e^{-T_{14}q^2-\delta{\bfs q}^2}&=&
{\pi^2(3T_{14}+\delta)\over 2T_{14}^2(T_{14}+\delta)^3}\eea
Changing variables from the $T_i$ to
three of the $x_i$ and $T_{14}=T$ reduces the integration measure
to $ d^4x\delta(1-\sum x_i) T^3dT$. In the first three cases it
is safe to set $\delta=0$, and the integral over $T$ gives
\bea
{\pi^2\over (x_1x_3p_{12}^2+x_2x_4p_{14}^2)^2},\qquad 
{\pi^2\over 2(x_1x_3p_{12}^2+x_2x_4p_{14}^2)},\qquad
{2\pi^2\over x_1x_3p_{12}^2+x_2x_4p_{14}^2}
\eea
respectively. In the last two cases we must evaluate integrals that
are log divergent for $\delta\to0$. We find, as $\delta\to0$,
\bea
\int_0^\infty {T^n dT\over (T+\delta)^{n+1}}e^{-TH}
&\sim& -\sum_{k=1}^n{1\over k}-\gamma-\ln(\delta H)\nonumber\\
\int_0^\infty {T^2dT\over (T+\delta)^3}e^{-TH}&\sim& -{3\over2}-\gamma
-\ln(\delta H),\qquad \int_0^\infty {T(3T+\delta)dT\over 
(T+\delta)^3}e^{-TH}\sim -4-3\gamma-3\ln(\delta H)
\eea
where $\gamma =-\Gamma^\prime(1)$ is Euler's constant.

We shall have use for the following combinations of momenta which
arise in the box integrand after shifting $q$ and
sending $\delta\to0$ 
\bea
K_0&=&x_2p_1+x_3(p_1+p_2)-x_4p_4\\
K_0-p_1&=&x_3p_2+x_4(p_2+p_3)-x_1p_1\\
K_0-p_1-p_2&=&x_4p_3+x_1(p_3+p_4)-x_2p_2\\
K_0+p_4&=&x_1p_4+x_2(p_1+p_4)-x_3p_3
\eea
Finally, we list the integrals over Feynman parameters that
arise in the box diagrams. We use the shorthand notation
$d^3x=\prod_{i=1}^4 dx_i \delta(1-\sum_i x_i)$.
\bea
L &=&\int d^3x \ln(x_1x_3A+x_2x_4B)=-{11\over18}+{B\ln B+A\ln A\over
6(A+B)}+{AB\over 12(A+B)^2}\left(\pi^2+\ln^2{A\over B}\right)\\
L_1 &=& \int d^3x {1\over x_1x_3A+x_2x_4B}=
{1\over 2(A+B)}\left(\pi^2+\ln^2{A\over B}\right)\\
L_{1A}&=&\int d^3x {(x_1,x_3)\over x_1x_3A+x_2x_4B}={\ln(A/B)\over2(A+B)} 
+{B\over 4(A+B)^2}\left(\pi^2+\ln^2{A\over B}\right)\\
L_{1B}&=&\int d^3x {(x_2,x_4)\over x_1x_3A+x_2x_4B}={\ln(B/A)\over2(A+B)} 
+{A\over 4(A+B)^2}\left(\pi^2+\ln^2{A\over B}\right)\\
L_A&=&\int d^3x {x_1x_3\over x_1x_3A+x_2x_4B}= {1\over6(A+B)}+{B\ln(A/B)\over
3(A+B)^2}+{B(B-A)\over 12(A+B)^3}\left(\pi^2+\ln^2{A\over B}\right)\\
L_B&=&\int d^3x {x_2x_4\over x_1x_3A+x_2x_4B}= {1\over6(A+B)}+{A\ln(B/A)\over
3(A+B)^2}+{A(A-B)\over 12(A+B)^3}\left(\pi^2+\ln^2{A\over B}\right)\\
L_C &=& \int d^3x {(x_1x_2,x_2x_3,x_3x_4,x_4x_1)
\over x_1x_3A+x_2x_4B}\nonumber\\
&=& -{1\over6(A+B)}+{(A-B)\ln(A/B)\over
6(A+B)^2}+{AB\over 6(A+B)^3}\left(\pi^2+\ln^2{A\over B}\right)\\
L_{2B}&=&\int d^3x {(x_2^2,x_4^2)\over x_1x_3A+x_2x_4B}
= {1-\ln(A/B)\over6(A+B)}-{A\ln(A/B)\over
3(A+B)^2}+{A^2\over 6(A+B)^3}\left(\pi^2+\ln^2{A\over B}\right)\\
L_{2A} &=& \int d^3x {(x_1^2,x_3^2)\over x_1x_3A+x_2x_4B}
= {1-\ln(B/A)\over6(A+B)}
-{B\ln(B/A)\over
3(A+B)^2}+{B^2\over 6(A+B)^3}\left(\pi^2+\ln^2{A\over B}\right)\\
L_{AB}&=&\int d^3x {x_1x_2x_3x_4\over (x_1x_3A+x_2x_4B)^2}\nonumber\\
&=&{1\over2(A+B)^2}+{(B-A)\ln(A/B)\over
2(A+B)^3}+{A^2+B^2-4AB\over 12(A+B)^4}\left(\pi^2+\ln^2{A\over B}\right)\\
L_{AA}&=&\int d^3x {x_1^2x_3^2\over (x_1x_3A+x_2x_4B)^2}\nonumber\\
&=&
{A-2B\over6A(A+B)^2}+{B(5A-B)\ln(A/B)\over
6A(A+B)^3}+{B(2B-A)\over 6(A+B)^4}\left(\pi^2+\ln^2{A\over B}\right)\\
L_{CA}&=&\int d^3x {(x_1x_2,x_2x_3,x_3x_4,x_4x_1)
x_1x_3\over (x_1x_3A+x_2x_4B)^2}\nonumber\\
&=& 
{B-2A\over6A(A+B)^2}-{(5B-A)\ln(A/B)\over
6(A+B)^3}+{B(2A-B)\over 6(A+B)^4}\left(\pi^2+\ln^2{A\over B}\right)\\
L_{2AB}&=&\int d^3x {(x_1^2,x_3^2)x_2x_4\over (x_1x_3A+x_2x_4B)^2}\nonumber\\
&=& 
{A+4B\over6B(A+B)^2}+{(A-5B)\ln(B/A)\over
6(A+B)^3}+{B(B-2A)\over 6(A+B)^4}\left(\pi^2+\ln^2{A\over B}\right)\\
L_{2AA}&=&\int d^3x {(x_1^2,x_3^2)x_1x_3\over (x_1x_3A+x_2x_4B)^2}\nonumber\\
&=& 
-{B\over2A(A+B)^2}+{(A^2+5AB+3B^2)\ln(A/B)\over
6A(A+B)^3}+{B^2\over 2(A+B)^4}\left(\pi^2+\ln^2{A\over B}\right)\\
L_{3A}&=&\int d^3x {x_2^2x_3\over x_1x_3A+x_2x_4B}\nonumber\\
&=& -{A\over8(A+B)^2}
+{(2B^2-5AB-A^2)\ln(A/B)\over
24(A+B)^2}+{A^2B\over 8(A+B)^4}\left(\pi^2+\ln^2{A\over B}\right)
\eea
\section{Evaluation of Eq(\ref{ceeterm})}
The numerator in the integrand, after introduction of Schwinger parameters and
the appropriate shift of $q$ can be replaced as
\bea
K_{61}^\land K_{25}^\lor K_{35}^\land K_{64}^\lor  
&\to& [p_3^+q^\land +x_2 K_{23}^\land + x_1 K_{34}^\land]
[p_4^+q^\lor +x_3 K_{34}^\lor + x_2 K_{41}^\lor] \nonumber\\
&&\times
[p_1^+q^\land +x_3 K_{12}^\land + x_4 K_{41}^\land]
[p_2^+q^\lor +x_4 K_{23}^\lor + x_1 K_{12}^\lor]\nonumber\\
&\to&p_1^+p_2^+p_3^+p_4^+q^\land q^\lor q^\land q^\lor 
-p_1^+p_2^+p_3^+p_4^+H^2/4\nonumber\\
&&+p_1^+p_2^+(q^\land q^\lor-x_1x_3p_{12}^2/2) 
[x_2 K_{23}^\land + x_1 K_{34}^\land]
[x_3 K_{34}^\lor + x_2 K_{41}^\lor]\nonumber\\
&&+p_1^+p_4^+(q^\land q^\lor-x_2x_4p_{14}^2/2)
[x_2 K_{23}^\land + x_1 K_{34}^\land][x_4 K_{23}^\lor + x_1 K_{12}^\lor]
\nonumber\\
&&+p_3^+p_4^+(q^\land q^\lor-x_1x_3p_{12}^2/2)
[x_3 K_{12}^\land + x_4 K_{41}^\land]
[x_4 K_{23}^\lor + x_1 K_{12}^\lor]\nonumber\\
&&
+p_2^+p_3^+(q^\land q^\lor-x_2x_4p_{14}^2/2)
[x_3 K_{12}^\land + x_4 K_{41}^\land][x_3 K_{34}^\lor + x_2 K_{41}^\lor]
\eea
The second term in the first line of the last equality
came from the quantity
\bea
-{1\over4}p_1^+p_2^+p_3^+p_4^+[(x_1x_3p_{12}^2)^2+(x_2x_4p_{14}^2)^2]
+x_1x_2x_3x_4[K_{41}^\land K_{12}^\lor K_{23}^\land K_{34}^\lor
+ K_{34}^\land  K_{41}^\lor K_{12}^\land K_{23}^\lor]\nonumber
\eea
which can be  greatly simplified using
\bea
K_{41}^\land K_{12}^\lor K_{23}^\land K_{34}^\lor
=-{p_3^+\over p_1^+}K_{41}^\land K_{12}^\lor K_{12}^\land K_{41}^\lor
=-{1\over4}p_1^+p_2^+p_3^+p_4^+p_{12}^2p_{14}^2
\eea
so it becomes
\bea
-{1\over4}p_1^+p_2^+p_3^+p_4^+(x_1x_3p_{12}^2+x_2x_4p_{14}^2)^2
\equiv-{1\over4}p_1^+p_2^+p_3^+p_4^+H^2
\eea
Putting in the rest of the factors and doing the $q$ integration
yields the $x$ integral
\bea
&&\hskip-.5in\int d^3x{1\over16\pi^2}
\bigg\{-(\ln(H\delta e^\gamma)+2)\nonumber\\
&&+x_2x_4p_{14}^2 
\left[{(x_2 K_{23}^\land + x_1 K_{34}^\land)
(x_3 K_{34}^\lor + x_2 K_{41}^\lor)\over p_3^+p_4^+H^2}
+{(x_3 K_{12}^\land + x_4 K_{41}^\land)
(x_4 K_{23}^\lor + x_1 K_{12}^\lor)\over p_1^+p_2^+H^2}\right]\nonumber\\
&&+x_1x_3p_{12}^2
\left[{(x_2 K_{23}^\land + x_1 K_{34}^\land)(x_4 K_{23}^\lor +
 x_1 K_{12}^\lor)\over p_2^+p_3^+H^2}
+{(x_3 K_{12}^\land + x_4 K_{41}^\land)(x_3 K_{34}^\lor + x_2 K_{41}^\lor)
\over p_1^+p_4^+H^2}\right]\bigg\}\nonumber\\
&=&{1\over16\pi^2}
\bigg\{-{1\over6}(\ln(\delta e^\gamma)+2)-L(p_{12}^2,p_{14}^2)\nonumber\\
&&-B{\partial\over\partial B}\int d^3x
\left[{(x_2 K_{23}^\land + x_1 K_{34}^\land)
(x_3 K_{34}^\lor + x_2 K_{41}^\lor)\over p_3^+p_4^+H}
+{(x_3 K_{12}^\land + x_4 K_{41}^\land)
(x_4 K_{23}^\lor + x_1 K_{12}^\lor)\over p_1^+p_2^+H}\right]\\
&&-A{\partial\over\partial A}\int d^3x
\left[{(x_2 K_{23}^\land + x_1 K_{34}^\land)(x_4 K_{23}^\lor +
 x_1 K_{12}^\lor)\over p_2^+p_3^+H}
+{(x_3 K_{12}^\land + x_4 K_{41}^\land)(x_3 K_{34}^\lor + x_2 K_{41}^\lor)
\over p_1^+p_4^+H}\right]\bigg\}\bigg|_{A=p_{12}^2\atop B=p_{14}^2}\nonumber\\
&=&{1\over16\pi^2}
\bigg\{-{1\over6}(\ln(\delta e^\gamma)+2)-L(p_{12}^2,p_{14}^2)
-2p_{12}^2p_{14}^2L_{AB}(p_{12}^2,p_{14}^2)\nonumber\\
&&\hskip-.5in
-B{\partial\over\partial B}\bigg\{L_{2B}(A,B)\left[{K_{23}^\land K_{41}^\lor
\over p_3^+p_4^+}+{K_{41}^\land K_{23}^\lor
\over p_1^+p_2^+}\right]
+L_C(A,B)\left[{K_{34}^\land K_{41}^\lor+K_{23}^\land K_{34}^\lor
\over p_3^+p_4^+}+{K_{12}^\land K_{23}^\lor+K_{41}^\land K_{12}^\lor
\over p_1^+p_2^+}\right]\bigg\}\bigg|_{A=p_{12}^2\atop B=p_{14}^2}\nonumber\\
&&\hskip-.5in -A{\partial\over\partial A}\bigg\{L_{2A}(A,B)
\left[{K_{34}^\land K_{12}^\lor
\over p_2^+p_3^+}+{K_{12}^\land K_{34}^\lor
\over p_1^+p_4^+}\right]
+L_C(A,B)\left[{K_{23}^\land K_{12}^\lor+K_{34}^\land K_{23}^\lor
\over p_2^+p_3^+}+{K_{12}^\land K_{41}^\lor+K_{41}^\land K_{34}^\lor
\over p_1^+p_4^+}\right]\bigg\}\bigg|_{A=p_{12}^2\atop B=p_{14}^2}\bigg\}
\nonumber\\
&=&{1\over16\pi^2}
\bigg\{-{1\over6}(\ln(\delta e^\gamma)+2)-L(p_{12}^2,p_{14}^2)
-2p_{12}^2p_{14}^2L_{AB}(p_{12}^2,p_{14}^2)\label{93integral}\\
&&\hskip-.5in
+p_{14}^2\bigg\{L_{2AA}(p_{14}^2,p_{12}^2)
\left[{K_{23}^\land K_{41}^\lor
\over p_3^+p_4^+}+{K_{41}^\land K_{23}^\lor
\over p_1^+p_2^+}\right]
+L_{CA}(p_{14}^2,p_{12}^2)
\left[{K_{34}^\land K_{41}^\lor+K_{23}^\land K_{34}^\lor
\over p_3^+p_4^+}+{K_{12}^\land K_{23}^\lor+K_{41}^\land K_{12}^\lor
\over p_1^+p_2^+}\right]\bigg\}\nonumber\\
&&\hskip-.5in 
+p_{12}^2\bigg\{L_{2AA}(p_{12}^2,p_{14}^2)
\left[{K_{34}^\land K_{12}^\lor
\over p_2^+p_3^+}+{K_{12}^\land K_{34}^\lor
\over p_1^+p_4^+}\right]
+L_{CA}(p_{12}^2,p_{14}^2)\left[{K_{23}^\land K_{12}^\lor
+K_{34}^\land K_{23}^\lor
\over p_2^+p_3^+}+{K_{12}^\land K_{41}^\lor+K_{41}^\land K_{34}^\lor
\over p_1^+p_4^+}\right]\bigg\}\bigg\}\nonumber
\eea
The various $L$'s in this formula are listed in Appendix B. 
Eq.~\ref{93integral} can be algebraically
rearranged and cast into a nicer form. Putting in the coupling constant
we find for the contribution of (\ref{ceeterm})
\bea
&&- {g^2\over8\pi^2}
A^{\rm tree}_{\land\lor\land\lor}
\bigg\{
\left[\log^2{\frac{p_{12}^2}{p_{14}^2}}+\pi^2\right]\left[-\frac{1}{32}{\frac
{{{{p_{12}^2}}}^{2}}{({{p_{14}^2}}+{{p_{12}^2}})^{2}}}+\frac{1}{16}{\frac
{{{{p_{12}^2}}}^{3}}{({{p_{14}^2}}+{{p_{12}^2}})^{3}}}-\frac{1}{32}{\frac
{{{{p_{12}^2}}}^{4}}{({{p_{14}^2}}+{{p_{12}^2}})^{4}}}\right]\nn\\
&&+\log{\frac{p_{12}^2}{p_{14}^2}}\left[\frac{1}{48}{\frac
{{{p_{12}^2}}}{{{p_{14}^2}}+{{p_{12}^2}}}}-{\frac {3}{32}}{\frac
{{{{p_{12}^2}}}^{2}}{({{p_{14}^2}}+{{p_{12}^2}})^{2}}}+\frac{1}{16}{\frac
{{{{p_{12}^2}}}^{3}}{({{p_{14}^2}}+{{p_{12}^2}})^{3}}}\right]
+\frac{1}{32}{\frac
{{{p_{12}^2}}}{{{p_{14}^2}}+{{p_{12}^2}}}}-\frac{1}{32}{\frac
{{{{p_{12}^2}}}^{2}}{({{p_{14}^2}}+{{p_{12}^2}})^{2}}}\bigg\}\nn\\
&&+{g^4\over8\pi^2}\bigg\{\log{\frac{p_{12}^2}{p_{14}^2}}
\bigg(\bigg[-\frac{1}{48}{\frac {{{p_{12}^2}}({\Pth}-{\Pf})
({\Pth}-{\Pt})}{({\Pf}-{\Pt}){{p_{14}^2}}({\Po}-{\Pt})}}+
\frac{1}{48}{\frac {{{p_{12}^2}}({\Pth}-{\Pf})({\Pth}-{\Pt})}
{({\Pf}-{\Pt}){{p_{14}^2}}({\Po}-{\Pf})}}
-\frac{1}{48}{\frac {{\Pth}-{\Pt}}{{\Po}-{\Pf}}}
\nn\\&&+\frac{1}{48}{\frac {{\Pth}-{\Po}}{{\Pth}-{\Pf}}}
+{\frac {2{\Pf}-3{\Pth}+{\Pt}}{48{\Pth}-48{\Pf}}}\bigg]+\frac{1}{24}
\bigg[{\frac
{1}{({\Pf}-{\Pt}){{p_{14}^2}}({\Po}-{\Pt})}}-{\frac
{{\Pth}-{\Pt}}{{{p_{14}^2}}({\Pth}-{\Pf})({\Pf}-{\Pt})({\Po}-{\Pf})}}\nn\\
&&+{\frac{{\Pth}-{\Po}}{{{p_{14}^2}}({\Pth}-{\Pf})^{2}({\Pth}-{\Pt})}}
-{\frac
{2{\Pth}-{\Pf}-{\Pt}}{{{p_{14}^2}}({\Pth}-{\Pf})^{2}({\Pth}-{\Pt})}}\bigg]
{\Kttw}\Kftv\bigg)-\frac{1}{24}\log{p_{14}^2\delta e^\gamma}+{1\over144}\bigg\}\nn\eea

\section{Complete Results of the Two Model Boxes}
There are basically two model integrands that has an adjacent pair
of vertices with the same helicity. \bea\mathcal{N}=K_{61}^\land
K_{25}^{\lor} K_{5,3}^{\lor}
K_{4,6}^\land\hspace{.3in}\mathrm{or}\hspace{.3in}\mathcal{N}=K_{61}^\land
K_{25}^{\land} K_{5,3}^{\lor} K_{4,6}^\lor\nn\eea Assume for the
moment that $\mathcal{R}=1$. These diagram can be reduced into
triangles using Eq.(\ref{92}). The first contribution is 
\bea
&&\hskip-.5in\pref g^4\Nc K_{61}^\land K_{25}^{\lor} K_{5,3}^{\lor} 
K_{4,6}^\land\times\nn\\ 
&&\phantom{a}\nn\\
&&\rleft\nn\\
&&\phantom{a}\nn\\
&-&\frac{1}{8}{\frac
{(\Kp-{\Pf})^{2}({\Pt}-{\Po})({\Pth}-{\Pt})}{{\Pt}-{\Pf}}}\log
({\it H_s}\delta e^{\gamma})+\frac{1}{8}{\frac {({\Pth}-\Kp)^{2}({\Po}-{\Pf})({\Pt}-{\Po})}{{\Pth}-{\Po}}}\log({\it H_d}\delta e^{\gamma})\nn\\
&+&\frac{1}{4}{\frac
{(\Kp-{\Po})({\Pth}-\Kp)}{({\Pth}-{\Po}){{p_{14}^2}}}}{\Kofw}\Kttv\log
({\it H_d}\delta e^{\gamma})+\frac{1}{8}{\frac {(\Kp-{\Po})^{2}({\Pth}-{\Pf})({\Pth}-{\Pt})}{{\Pth}-{\Po}}}\log ({\it H_u}\delta e^{\gamma})\nn\\
&-&\frac{1}{4}{\frac
{(\Kp-{\Po})({\Pth}-\Kp)}{({\Pth}-{\Po}){{p_{14}^2}}}}{\Kofw}\Kttv\log
({\it H_u}\delta e^{\gamma})\nn\\
&&\phantom{a}\nn\\
& &\rmid\nn\\
&&\phantom{a}\nn\\
&+&\bigg[\frac{1}{8}{\frac {({\Po}-{\Pf})({\Pt}-{\Po})^{2}({\Pth}-{\Pf})}{{\Pt}-{\Pf}}}-\frac{1}{4}{\frac {({\Po}-{\Pf})({\Pth}-{\Pf})({\Pt}-{\Po})(\Kp-{\Po})}{{\Pt}-{\Pf}}}\nn\\
&-&\frac{1}{8}{\frac {({\Pt}-{\Po})({\Pth}-{\Pf})({\Pth}-{\Pt}-{\Po}+{\Pf})(\Kp-{\Po})^{2}}{({\Pt}-{\Pf})({\Pth}-{\Po})}}\bigg]\log ({\it H_s}\delta e^{\gamma})\nn\\
&+&\frac{1}{4}{\frac
{(\Kp-{\Po})({\Pth}-\Kp)}{({\Pth}-{\Po}){{p_{14}^2}}}}{\Kofw}\Kttv\log
({\it H_s}\delta e^{\gamma})+\frac{1}{4}{\frac {(\Kp-{\Po})({\Pth}-\Kp)}{({\Pth}-{\Po}){{p_{14}^2}}}}{\Kofw}\Kttv\log ({\it H_t}\delta e^{\gamma})\nn\\
&-&\frac{1}{8}{\frac
{({\Pth}-\Kp)^{2}({\Po}-{\Pf})({\Pt}-{\Po})}{{\Pth}-{\Po}}}\log
({\it H_d}\delta e^{\gamma})-\frac{1}{4}{\frac {(\Kp-{\Po})({\Pth}-\Kp)}{({\Pth}-{\Po}){{p_{14}^2}}}}{\Kofw}\Kttv\log ({\it H_d}\delta e^{\gamma})\nn\\
&+&\frac{1}{8}{\frac
{(\Kp-{\Po})^{2}({\Pth}-{\Pf})({\Pth}-{\Pt})}{{\Pth}-{\Po}}}\log
({\it H_u}\delta e^{\gamma})-\frac{1}{4}{\frac
{(\Kp-{\Po})({\Pth}-\Kp)}{({\Pth}-{\Po}){{p_{14}^2}}}}{\Kofw}\Kttv\log
({\it H_u}\delta e^{\gamma})\nn\\
&&\phantom{a}\nn\\
& &\rright\nn\\
&&\phantom{a}\nn\\
&+&\frac{1}{8}{\frac
{({\Po}-{\Pf})({\Pt}-\Kp)^{2}({\Pth}-{\Pf})}{{\Pt}-{\Pf}}}\log
({\it H_s}\delta e^{\gamma})-\frac{1}{8}{\frac {({\Pth}-\Kp)^{2}({\Po}-{\Pf})({\Pt}-{\Po})}{{\Pth}-{\Po}}}\log ({\it H_d}\delta e^{\gamma})\nn\\
&-&\frac{1}{4}{\frac
{({\Pth}-\Kp)(\Kp-{\Po})}{({\Pth}-{\Po}){{p_{14}^2}}}}{\Kofw}\Kttv\log
({\it H_d}\delta e^{\gamma})-\frac{1}{8}{\frac {(\Kp-{\Po})^{2}({\Pth}-{\Pf})({\Pth}-{\Pt})}{{\Pth}-{\Po}}}\log ({\it H_u}\delta e^{\gamma})\nn\\
&+&\frac{1}{4}{\frac
{({\Pth}-\Kp)(\Kp-{\Po})}{({\Pth}-{\Po}){{p_{14}^2}}}}{\Kofw}\Kttv\log
({\it H_u}\delta e^{\gamma})\nn\eea
Assuming its coefficient to be
$-C-A\frac{1}{(\Kp-\Po)^2}+B\frac{1}{(\Kp-\Po)}$ or
$-C-A\frac{1}{(\Kp-\Pth)^2}+B\frac{1}{(\Kp-\Pth)}$, its
contribution
will be 
\bea
&&\hskip-.5in\pref g^4 \Nc\bigg\{ \left[\frac{C}{36}-\frac{C}{24}\log{(p_{12}^2\delta e^{\gamma})}\right](-{\Pth}+{\Pt})(-{\Pt}+{\Po})({\Pf}-{\Pth})({\Pf}-{\Po})\nn\\
&+&\bigg[\frac{C}{24}{\frac {2 {\Pth} {\Po}-{\Pt} {\Po}-{\Pf} {\Po}-{\Pth} {\Pt}-{\Pth} {\Pf}+{{\Pf}}^{2}+{{\Pt}}^{2}}{{p_{14}^2}}}-\frac{C}{24}\frac{(\Po-\Pth)^2}{p_{14}^2}\log{\frac{p_{12}^2}{p_{14}^2}}\nn\\
&+&\frac{B}{8}\frac
{{\Po}+{\Pth}-{\Pt}-{\Pf}}{p_{14}^2}+\frac{A}{4p_{14}^2}\log{\frac{p_{12}^2}{p_{14}^2}}\mp\frac{B}{8}
\frac
{{\Po}-{\Pth}}{p_{14}^2}\log{\frac{p_{12}^2}{p_{14}^2}}\bigg]\Kofw\Kttv+\mathrm{infra\
red\ terms}\bigg\}\nn\eea 
And it's impossible for this diagram to have
poles at $\Kp=\Pt\ \mathrm{or}\ \Pf$. The reader can work out the
infrared terms by partial fractioning the coefficient of each
logarithm, and taking only the pole terms (but be very careful when
applying these results to actual boxes, you might need to take a
conjugation occasionally).

The second case is very similar:
\bea
&&\hskip-.5in\pref g^4\Nc K_{61}^\land K_{25}^{\land} K_{5,3}^{\lor} 
K_{4,6}^\lor\times\nn\\ 
&&\phantom{a}\nn\\
&&\rleft\nn\\
&&\phantom{a}\nn\\
&+&\frac{1}{4}{\frac
{(\Kp-{\Pf})({\Pt}-\Kp)}{({\Pt}-{\Pf}){{p_{12}^2}}}}{\Ktow}\Kftv\log
({\it H_s}\delta e^{\gamma})+\frac{1}{8}{\frac {({\Pt}-\Kp)^{2}({\Po}-{\Pf})({\Pth}-{\Pf})}{{\Pt}-{\Pf}}}\log ({\it H_r}\delta e^{\gamma})\nn\\
&-&\frac{1}{4}{\frac {(\Kp-{\Pf})({\Pt}-\Kp)}{({\Pt}-{\Pf}){{p_{12}^2}}}}{\Ktow}\Kftv\log ({\it H_r}\delta e^{\gamma})\nn\\
\nn\\
& &\rmid\nn\\&+&\frac{1}{4}{\frac {(\Kp-{\Pf})(-\Kp+{\Pt})}{({\Pt}-{\Pf}){{p_{12}^2}}}}{\Ktow}\Kftv\log ({\it H_s}\delta e^{\gamma})\nn\\
&+&\bigg[\frac{1}{8}{\frac {({\Pt}-{\Po})^{2}({\Pth}-{\Pt})({\Pth}-{\Pf})}{{\Pth}-{\Po}}}-\frac{1}{4}{\frac {({\Pth}-{\Pt})({\Pth}-{\Pf})({\Pt}-{\Po})(-\Kp+{\Pt})}{{\Pth}-{\Po}}}\nn\\
&+&\frac{1}{8}{\frac {({\Pt}-{\Po})({\Pth}-{\Pf})({\Pth}-{\Po}-{\Pt}+{\Pf})(-\Kp+{\Pt})^{2}}{({\Pt}-{\Pf})({\Pth}-{\Po})}}\bigg]\log ({\it H_t}\delta e^{\gamma})\nn\\
&+&\frac{1}{4}{\frac
{(\Kp-{\Pf})(-\Kp+{\Pt})}{({\Pt}-{\Pf}){{p_{12}^2}}}}{\Ktow}\Kftv\log
({\it H_t}\delta e^{\gamma})-\frac{1}{8}{\frac {(\Kp-{\Pf})^{2}({\Pt}-{\Po})({\Pth}-{\Pt})}{{\Pt}-{\Pf}}}\log ({\it H_l}\delta e^{\gamma})\nn\\
&-&\frac{1}{4}{\frac
{(\Kp-{\Pf})(-\Kp+{\Pt})}{({\Pt}-{\Pf}){{p_{12}^2}}}}{\Ktow}\Kftv\log
({\it H_l}\delta e^{\gamma})+\frac{1}{8}{\frac {(-\Kp+{\Pt})^{2}({\Po}-{\Pf})({\Pth}-{\Pf})}{{\Pt}-{\Pf}}}\log ({\it H_r}\delta e^{\gamma})\nn\\
&-&\frac{1}{4}{\frac
{(\Kp-{\Pf})(-\Kp+{\Pt})}{({\Pt}-{\Pf}){{p_{12}^2}}}}{\Ktow}\Kftv\log
({\it H_r}\delta e^{\gamma})\nn\\
\nn\\
& &\rright\nn\\&+&\frac{1}{4}{\frac
{(\Kp-{\Pf})(-\Kp+{\Pt})}{({\Pt}-{\Pf}){{p_{12}^2}}}}{\Ktow}\Kftv\log
({\it H_s}\delta e^{\gamma})-\frac{1}{8}{\frac {(\Kp-{\Pf})^{2}({\Pt}-{\Po})({\Pth}-{\Pt})}{{\Pt}-{\Pf}}}\log ({\it H_l}\delta e^{\gamma})\nn\\
&-&\frac{1}{4}{\frac
{(\Kp-{\Pf})(-\Kp+{\Pt})}{({\Pt}-{\Pf}){{p_{12}^2}}}}{\Ktow}\Kftv\log
({\it H_l}\delta e^{\gamma})\nn\eea

Assuming its coefficient to be
$-C-A\frac{1}{(\Kp-\Pt)^2}+B\frac{1}{(\Kp-\Pt)}$ or
$-C-A\frac{1}{(\Kp-\Pf)^2}+B\frac{1}{(\Kp-\Pf)}$, 
its contribution will be:
\bea&&
\hskip-.5in\pref g^4 \Nc
\bigg\{
\left[\frac{C}{36}-\frac{C}{24}\log{(p_{14}^2\delta e^{\gamma})}\right]({\Pth}-{\Pf})({\Po}-{\Pf})({\Pt}-{\Po})({\Pth}-{\Pt})\nn\\
&+&\bigg[-\frac{C}{24} {\frac {(-{{\Po}}^{2}-{{\Pth}}^{2}+{\Po} {\Pt}+{\Pt} {\Pth}+{\Po} {\Pf}+{\Pf} {\Pth}-2 {\Pt} {\Pf})}{p_{12}^2}}+\frac{1}{24}\frac{(\Pt-\Pf)^2}{p_{12}^2}\log{\frac{p_{12}^2}{p_{14}^2}}\nn\\
&-&\frac{B}{8}{\frac
{-{\Pt}+{\Po}+{\Pth}-{\Pf}}{p_{12}^2}}-\frac{1}{4p_{12}^2}\log{\frac{p_{12}^2}{p_{14}^2}}\pm\frac{1}{8}\cdot{\frac
{{\Pt}-{\Pf}}{p_{12}^2}}\log{\frac{p_{12}^2}{p_{14}^2}}\bigg]{\Ktow}\Kftv+\mathrm{infra\
red\ terms}\nn\bigg\}
\eea And it's impossible for this diagram to have
poles at $\Kp=\Po\ \mathrm{or}\ \Pth$.
\vskip14pt
\section{Complete Results of the Second Term of Eq.(\ref{104})}
\bea
&&\pref g^4\Nc A \log{\frac{p_{12}^2}{p_{14}^2}}\times\bigg\{\nn\\
&&-\frac{1}{8}{\frac
{({\Pth}-{\Pf})({\Pth}-{\Pt})(({\Pf}-{\Po}){{p_{14}^2}}+{{p_{12}^2}}{\Pth}-{{p_{12}^2}}{\Pt})}{({\Pth}-{\Po})({{p_{14}^2}}+{{p_{12}^2}})}}
+\frac{1}{4}{\frac
{({\Pth}-{\Pf}){{p_{14}^2}}+{{p_{12}^2}}{\Pt}-{\Po}{{p_{12}^2}}}{({\Pth}-{\Po}){{p_{14}^2}}({{p_{14}^2}}+{{p_{12}^2}})}}{\Kofw}\Kttv\bigg\}\nn\\
\nn\\
&&\pref g^4\Nc A
\bigg[\log^2{\frac{p_{12}^2}{p_{14}^2}}+\pi^2\bigg]\times\bigg\{-\frac{1}{16}\frac
{{{p_{14}^2}}{{p_{12}^2}}(-{\Pt}+{\Pth})({\Pth}-{\Pf})({\Pth}-{\Pt}+{\Po}-{\Pf})}{({{p_{14}^2}}+{{p_{12}^2}})^{2}({\Pth}-{\Po})}\nn\\
&&-\frac{1}{8}{\frac
{{{p_{12}^2}}((-{\Pt}+2{\Pth}-{\Pf}){{p_{14}^2}}+{{p_{12}^2}}(\Pth-\Po))}{({\Pth}-{\Po})({{p_{14}^2}}+{{p_{12}^2}})^{2}{{p_{14}^2}}}}{\Kofw}\Kttv\bigg\}\nn\\
\nn\\
&&\pref g^4\Nc B \log{\frac{p_{12}^2}{p_{14}^2}}\times\bigg\{\nn\\
&&-\frac{1}{16}{\frac {{{p_{12}^2}}({\Pf}-{\Po})(3{{\Pf}}^{2}-6{\Pth}{\Pf}-2{\Po}{\Pf}+2{\Pt}{\Pf}+3{{\Pth}}^{2}-4{\Pt}{\Pth}+4{\Pth}{\Po}-{{\Po}}^{2}+{{\Pt}}^{2})}{{{p_{14}^2}}+{{p_{12}^2}}}}\nn\\
&&+\frac{1}{8}{\frac {{{\Pth}}^{2}+2{\Pth}{\Po}-2{\Pt}{\Pth}-2{\Pth}{\Pf}-{{\Po}}^{2}+{{\Pf}}^{2}+{{\Pt}}^{2}}{({\Pth}-{\Pf})({{p_{14}^2}}+{{p_{12}^2}})}}{\Kttw}\Kftv\nn\\
&&+\frac{1}{8}{\frac
{{{{p_{12}^2}}}^{2}({\Pf}-{\Po})({\Pth}-{\Pt}+{\Po}-{\Pf})^{2}}{({{p_{14}^2}}+{{p_{12}^2}})^{2}}}-\frac{1}{4}{\frac {{{p_{12}^2}}({\Pth}-{\Pt}+{\Po}-{\Pf})^{2}}{({\Pth}-{\Pf})({{p_{14}^2}}+{{p_{12}^2}})^{2}}}{\Kttw}\Kftv\nn\\
&&+\frac{1}{16}({\Pth}-{\Po})^{2}({\Pth}-{\Pf})-\frac{1}{16}({\Pth}-{\Po})({\Pth}-{\Pt})({\Pth}-{\Pf})\nn\\
&&+\frac{1}{8}{\frac
{{\Po}-{\Pth}}{{{p_{14}^2}}}}{\Kofw}\Kttv+\frac{1}{16}({\Pth}-{\Pf})^{2}(-{\Pt}+{\Pf})\bigg\}\nn\\
\nn\\
&&\pref g^4\Nc B
\left[\log^2{\frac{p_{12}^2}{p_{14}^2}}+\pi^2\right]\times\bigg\{-\frac{1}{16}{\frac {{{p_{12}^2}}({\Pth}-{\Pf})^{2}({\Pf}-{\Po})}{{{p_{14}^2}}+{{p_{12}^2}}}}\nn\\
&&+\frac{1}{16}{\frac {{{{p_{12}^2}}}^{2}({\Pf}-{\Po})(2{{\Pf}}^{2}+2{\Pt}{\Pf}-4{\Pth}{\Pf}-2{\Po}{\Pf}-3{\Pt}{\Pth}+3{\Pth}{\Po}+2{{\Pth}}^{2}+{{\Pt}}^{2}-{\Pt}{\Po})}{({{p_{14}^2}}+{{p_{12}^2}})^{2}}}\nn\\
&&-\frac{1}{8}{\frac {(-{\Po}{\Pf}+{\Pt}{\Pf}-{\Pt}{\Po}+{{\Pth}}^{2}-2{\Pt}{\Pth}+2{\Pth}{\Po}-2{\Pth}{\Pf}+{{\Pt}}^{2}+{{\Pf}}^{2}){{p_{12}^2}}}{({\Pth}-{\Pf})({{p_{14}^2}}+{{p_{12}^2}})^{2}}}{\Kttw}\Kftv\nn\\
&&-\frac{1}{16}{\frac
{{{{p_{12}^2}}}^{3}({\Pf}-{\Po})({\Pth}-{\Pt}+{\Po}-{\Pf})^{2}}{({{p_{14}^2}}+{{p_{12}^2}})^{3}}}+\frac{1}{8}{\frac
{{{{p_{12}^2}}}^{2}({\Pth}-{\Pt}+{\Po}-{\Pf})^{2}}{({\Pth}-{\Pf})({{p_{14}^2}}+{{p_{12}^2}})^{3}}}{\Kttw}\Kftv\bigg\}\nn\\
\nn\\
&&\pref g^4\Nc B\times\bigg\{-\frac{1}{16}{\frac
{{{p_{12}^2}}({\Pf}-{\Po})({\Pth}-{\Pt}+{\Po}-{\Pf})^{2}}{{{p_{14}^2}}+{{p_{12}^2}}}}+\frac{1}{8}{\frac {({\Pth}-{\Pt}+{\Po}-{\Pf})^{2}}{({\Pth}-{\Pf})({{p_{14}^2}}+{{p_{12}^2}})}}{\Kttw}\Kftv\nn\\
&&-\frac{1}{16}({\Pth}-{\Po})^{2}({\Pth}-{\Pf})+\frac{1}{16}({\Pth}-{\Po})({\Pth}-{\Pf})(3{\Pth}-{\Pt}-2{\Pf})\nn\\
&&+\frac{1}{8}{\frac
{{\Pth}-{\Po}}{{{p_{14}^2}}}}{\Kofw}\Kttv-\frac{1}{16}({\Pth}-{\Pf})^{2}(2{\Pth}-{\Pt}-{\Pf})-\frac{1}{8}{\frac
{2{\Pth}-{\Pt}-{\Pf}}{{{p_{14}^2}}}}{\Kofw}\Kttv\nn\bigg\}\nn
\eea
The results for other pole locatioins can be obtained from this by
rotational symmetry and conjugation.

If we put these terms from each diagram together, there is usually
some nice simplifications.\newline
For the fifth box of fig.\ref{boxreduction3}:
\bea
-{g^2\over8\pi^2}\Nc\Aalt^{tree}
\bigg\{\left[\log^2{\frac{p_{12}^2}{p_{14}^2}}+\pi^2\right]
\left[\frac{1}{2}{\frac
{{{{p_{12}^2}}}^{2}}{({{p_{14}^2}}+{{p_{12}^2}})^{2}}}-{\frac
{{{{p_{12}^2}}}^{3}}{({{p_{14}^2}}+{{p_{12}^2}})^{3}}}\right]+\log{\frac{p_{12}^2}{p_{14}^2}}\left[2{\frac
{{{{p_{12}^2}}}^{2}}{({{p_{14}^2}}+{{p_{12}^2}})^{2}}}\right]-{\frac
{{{p_{12}^2}}}{{{p_{14}^2}}+{{p_{12}^2}}}}\bigg\}\nn\eea And also a piece
that doesn't fall into a tree: \bea&&\pref
g^4\Nc\log{\frac{p_{12}^2}{p_{14}^2}}\times\bigg\{\bigg[2{\frac
{{{p_{12}^2}}(-{\Pt}+{\Pf})}{{{p_{14}^2}}({\Pt}-{\Po})}}+2{\frac
{{{p_{12}^2}}({\Pth}-{\Pf})}{{{p_{14}^2}}({\Pth}-{\Po})}}+2{\frac {{{p_{12}^2}}({\Pth}-{\Pf})}{{{p_{14}^2}}({\Pf}-{\Po})}}-2{\frac {({\Pth}-{\Pt})(-{\Pt}+{\Pf})}{({\Pth}-{\Po})^{2}}}\nn\\&&+2{\frac {(-{\Pt}+{\Pf})(2{\Pth}-{\Pt}-{\Pf})}{({\Pth}-{\Pf})({\Pth}-{\Po})}}-2{\frac {-{\Pt}+{\Pf}}{{\Pf}-{\Po}}}\bigg]\nn\\
&&+\bigg[-4{\frac
{-{\Pt}+{\Pf}}{{{p_{14}^2}}({\Pth}-{\Pf})({\Pth}-{\Pt})({\Pt}-{\Po})}}-4{\frac
{2{\Pth}-{\Pt}-{\Pf}}{{{p_{14}^2}}({\Pth}-{\Pf})({\Pth}-{\Po})^{2}}}+4{\frac
{1}{{{p_{14}^2}}({\Pth}-{\Pf})({\Pth}-{\Pt})}}\nn\\
&&+4{\frac
{-{{\Pf}}^{2}+{\Pt}{\Pf}+{\Pth}{\Pf}-3{\Pt}{\Pth}+{{\Pt}}^{2}+{{\Pth}}^{2}}{({\Pth}-{\Pt})({\Pth}-{\Pf})^{2}{{p_{14}^2}}({\Pth}-{\Po})}}-4{\frac
{1}{{{p_{14}^2}}({\Pth}-{\Pf})({\Pf}-{\Po})}}\bigg]{\Kttw}\Kftv\bigg\}\nn\\
\nn\\
&&\pref g^4\Nc\times\bigg\{\bigg[2{\frac {(2{\Pth}-{\Pt}-{\Pf}){{p_{12}^2}}}{({\Pth}-{\Po}){{p_{14}^2}}}}-2{\frac {(-{\Pt}+{\Pf})(2{\Pth}-{\Pt}-{\Pf})}{({\Pth}-{\Po})^{2}}}+2{\frac {-{\Pt}+{\Pf}}{{\Pth}-{\Po}}}+2\bigg]\nn\\
&&+\bigg[-8{\frac
{2{\Pth}-{\Pt}-{\Pf}}{{{p_{14}^2}}({\Pth}-{\Po})^{2}({\Pth}-{\Pf})}}+8{\frac
{1}{{{p_{14}^2}}({\Pth}-{\Po})({\Pth}-{\Pf})}}\bigg]{\Kttw}\Kftv\bigg\}\nn\eea
For the fourth box of fig.\ref{boxreduction3}:
\bea 
&&-\pref g^2\Nc\Aalt^{tree}
\bigg\{\left[\log^2{\frac{p_{12}^2}{p_{14}^2}}+\pi^2\right]\left[{\frac
{{{p_{12}^2}}}{{{p_{14}^2}}+{{p_{12}^2}}}}-\frac{5}{2}{\frac
{{{{p_{12}^2}}}^{2}}{({{p_{14}^2}}+{{p_{12}^2}})^{2}}}+{\frac
{{{{p_{12}^2}}}^{3}}{({{p_{14}^2}}+{{p_{12}^2}})^{3}}}-\frac{1}{2}\right]\nn\\
&&\hskip1in+\log{\frac{p_{12}^2}{p_{14}^2}}\left[4{\frac
{{{p_{12}^2}}}{{{p_{14}^2}}+{{p_{12}^2}}}}-2{\frac
{{{{p_{12}^2}}}^{2}}{({{p_{14}^2}}+{{p_{12}^2}})^{2}}}\right]+\frac
{{{p_{12}^2}}}{{{p_{14}^2}}+{{p_{12}^2}}}\bigg\}\nn
\eea 
And also a piece
that doesn't fall into a tree: \bea&&\pref g^4\Nc\log{\frac{p_{12}^2}{p_{14}^2}}\times\bigg\{\nn\\&&+\bigg[4{\frac {{{p_{12}^2}}({\Pth}-{\Pt})({\Pth}-{\Pf})}{({\Pf}-{\Pt}){{p_{14}^2}}({\Po}-{\Pt})}}-4{\frac {{{p_{12}^2}}({\Pth}-{\Pt})({\Pth}-{\Pf})}{{{p_{14}^2}}({\Pf}-{\Pt})({\Po}-{\Pf})}}+4{\frac {{\Pth}-{\Pt}}{{\Po}-{\Pf}}}-4{\frac {{\Pth}-{\Po}}{{\Pth}-{\Pf}}}+4{\frac {2{\Pth}-{\Pf}-{\Pt}}{{\Pth}-{\Pf}}}\bigg]\nn\\
&&+\bigg[-4{\frac {{\Pth}-{\Pf}}{({\Pf}-{\Pt})({\Pth}-{\Pt})({\Po}-{\Pt}){{p_{12}^2}}}}+4{\frac {{\Pth}-{\Pf}}{({\Pf}-{\Pt})^{2}({\Pth}-{\Pt}){{p_{12}^2}}}}\bigg]{\Ktow}\Kofv\nn\\
&&+\bigg[4{\frac {-2{\Pf}+{\Pt}+{\Pth}}{({\Pf}-{\Pt})({\Pth}-{\Pt})({\Po}-{\Pt}){{p_{12}^2}}}}+4{\frac {1}{({\Pf}-{\Pt})({\Pth}-{\Pt}){{p_{12}^2}}}}\bigg]{\Ktow}\Kftv\nn\\
&&+\bigg[-4{\frac {-2{\Pt}+{\Pf}+{\Pth}}{({\Pf}-{\Pt})({\Pth}-{\Pf})({\Po}-{\Pf}){{p_{12}^2}}}}-4{\frac {1}{({\Pf}-{\Pt})({\Pth}-{\Pf}){{p_{12}^2}}}}\bigg]{\Kftw}\Ktov\nn\\
&&+\bigg[-8{\frac {1}{{{p_{14}^2}}({\Pf}-{\Pt})({\Po}-{\Pt})}}+8{\frac {{\Pth}-{\Pt}}{{{p_{14}^2}}({\Pth}-{\Pf})({\Pf}-{\Pt})({\Po}-{\Pf})}}-8{\frac {{\Pth}-{\Po}}{({\Pth}-{\Pt})({\Pth}-{\Pf})^{2}{{p_{14}^2}}}}\nn\\
&&+8{\frac {2{\Pth}-{\Pf}-{\Pt}}{({\Pth}-{\Pt})({\Pth}-{\Pf})^{2}{{p_{14}^2}}}}\bigg]{\Kttw}\Kftv\nn\\
&&+\bigg[4{\frac
{{\Pth}-{\Pt}}{({\Pf}-{\Pt})({\Pth}-{\Pf})({\Po}-{\Pf}){{p_{12}^2}}}}+4{\frac
{{\Pth}-{\Pt}}{({\Pf}-{\Pt})^{2}({\Pth}-{\Pf}){{p_{12}^2}}}}\bigg]{\Kftw}\Kttv\bigg\}\nn\\
\nn\\
&&\pref g^4\Nc\times\bigg\{\bigg[2{\frac {{\Pth}-{\Pt}}{{\Po}-{\Pf}}}-2{\frac {{\Pth}-{\Po}}{{\Pth}-{\Pf}}}+2{\frac {2{\Pth}-{\Pf}-{\Pt}}{{\Pth}-{\Pf}}}-2{\frac {{{p_{12}^2}}({\Pth}-{\Pt})({\Pth}-{\Pf})}{{{p_{14}^2}}({\Po}-{\Pf})({\Po}-{\Pt})}}\bigg]\nn\\
&&+\bigg[4{\frac {{\Pth}-{\Pf}}{({\Pf}-{\Pt})({\Pth}-{\Pt})({\Po}-{\Pt}){{p_{12}^2}}}}+4{\frac {{\Pth}-{\Po}}{({\Pf}-{\Pt})^{3}{{p_{12}^2}}}}-4{\frac {({\Pth}-{\Pf})({\Pf}-3{\Pt}+2{\Pth})}{({\Pf}-{\Pt})^{3}({\Pth}-{\Pt}){{p_{12}^2}}}}\bigg]{\Ktow}\Kftv\nn\\
&&+\bigg[-4{\frac {{\Pth}-{\Pt}}{({\Pf}-{\Pt})({\Pth}-{\Pf})({\Po}-{\Pf}){{p_{12}^2}}}}-4{\frac {{\Pth}-{\Po}}{({\Pf}-{\Pt})^{3}{{p_{12}^2}}}}+4{\frac {({\Pth}-{\Pt})(2{\Pth}+{\Pt}-3{\Pf})}{({\Pf}-{\Pt})^{3}({\Pth}-{\Pf}){{p_{12}^2}}}}\bigg]{\Kftw}\Ktov\nn\\
&&+4{\frac
{(-{\Pf}+{\Pth}-{\Pt}+{\Po})({{\Po}}^{2}-{\Pf}{\Po}-{\Pt}{\Po}-{\Pth}{\Pf}+2{\Pf}{\Pt}-{\Pth}{\Pt}+{{\Pth}}^{2})}{({\Pth}-{\Pt})({\Po}-{\Pf})({\Pth}-{\Pf})^{2}({\Po}-{\Pt}){{p_{14}^2}}}}{\Kttw}\Kftv\bigg\}\nn\eea

For the seventh box of fig.\ref{boxreduction3}:

\bea -\frac{1}{2}\cdot\left[-\pref g^2
\Nc\Aalt^{tree}\right]\nn\eea

And:
\bea&&\pref g^4\Nc\log{\frac{p_{12}^2}{p_{14}^2}}\times\bigg\{\nn\\
&&+4{\frac
{({\Pth}-{\Pt})({\Po}-{\Pt})}{({\Pf}-{\Pt})^{2}({\Pth}-{\Pf})({\Po}-{\Pf}){{p_{12}^2}}}}{\Kofw}\Ktov+4{\frac {({\Pth}-{\Pt})({\Pth}-{\Pf})(-{\Pf}-{\Pt}+2{\Po})}{({\Po}-{\Pf})^{2}({\Pth}-{\Po})({\Po}-{\Pt})^{2}{{p_{14}^2}}}}{\Kofw}\Kttv\nn\\
&&-4{\frac
{(2{\Pth}-{\Pf}-{\Pt})({\Po}-{\Pt})({\Po}-{\Pf})}{({\Pth}-{\Pt})^{2}({\Pth}-{\Pf})^{2}{{p_{14}^2}}({\Pth}-{\Po})}}{\Kttw}\Kofv-4{\frac {({\Po}-{\Pt})({\Po}-{\Pf})}{({\Pth}-{\Pt})({\Pth}-{\Pf})({\Pth}-{\Po})^{2}{{p_{14}^2}}}}{\Kttw}\Ktov\nn\\
&&-4{\frac
{({\Pth}-{\Pf})({\Pth}-{\Pt})}{({\Po}-{\Pt})({\Pth}-{\Po})^{2}({\Po}-{\Pf}){{p_{14}^2}}}}{\Kofw}\Kftv+4{\frac {({\Pth}-{\Pf})({\Po}+{\Pth}-2{\Pt})({\Po}-{\Pf})}{({\Pth}-{\Pt})^{2}({\Po}-{\Pt})^{2}{{p_{12}^2}}({\Pf}-{\Pt})}}{\Ktow}\Kftv\nn\\
&&-4{\frac
{({\Pth}-{\Pt})({\Po}-{\Pt})(-2{\Pf}+{\Po}+{\Pth})}{({\Pth}-{\Pf})^{2}({\Po}-{\Pf})^{2}{{p_{12}^2}}({\Pf}-{\Pt})}}{\Kftw}\Ktov+4{\frac
{({\Pth}-{\Pf})({\Po}-{\Pf})}{({\Pf}-{\Pt})^{2}({\Pth}-{\Pt})({\Po}-{\Pt}){{p_{12}^2}}}}{\Kttw}\Kftv\bigg\}\nn\\
\nn\\
&&\pref g^4\Nc\times\bigg\{\nn\\
&+&4{\frac {(-{\Pf}+{\Pth}-{\Pt}+{\Po})({\Pth}-{\Pf})({\Pth}-{\Pt})}{{{p_{14}^2}}({\Po}-{\Pf})^{2}({\Pth}-{\Po})^{3}({\Po}-{\Pt})^{2}}}\cdot\nn\\
&&(-3{{\Po}}^{2}+2{\Pth}{\Po}+2{\Po}{\Pf}+2{\Pt}{\Po}-{\Pth}{\Pt}-{\Pth}{\Pf}-{\Pf}{\Pt}){\Kofw}\Kttv\nn\\
&&+4{\frac {(-{\Pf}+{\Pth}-{\Pt}+{\Po})({\Po}-{\Pf})({\Po}-{\Pt})}{{{p_{14}^2}}({\Pth}-{\Po})^{3}({\Pth}-{\Pf})^{2}({\Pth}-{\Pt})^{2}}}\cdot\nn\\
&&({\Pf}{\Pt}-2{\Pth}{\Pt}+{\Pt}{\Po}+3{{\Pth}}^{2}-2{\Pth}{\Po}-2{\Pth}{\Pf}+{\Po}{\Pf}){\Kttw}\Kofv\nn\\
&&+4{\frac
{({\Pth}-{\Pf})({\Po}-{\Pf})(-{\Pf}+{\Pth}-{\Pt}+{\Po})}{({\Pth}-{\Pt})^{2}({\Pf}-{\Pt})^{3}({\Po}-{\Pt})^{2}{{p_{12}^2}}}}\cdot\nn\\
&&(-2{\Pf}{\Pt}+{\Po}{\Pf}+{\Pth}{\Pf}+3{{\Pt}}^{2}-2{\Pt}{\Po}-2{\Pth}{\Pt}+{\Pth}{\Po}){\Ktow}\Kftv\nn\\
&&-4{\frac
{({\Pth}-{\Pt})({\Po}-{\Pt})(-{\Pf}+{\Pth}-{\Pt}+{\Po})}{({\Pth}-{\Pf})^{2}({\Pf}-{\Pt})^{3}({\Po}-{\Pf})^{2}{{p_{12}^2}}}}\nn\cdot\\
&&(3{{\Pf}}^{2}-2{\Po}{\Pf}-2{\Pth}{\Pf}-2{\Pf}{\Pt}+{\Pt}{\Po}+{\Pth}{\Po}+{\Pth}{\Pt}){\Kftw}\Ktov\bigg\}\nn\eea
\vskip14pt
\section{Details on Triangle-like Diagrams
with Collinear Divergences}
After ascertaining that the combination of
\bea &&-\frac{A}{2}\frac{K_{3,5}^\lor K_{4,3}^\land
p_{12}^2}{q_1^2 q_2^2 q_3^2
p_{12}^2}-\frac{A}{2}\frac{K_{3,5}^\lor
K_{1,4}^\land p_{12}^2}{q_1^2 q_2^2q_0^2 p_{14}^2}\nn\\
&&-\frac{A}{2}\frac{K_{3,5}^\lor
K_{6,4}^\land}{(\Kp-\Po)^2}\left[\frac{(-{\Po}+{\Pt})(-{\Po}+\Kp)}{q_1^2q_3^2q_0^2}+\frac{(-{\Pf}+{\Po})({\Po}-\Kp)}{q_1^2q_2^2q_3^2}\right]\nn\eea
doesn't have collinear divergence, we simply list its contribution
here. Note in the results below, we have put back in the term
\bea-\frac{A}{2}\frac{K_{3,5}^\lor
K_{6,4}^\land}{(\Kp-\Po)^2}\frac{(\Po+\Pt)({\Kp}-\Pf)+({\Pf}-{\Po})(\Kp-\Pf)}{q_0^2q_3^2q_2^2}-A\frac{K_{3,5}^\lor
K_{6,4}^\land}{(\Kp-\Po)^2}\frac{K_{16}^{\land}K_{5,2}^{\lor}}{q_0^2q_1^2q_3^2q_2^2}\nn\eea
(more specifically, the above contains all the triangle-like terms
in Eq.(\ref{104}) and the two subtractions from the first term of
Eq.(\ref{104})).

\bea&&\pref g^4\Nc\times\nn\\
&&\rleft\nn\\
&-&\frac{1}{8}{\frac
{({\Pth}-{\Pt})({\Pt}-{\Po})(\Kp-{\Pf})^{2}}{(\Kp-{\Po})^{2}({\Pt}-{\Pf})}}\log
({\it H_s})+\frac{1}{4}{\frac {(\Kp-{\Pf})({\Pth}-{\Po})}{(\Kp-{\Po})({\Po}-{\Pf}){{p_{14}^2}}({\Pth}-{\Pf})}}{\Kftw}\Kttv\log ({\it H_s})\nn\\
&+&\frac{1}{8}{\frac
{({\Pth}-\Kp)(\Kp-{\Pf})({\Pt}-{\Po})}{(\Kp-{\Po})^{2}}}\log ({\it
H_d})-\frac{1}{4}{\frac {{\Pth}-\Kp}{(\Kp-{\Po})({\Pth}-{\Pf}){{p_{14}^2}}}}{\Kftw}\Kttv\log ({\it H_d})\nn\\
&-&\frac{1}{4}{\frac
{\Kp-{\Pf}}{{{p_{14}^2}}({\Po}-{\Pf})(\Kp-{\Po})}}{\Kofw}\Kttv\log
({\it H_u})-\frac{1}{8}{\frac {({\Pth}-{\Pf})({\Pth}-\Kp)({\Pt}-{\Po})}{(\Kp-{\Po})({\Pth}-{\Po})}}\log ({\it H_r})\nn\\
&+&\frac{1}{4}{\frac {{\Pth}-\Kp}{({\Pth}-{\Po})(\Kp-{\Po}){{p_{14}^2}}}}{\Kofw}\Kttv\log ({\it H_r})\nn\\
\nn\\
&&\rmid\nn\\&+&\frac{1}{8}{\frac
{(\Kp-{\Pf})(-\Kp+{\Pt})({\Pt}-{\Po})({\Pth}-{\Pf})}{({\Pt}-{\Pf})(\Kp-{\Po})^{2}}}\log
({\it H_s})+\frac{1}{4}{\frac {1}{({\Po}-{\Pf}){{p_{14}^2}}}}{\Kftw}\Kttv\log ({\it H_s})\nn\\
&+&\frac{1}{8}{\frac
{({\Pth}-{\Pf})({\Pth}-\Kp)({\Pt}-{\Po})}{(\Kp-{\Po})({\Pth}-{\Po})}}\log
({\it H_t})+\frac{1}{4}{\frac {1}{({\Po}-{\Pf}){{p_{14}^2}}}}{\Kftw}\Kttv\log ({\it H_t})\nn\\
&-&\frac{1}{8}{\frac
{(\Kp-{\Pf})({\Pth}-\Kp)({\Pt}-{\Po})}{(\Kp-{\Po})^{2}}}\log ({\it
H_d})+\frac{1}{4}{\frac {{\Pth}-\Kp}{(\Kp-{\Po})({\Pth}-{\Pf}){{p_{14}^2}}}}{\Kftw}\Kttv\log ({\it H_d})\nn\\
&-&\frac{1}{4}{\frac
{(\Kp-{\Pf})({\Pth}-{\Po})}{({\Po}-{\Pf}){{p_{14}^2}}(\Kp-{\Po})({\Pth}-{\Pf})}}{\Kftw}\Kttv\log
({\it H_l})-\frac{1}{4}{\frac {\Kp-{\Pf}}{(\Kp-{\Po}){{p_{14}^2}}({\Po}-{\Pf})}}{\Kofw}\Kttv\log ({\it H_u})\nn\\
&-&\frac{1}{8}{\frac
{({\Pth}-{\Pf})({\Pth}-\Kp)({\Pt}-{\Po})}{(\Kp-{\Po})({\Pth}-{\Po})}}\log
({\it H_r})+\frac{1}{4}{\frac {{\Pth}-\Kp}{({\Pth}-{\Po})(\Kp-{\Po}){{p_{14}^2}}}}{\Kofw}\Kttv\log ({\it H_r})\nn\\
\nn\\
&&\rright\nn\\
&+&\frac{1}{8}{\frac
{({\Pth}-{\Pf})(-\Kp+{\Pt})(\Kp-{\Pf})({\Pt}-{\Po})}{(\Kp-{\Po})^{2}({\Pt}-{\Pf})}}\log
({\it H_s})-\frac{1}{4}{\frac {-\Kp+{\Pt}}{{{p_{14}^2}}({\Pt}-{\Po})(\Kp-{\Po})}}{\Kofw}\Kttv\log ({\it H_s})\nn\\
&-&\frac{1}{4}{\frac
{1}{({\Pt}-{\Po}){{p_{14}^2}}}}{\Kofw}\Kftv\log ({\it
H_s})-\frac{1}{8}{\frac {({\Pth}-\Kp)(\Kp-{\Pf})({\Pt}-{\Po})}{(\Kp-{\Po})^{2}}}\log ({\it H_d})\nn\\
&+&\frac{1}{4}{\frac
{{\Pth}-\Kp}{(\Kp-{\Po})({\Pth}-{\Pf}){{p_{14}^2}}}}{\Kftw}\Kttv\log
({\it H_d})-\frac{1}{4}{\frac {(\Kp-{\Pf})({\Pth}-{\Po})}{{{p_{14}^2}}(\Kp-{\Po})({\Po}-{\Pf})({\Pth}-{\Pf})}}{\Kftw}\Kttv\log ({\it H_l})\nn\\
&+&\frac{1}{4}{\frac
{\Kp-{\Pf}}{({\Po}-{\Pf})(\Kp-{\Po}){{p_{14}^2}}}}{\Kofw}\Kttv\log
({\it H_u})\nn\eea

Again, we can integrate out whatever can be integrated out, and
sweep the rest underneath the infra red term rug. Thus we have:
\bea\pref
g^4\Nc\log{\frac{p_{12}^2}{p_{14}^2}}\left[-\frac{1}{8}{\frac
{({\Pf}-{\Pth})({\Pt}-{\Pth})({\Po}-{\Pt})}{{\Po}-{\Pth}}}-\frac{1}{4}{\frac
{{\Po}-{\Pt}}{({\Po}-{\Pth}){{p_{14}^2}}}}{\Kofw}\Kttv\right]\nn\eea

The combination of:\bea &&\frac{A}{2}\frac{K_{5,2}^\land
K_{3,2}^\lor
p_{14}^2}{q_0^2q_1^2q_2^2p_{14}^2}+\frac{A}{2}\frac{K_{5,2}^\land
K_{4,3}^\lor p_{14}^2}{q_0^2q_1^2q_3^2 p_{12}^2}\nn\\
&&+\frac{A}{2}\frac{K_{5,2}^\land
K_{3,5}^\lor}{(\Kp-\Pt)^2}\left[\frac{(-{\Pt}+{\Pth})(-{\Pt}+\Kp)}{q_0^2q_2^2q_3^2}+\frac{(-{\Po}+{\Pt})({\Pt}-\Kp)}{q_0^2q_1^2q_2^2}\right]\nn\\
&&+\frac{A}{2}\frac{K_{5,2}^\land
K_{3,5}^\lor}{(\Kp-\Pt)^2}\frac{(\Pt-\Pth)({\Kp}-\Po)+({\Po}-{\Pt})(\Pth-\Kp)}{q_3^2q_2^2q_1^2}-A\frac{K_{5,2}^\land
K_{3,5}^\lor}{(\Kp-\Pt)^2}\frac{K_{6,4}^{\lor}K_{1,6}^{\land}}{q_3^2q_0^2q_2^2q_1^2}\nn\eea
This is the rotation by 90 degrees clockwise of the previous one.
Its contribution is: 
\bea
&&\pref
g^4\Nc\times\nn\\
&&\rleft\nn\\
&-&\frac{1}{8}{\frac
{(-{\Pth}+{\Pf})(-{\Pf}+\Kp)({\Po}-{\Pt})}{(-{\Pt}+\Kp)(-{\Pf}+{\Pt})}}\log
({\it H_s})-\frac{1}{4}{\frac {-{\Pf}+\Kp}{(-{\Pt}+\Kp){{p_{12}^2}}(-{\Pth}+{\Pf})}}{\Kttw}\Kftv\log ({\it H_s})\nn\\
&+&\frac{1}{4}{\frac
{-{\Pf}+\Kp}{(-{\Pt}+\Kp){{p_{12}^2}}(-{\Pth}+{\Pf})}}{\Kttw}\Kftv\log
({\it H_d})+\frac{1}{8}{\frac {(-{\Pf}+{\Pt})(-{\Pth}+{\Po})(-{\Po}+\Kp)}{(-{\Pt}+\Kp)({\Po}-{\Pt})}}\log ({\it H_u})\nn\\
&-&\frac{1}{4}{\frac
{(-{\Po}+\Kp)(-{\Pf}+{\Pt})}{(-{\Pt}+\Kp)({\Po}-{\Pt}){{p_{12}^2}}(-{\Pth}+{\Pf})}}{\Kttw}\Kftv\log
({\it H_u})\nn\\
&-&\frac{1}{8}{\frac {({\Pt}-{\Pth})(-{\Po}+\Kp)(-{\Pf}+{\Po})}{({\Po}-{\Pt})(-{\Pt}+\Kp)}}\log ({\it H_r})\nn\\
&+&\frac{1}{4}{\frac
{(-{\Pf}+{\Pt})(-{\Po}+\Kp)}{(-{\Pt}+\Kp)({\Po}-{\Pt}){{p_{12}^2}}(-{\Pth}+{\Pf})}}{\Kttw}\Kftv\log
({\it H_r})\nn\\
&&\rmid\nn\\
\nn\\
&-&\frac{1}{8}{\frac
{(\Kp-{\Pf})({\Po}-{\Pt})(-{\Pth}+{\Pf})}{(-{\Pt}+\Kp)(-{\Pf}+{\Pt})}}\log
({\it H_s})+\frac{1}{8}{\frac {(-{\Pth}+{\Pf})(-{\Pth}+\Kp)(-{\Po}+\Kp)({\Po}-{\Pt})}{(-{\Pth}+{\Po})(-{\Pt}+\Kp)^{2}}}\log ({\it H_t})\nn\\
&-&\frac{1}{4}{\frac
{\Kp-{\Pf}}{(-{\Pth}+{\Pf}){{p_{12}^2}}(-{\Pt}+\Kp)}}{\Kttw}\Kftv\log
({\it H_d})-\frac{1}{8}{\frac {(\Kp-{\Pf})({\Po}-{\Pt})(-{\Pth}+\Kp)}{(-{\Pt}+\Kp)^{2}}}\log ({\it H_l})\nn\\
&+&\frac{1}{4}{\frac
{\Kp-{\Pf}}{(-{\Pth}+{\Pf}){{p_{12}^2}}(-{\Pt}+\Kp)}}{\Kttw}\Kftv\log
({\it H_l})+\frac{1}{8}{\frac {(-{\Pf}+{\Pt})(-{\Pth}+{\Po})(-{\Po}+\Kp)}{({\Po}-{\Pt})(-{\Pt}+\Kp)}}\log ({\it H_u})\nn\\
&-&\frac{1}{4}{\frac
{(-{\Pf}+{\Pt})(-{\Po}+\Kp)}{(-{\Pth}+{\Pf}){{p_{12}^2}}({\Po}-{\Pt})(-{\Pt}+\Kp)}}{\Kttw}\Kftv\log
({\it H_u})-\frac{1}{8}{\frac {({\Pt}-{\Pth})(-{\Po}+\Kp)(-{\Pf}+{\Po})}{({\Po}-{\Pt})(-{\Pt}+\Kp)}}\log ({\it H_r})\nn\\
&+&\frac{1}{4}{\frac
{(-{\Pf}+{\Pt})(-{\Po}+\Kp)}{(-{\Pth}+{\Pf}){{p_{12}^2}}({\Po}-{\Pt})(-{\Pt}+\Kp)}}{\Kttw}\Kftv\log
({\it H_r})\nn\\
&&\rright\nn\\
&+&\bigg[-\frac{1}{4}{\frac {(-{\Pf}+{\Po})(-{\Pth}+{\Pf})}{-{\Pf}+{\Pt}}}+\frac{1}{8}{\frac {({\Pt}-{\Pth})(-{\Pf}+{\Po})(-{\Pf}+\Kp)}{(-{\Pt}+\Kp)(-{\Pf}+{\Pt})}}\bigg]\log ({\it H_s})\nn\\
&-&\frac{1}{4}{\frac {{\Pt}{\Po}-2{\Pth}{\Po}-2{\Pt}\Kp+\Kp{\Pth}+{\Pth}{\Pt}+\Kp{\Po}}{({\Po}-{\Pt})(-{\Pt}+\Kp)({\Pt}-{\Pth}){{p_{12}^2}}}}{\Ktow}\Kftv\log ({\it H_s})\nn\\
&-&\frac{1}{4}{\frac
{(-{\Pf}+{\Pt})(-{\Pth}+\Kp)}{({\Pt}-{\Pth})(-{\Pt}+\Kp)(-{\Pth}+{\Pf}){{p_{12}^2}}}}{\Kttw}\Kftv\log
({\it H_s})-\frac{1}{4}{\frac {-{\Pf}+\Kp}{(-{\Pt}+\Kp){{p_{12}^2}}(-{\Pth}+{\Pf})}}{\Kttw}\Kftv\log ({\it H_d})\nn\\
&-&\frac{1}{8}{\frac
{(-{\Pf}+\Kp)({\Po}-{\Pt})(-{\Pth}+\Kp)}{(-{\Pt}+\Kp)^{2}}}\log
({\it H_l})+\frac{1}{4}{\frac {-{\Pf}+\Kp}{(-{\Pt}+\Kp){{p_{12}^2}}(-{\Pth}+{\Pf})}}{\Kttw}\Kftv\log ({\it H_l})\nn\\
&-&\frac{1}{8}{\frac
{(-{\Pf}+{\Pt})(-{\Pth}+{\Po})(-{\Po}+\Kp)}{(-{\Pt}+\Kp)({\Po}-{\Pt})}}\log
({\it H_u})\nn\\
&+&\frac{1}{4}{\frac
{(-{\Po}+\Kp)(-{\Pf}+{\Pt})}{({\Po}-{\Pt}){{p_{12}^2}}(-{\Pt}+\Kp)(-{\Pth}+{\Pf})}}{\Kttw}\Kftv\log
({\it H_u})\nn
\eea 
Its finite contribution is:
\bea\pref
g^4\Nc\log{\frac{p_{12}^2}{p_{14}^2}}\left[-\frac{1}{8}{\frac
{({\Pf}-{\Pth})({\Pt}-{\Pth})({\Po}-{\Pf})}{{\Pt}-{\Pf}}}+\frac{1}{4}{\frac
{{\Pt}-{\Pth}}{({\Pt}-{\Pf}){{p_{12}^2}}}}{\Ktow}\Kftv\right]\nn\eea
\\\\
The combination of:
\bea &&\frac{A}{2}\frac{K_{1,6}^\lor
K_{2,1}^\land
p_{12}^2}{q_3^2q_0^2q_1^2p_{12}^2}+\frac{A}{2}\frac{K_{1,6}^\lor
K_{3,2}^\land p_{12}^2}{q_3^2q_0^2q_2^2 p_{14}^2}\nn\\
&&-\frac{A}{2}\frac{K_{1,6}^\lor
K_{5,2}^\land}{(\Kp-\Pth)^2}\left[\frac{(-{\Pth}+{\Pf})(-{\Pth}+\Kp)}{q_3^2q_1^2q_2^2}+\frac{(-{\Pt}+{\Pth})({\Pth}-\Kp)}{q_3^2q_0^2q_1^2}\right]\nn\\
&&-\frac{A}{2}\frac{K_{1,6}^\lor
K_{5,2}^\land}{(\Kp-\Pth)^2}\frac{(\Pth-\Pf)({\Kp}-\Pt)+({\Pt}-{\Pth})(\Pf-\Kp)}{q_2^2q_1^2q_0^2}-A\frac{K_{1,6}^\lor
K_{5,2}^\land}{(\Kp-\Pth)^2}\frac{K_{3,5}^{\land}K_{6,4}^{\lor}}{q_2^2q_3^2q_1^2q_0^2}\nn\eea
gives: 
\bea &&A\pref g^4\Nc\times\nn\\
&&\rleft\nn\\
&-&\frac{1}{8}{\frac
{(-{\Pf}+\Kp)({\Po}-{\Pt})(-{\Pth}+{\Pf})(-{\Pt}+\Kp)}{(-{\Pf}+{\Pt})(-{\Pth}+\Kp)^{2}}}\log
({\it H_s})\nn\\
&-&\frac{1}{4}{\frac {-{\Pf}+\Kp}{(-{\Pth}+\Kp)(-{\Pth}+{\Pf}){{p_{14}^2}}}}{\Kttw}\Kofv\log ({\it H_s})\nn\\
&+&\frac{1}{4}{\frac
{1}{{{p_{14}^2}}(-{\Pth}+{\Pf})}}{\Kttw}\Ktov\log
({\it H_s})+\frac{1}{8}{\frac {(-{\Pf}+{\Po})(\Kp-{\Pt})}{-{\Pth}+\Kp}}\log ({\it H_d})\nn\\
&+&\frac{1}{4}{\frac
{(-{\Pth}+{\Po})(\Kp-{\Pt})}{({\Pt}-{\Pth}){{p_{14}^2}}(-{\Pth}+{\Pf})(-{\Pth}+\Kp)}}{\Kttw}\Kftv\log
({\it H_d})-\frac{1}{8}{\frac {(-{\Po}+\Kp)(-{\Pth}+{\Pf})(\Kp-{\Pt})}{(-{\Pth}+\Kp)^{2}}}\log ({\it H_u})\nn\\
&+&\frac{1}{4}{\frac
{-{\Po}+\Kp}{{{p_{14}^2}}(-{\Pth}+{\Pf})(-{\Pth}+\Kp)}}{\Kttw}\Kftv\log
({\it H_u})-\frac{1}{4}{\frac {(-{\Pt}+\Kp)(-{\Pth}+{\Po})}{{{p_{14}^2}}(-{\Pth}+\Kp)({\Pt}-{\Pth})(-{\Pth}+{\Pf})}}{\Kttw}\Kftv\log ({\it H_r})\nn\\
\nn\\
&&\rmid\nn\\
&-&\frac{1}{8}{\frac
{(-{\Pth}+{\Pf})(\Kp-{\Pf})(-{\Pt}+\Kp)({\Po}-{\Pt})}{(-{\Pf}+{\Pt})(-{\Pth}+\Kp)^{2}}}\log
({\it H_s})+\frac{1}{4}{\frac {{\Po}-{\Pt}}{(-{\Pth}+{\Pf}){{p_{14}^2}}({\Pt}-{\Pth})}}{\Kttw}\Kftv\log ({\it H_s})\nn\\
&+&\frac{1}{8}{\frac
{({\Po}-{\Pt})(-{\Pth}+{\Pf})(-{\Po}+\Kp)}{(-{\Pth}+{\Po})(-{\Pth}+\Kp)}}\log
({\it H_t})+\frac{1}{4}{\frac {{\Po}-{\Pt}}{(-{\Pth}+{\Pf}){{p_{14}^2}}({\Pt}-{\Pth})}}{\Kttw}\Kftv\log ({\it H_t})\nn\\
&-&\frac{1}{8}{\frac
{(-{\Pf}+{\Po})(-{\Pt}+\Kp)}{-{\Pth}+\Kp}}\log ({\it
H_d})-\frac{1}{4}{\frac {(-{\Pt}+\Kp)(-{\Pth}+{\Po})}{(-{\Pth}+\Kp){{p_{14}^2}}({\Pt}-{\Pth})(-{\Pth}+{\Pf})}}{\Kttw}\Kftv\log ({\it H_d})\nn\\
&+&\frac{1}{8}{\frac
{(-{\Pf}+{\Pt})(-{\Po}+\Kp)}{-{\Pth}+\Kp}}\log ({\it
H_l})+\frac{1}{4}{\frac {-{\Po}+\Kp}{{{p_{14}^2}}(-{\Pth}+{\Pf})(-{\Pth}+\Kp)}}{\Kttw}\Kftv\log ({\it H_l})\nn\\
&-&\frac{1}{8}{\frac
{(-{\Po}+\Kp)(-{\Pth}+{\Pf})(-{\Pt}+\Kp)}{(-{\Pth}+\Kp)^{2}}}\log
({\it H_u})+\frac{1}{4}{\frac {-{\Po}+\Kp}{{{p_{14}^2}}(-{\Pth}+{\Pf})(-{\Pth}+\Kp)}}{\Kttw}\Kftv\log ({\it H_u})\nn\\
&-&\frac{1}{4}{\frac
{(-{\Pt}+\Kp)(-{\Pth}+{\Po})}{(-{\Pth}+\Kp){{p_{14}^2}}({\Pt}-{\Pth})(-{\Pth}+{\Pf})}}{\Kttw}\Kftv\log
({\it H_r})\nn\\
\nn\\
&&\rright\nn\\
&-&\frac{1}{8}{\frac
{(-{\Pth}+{\Pf})(-{\Pf}+{\Po})(-{\Pt}+\Kp)^{2}}{(-{\Pth}+\Kp)^{2}(-{\Pf}+{\Pt})}}\log
({\it H_s})\nn\\
&+&\frac{1}{4}{\frac {(-{\Pt}+\Kp)(-{\Pth}+{\Po})}{({\Pt}-{\Pth}){{p_{14}^2}}(-{\Pth}+\Kp)(-{\Pth}+{\Pf})}}{\Kttw}\Kftv\log ({\it H_s})\nn\\
&-&\frac{1}{8}{\frac
{(-{\Pf}+{\Po})(-{\Pt}+\Kp)}{-{\Pth}+\Kp}}\log ({\it
H_d})-\frac{1}{4}{\frac {(-{\Pt}+\Kp)(-{\Pth}+{\Po})}{({\Pt}-{\Pth}){{p_{14}^2}}(-{\Pth}+\Kp)(-{\Pth}+{\Pf})}}{\Kttw}\Kftv\log ({\it H_d})\nn\\
&+&\frac{1}{8}{\frac
{(-{\Pf}+{\Pt})(-{\Po}+\Kp)}{-{\Pth}+\Kp}}\log ({\it
H_l})+\frac{1}{4}{\frac {-{\Po}+\Kp}{(-{\Pth}+\Kp)(-{\Pth}+{\Pf}){{p_{14}^2}}}}{\Kttw}\Kftv\log ({\it H_l})\nn\\
&+&\frac{1}{8}{\frac
{(-{\Po}+\Kp)(-{\Pth}+{\Pf})(-{\Pt}+\Kp)}{(-{\Pth}+\Kp)^{2}}}\log
({\it H_u})-\frac{1}{4}{\frac
{-{\Po}+\Kp}{(-{\Pth}+\Kp)(-{\Pth}+{\Pf}){{p_{14}^2}}}}{\Kttw}\Kftv\log
({\it H_u})\nn\eea 
Its finite contribution is:\bea\pref
g^4\Nc\log{\frac{p_{12}^2}{p_{14}^2}}\left[-\frac{1}{8}{\frac
{({\Pf}-{\Pth})({\Po}-{\Pf})({\Po}-{\Pt})}{{\Po}-{\Pth}}}-\frac{1}{4}{\frac
{{\Pf}-{\Pth}}{({\Po}-{\Pth}){{p_{14}^2}}}}{\Kttw}\Kofv
\right]\nn\eea
\\\\
The combination of: \bea &&-\frac{A}{2}\frac{K_{6,4}^\land
K_{1,4}^\lor
p_{14}^2}{q_2^2q_3^2q_0^2p_{14}^2}-\frac{A}{2}\frac{K_{6,4}^\land
K_{2,1}^\lor p_{14}^2}{q_2^2q_3^2q_1^2p_{12}^2}\nn\\
&&+\frac{A}{2}\frac{K_{6,4}^\land
K_{1,6}^\lor}{(\Kp-\Pf)^2}\left[\frac{(-{\Pf}+{\Po})(-{\Pf}+\Kp)}{q_2^2q_0^2q_1^2}+\frac{(-{\Pth}+{\Pf})({\Pf}-\Kp)}{q_2^2q_3^2q_0^2}\right]\nn\\
&&+\frac{A}{2}\frac{K_{6,4}^\land
K_{1,6}^\lor}{(\Kp-\Pf)^2}\frac{(\Pf+\Po)({\Kp}-\Pth)+({\Pth}-{\Pf})(\Kp-\Pth)}{q_1^2q_0^2q_3^2}-A\frac{K_{6,4}^\land
K_{1,6}^\lor}{(\Kp-\Pf)^2}\frac{K_{5,2}^{\lor}K_{3,5}^{\land}}{q_1^2q_2^2q_0^2q_1^2}\nn\eea
gives: \bea &&A\pref g^4\Nc\times\nn\\
&&\rleft\nn\\
&+&\bigg[-\frac{1}{4}{\frac {({\Pt}-{\Pth})({\Po}-{\Pt})}{{\Pt}-{\Pf}}}+\frac{1}{8}{\frac {({\Po}-{\Pf})({\Pt}-{\Pth})({\Pt}-\Kp)}{({\Pf}-\Kp)({\Pt}-{\Pf})}}\bigg]\log ({\it H_s})\nn\\
&+&\frac{1}{4}{\frac
{1}{({\Po}-{\Pf}){{p_{12}^2}}}}{\Kofw}\Ktov\log ({\it
H_s})-\frac{1}{4}{\frac {{\Po}{\Pf}-2{\Po}{\Pth}-2\Kp{\Pf}+\Kp{\Pth}+\Kp{\Po}+{\Pf}{\Pth}}{({\Po}-{\Pf})({\Pf}-\Kp)({\Pf}-{\Pth}){{p_{12}^2}}}}{\Kftw}\Ktov\log ({\it H_s})\nn\\
&-&\frac{1}{4}{\frac
{{\Pt}-\Kp}{({\Pf}-\Kp)({\Pf}-{\Pth}){{p_{12}^2}}}}{\Kftw}\Kttv\log
({\it H_s})-\frac{1}{8}{\frac {({\Pt}-{\Pf})({\Po}-{\Pth})(-\Kp+{\Pth})}{({\Pf}-{\Pth})({\Pf}-\Kp)}}\log ({\it H_d})\nn\\
&+&\frac{1}{4}{\frac
{({\Pt}-{\Pf})(-\Kp+{\Pth})}{({\Pf}-\Kp){{p_{12}^2}}({\Pf}-{\Pth})^{2}}}{\Kftw}\Kttv\log
({\it H_d})-\frac{1}{8}{\frac {({\Pt}-{\Pth})({\Po}-{\Pf})({\Pt}-\Kp)}{({\Pf}-\Kp)({\Pt}-{\Pf})}}\log ({\it H_u})\nn\\
&+&\frac{1}{4}{\frac
{{\Pt}-\Kp}{({\Pt}-{\Pf})({\Pf}-\Kp){{p_{12}^2}}}}{\Kftw}\Ktov\log
({\it H_u})-\frac{1}{8}{\frac {({\Pf}-{\Pth})({\Pt}-\Kp)({\Po}-\Kp)}{({\Pf}-\Kp)^{2}}}\log ({\it H_r})\nn\\
&+&\frac{1}{4}{\frac
{{\Pt}-\Kp}{({\Po}-{\Pt})({\Pf}-\Kp){{p_{12}^2}}}}{\Kofw}\Ktov\log
({\it H_r})\nn\\
&&\rmid\nn\\
&-&\frac{1}{8}{\frac
{({\Pt}-\Kp)({\Pf}-{\Pth})({\Po}-{\Pt})}{({\Pf}-\Kp)({\Pt}-{\Pf})}}\log
({\it H_s})+\frac{1}{8}{\frac {({\Pth}-\Kp)({\Pf}-{\Pth})(-\Kp+{\Po})({\Po}-{\Pt})}{({\Po}-{\Pth})({\Pf}-\Kp)^{2}}}\log ({\it H_t})\nn\\
&+&\frac{1}{8}{\frac
{({\Pt}-{\Pf})({\Po}-{\Pth})({\Pth}-\Kp)}{({\Pf}-{\Pth})({\Pf}-\Kp)}}\log
({\it H_d})-\frac{1}{4}{\frac {({\Pth}-\Kp)({\Pt}-{\Pf})}{{{p_{12}^2}}({\Pf}-{\Pth})^{2}({\Pf}-\Kp)}}{\Kftw}\Kttv\log ({\it H_d})\nn\\
&-&\frac{1}{4}{\frac
{{\Pth}-\Kp}{{{p_{12}^2}}({\Pf}-\Kp)({\Pf}-{\Pth})}}{\Kftw}\Ktov\log
({\it H_l})-\frac{1}{8}{\frac {({\Pt}-\Kp)({\Pt}-{\Pth})({\Po}-{\Pf})}{({\Pf}-\Kp)({\Pt}-{\Pf})}}\log ({\it H_u})\nn\\
&+&\frac{1}{4}{\frac
{{\Pt}-\Kp}{({\Pt}-{\Pf})({\Pf}-\Kp){{p_{12}^2}}}}{\Kftw}\Ktov\log
({\it H_u})-\frac{1}{8}{\frac {({\Pf}-{\Pth})({\Pt}-\Kp)(-\Kp+{\Po})}{({\Pf}-\Kp)^{2}}}\log ({\it H_r})\nn\\
&+&\frac{1}{4}{\frac {{\Pt}-\Kp}{({\Po}-{\Pt})({\Pf}-\Kp){{p_{12}^2}}}}{\Kofw}\Ktov\log ({\it H_r})\nn\\
\nn\\
&&\rright\nn\\
&-&\frac{1}{8}{\frac
{({\Po}-{\Pt})({\Pf}-{\Pth})({\Pt}-\Kp)}{({\Pt}-{\Pf})({\Pf}-\Kp)}}\log
({\it H_s})-\frac{1}{4}{\frac {{\Pt}-\Kp}{({\Pf}-\Kp)({\Pf}-{\Pth}){{p_{12}^2}}}}{\Kftw}\Kttv\log ({\it H_s})\nn\\
&+&\frac{1}{8}{\frac
{({\Pt}-{\Pf})({\Po}-{\Pth})({\Pth}-\Kp)}{({\Pf}-{\Pth})({\Pf}-\Kp)}}\log
({\it H_d})-\frac{1}{4}{\frac
{({\Pt}-{\Pf})({\Pth}-\Kp)}{({\Pf}-\Kp)({\Pf}-{\Pth})^{2}{{p_{12}^2}}}}{\Kftw}\Kttv\log
({\it H_d})\nn\\
&-&\frac{1}{4}{\frac
{{\Pth}-\Kp}{({\Pf}-\Kp)({\Pf}-{\Pth}){{p_{12}^2}}}}{\Kftw}\Ktov\log
({\it H_l})+\frac{1}{8}{\frac
{({\Pt}-\Kp)({\Pt}-{\Pth})({\Po}-{\Pf})}{({\Pf}-\Kp)({\Pt}-{\Pf})}}\log
({\it H_u})\nn\\
&-&\frac{1}{4}{\frac
{{\Pt}-\Kp}{({\Pt}-{\Pf})({\Pf}-\Kp){{p_{12}^2}}}}{\Kftw}\Ktov\log
({\it H_u})\nn\eea Its finite contribution is:\bea\pref
g^4\Nc\log{\frac{p_{12}^2}{p_{14}^2}}\left[-\frac{1}{8}{\frac
{({\Pt}-{\Pth})({\Po}-{\Pt})({\Po}-{\Pf})}{{\Pt}-{\Pf}}}+\frac{1}{4}{\frac
{{\Po}-{\Pf}}{({\Pt}-{\Pf}){{p_{12}^2}}}}{\Kftw}\Ktov\right]\nn\eea

\end{document}